\documentclass[runningheads]{llncs}

\usepackage{booktabs}   
 
\usepackage{graphicx} 
\usepackage{verbatim} 
\usepackage{algorithm}
\usepackage[noend]{algpseudocode}
\usepackage{amsmath}
\usepackage{amssymb}
\usepackage{listings}
\usepackage[normalem]{ulem}
\usepackage{url}
\usepackage{xcolor}
\usepackage{marvosym}

\makeatletter
\RequirePackage[bookmarks,unicode,colorlinks=true]{hyperref}%
\def\@citecolor{blue}%
\def\@urlcolor{blue}%
\def\@linkcolor{blue}%

\def\orcidID#1{\smash{\href{http://orcid.org/#1}{\protect\raisebox{-1.25pt}{\protect\includegraphics{orcid_color.eps}}}}}
\makeatother

\newcommand{\supplypd}[2]{\ensuremath{\mathit{supply}({#1}, {#2})}} 

\newcommand{\vcfgtrans}[2]{\xrightarrow{#1,#2}}
\newcommand{\vstatetrans}{\Rightarrow}
\newcommand{\natzero}{\mathbb{N}}

\newcommand{\vcfggraph}{\ensuremath{\mathcal{G}}}

\newcommand{\callsite}[1]{\ensuremath{\textit{call-sites}\:(#1)}}

\newcommand{\entrynode}[1]{\ensuremath{\textit{en}_{#1}}}
\newcommand{\exitnode}[1]{\ensuremath{\textit{ex}_{#1}}}
\newcommand{\main}{\texttt{main }}

\newcommand{\sIVCPathsName}{\ensuremath{\textit{sIVCPaths}}}
\newcommand{\sIVCPaths}[2]{\ensuremath{\sIVCPathsName({#1},{#2})}}
\newcommand{\funcs}{\ensuremath{\mathit{Funcs}}}

\newcommand{\ptf}[1]{\ensuremath{\mathit{ptf}({#1})}} 
\newcommand{\demandp}[1]{\ensuremath{\mathit{demand}({#1})}} 
\newcommand{\demandpd}[2]{\ensuremath{\mathit{demand}({#1}, {#2})}} 
\newcommand{\join}{\ensuremath{\sqcup}}
\newcommand{\bigjoin}{\ensuremath{\bigsqcup}}
\newcommand{\dlattice}{\ensuremath{\mathcal L}}

\newcommand{\dominates}{\ensuremath{\sqsupseteq}}

\newcommand{\nodes}{\ensuremath{\mathit{Nodes}}}

\newcommand{\send}[2]{\ensuremath{\mathrm{#1}\,!\,\mathrm{#2}}}
\newcommand{\recv}[2]{\ensuremath{\mathrm{#1}\,?\,\mathrm{#2}}}

\newcommand{\sPathsName}{\ensuremath{\textit{sPaths}}}
\newcommand{\sPaths}[1]{\ensuremath{\sPathsName({#1}})}
\newcommand{\worklist}{\ensuremath{\mathit{workList}}}

\newcommand{\startnode}{\ensuremath{\mathit{start}}}
\newcommand{\targetnode}{\ensuremath{\mathit{target}}}
\newcommand{\fls}{\ensuremath{\mathit{false}}}
\newcommand{\tru}{\ensuremath{\mathit{true}}}
\newcommand{\concat}{\ensuremath{.}}
\newcommand{\zerovector}{\ensuremath{\overline{0}}}

\newcommand{\forwthresh}{\kappa}

\newcommand{\widen}{\triangledown}
\newcommand{\boundedmovei}{\mathit{boundedMove1}}
\newcommand{\boundedmove}{\mathit{boundedMove}}

\newcommand{\fun}{\mathit{fun}}
\newcommand{\funconc}{\mathit{fun\_conc}}

\newcommand{\vpath}[1]{\ensuremath{\mathit{#1}}}
\newcommand{\leaderMesg}[2]{\ensuremath{\langle {#1},{#2} \rangle}}

\newcommand{\ol}[1]{\ensuremath{\overline{#1}}}
\newcommand{\Path}[1]{\ensuremath{\stackrel{#1}{\longrightarrow}}}
\newcommand{\Move}[1]{\ensuremath{\stackrel{#1}{\rightarrow}}}
\newcommand{\below}{\ensuremath{\leq}}
\newcommand{\latbel}{\ensuremath{\sqsubseteq}}
\newcommand{\latabv}{\ensuremath{\sqsupseteq}}
\newcommand{\dist}{\ensuremath{dist}}
\newcommand{\length}{\ensuremath{\ell}}
\newcommand{\elength}{\ensuremath{e\mathbb{\ell}}}
\newcommand{\lbound}{\ensuremath{\mathbf{L}}}
\newcommand{\jop}{\ensuremath{\mathrm{JOP}}}

\newcommand{\compComp}{\ensuremath{\mathbf{A}}}
\newcommand{\compJoin}{\ensuremath{\mathrm{B}}}
\newcommand{\compCompare}{\ensuremath{\mathrm{C}}}

\newcommand{\cover}[1]{\ensuremath{\mathit{Cover(#1)}}}
\newcommand{\stmtno}[1]{\ensuremath{\hspace{15pt} - \hspace{3pt} (#1)}}

\begin{document}

\title{Data Flow Analysis of Asynchronous Systems using Infinite Abstract Domains}        

\author{Snigdha Athaiya(\Letter)\inst{1} \and Raghavan Komondoor\inst{1}  \and K. Narayan Kumar\inst{2} }

\institute{Indian Institute of Science, Bengaluru, India \\ 
	\email{\{snigdha,raghavan\}@iisc.ac.in} \and
	Chennai Mathematical Institute, Chennai, India \\
	\email{kumar@cmi.ac.in}}
%
\titlerunning{Data Flow Analysis of Async. Systems using Inf. Abstract Domains}
%
%

\authorrunning{Athaiya S. et al.}

\maketitle              

\begin{abstract}
 Asynchronous message-passing systems are employed frequently to implement
 distributed mechanisms, protocols, and processes. This paper addresses the
 problem of precise data flow analysis for such systems. To obtain good
 precision, data flow analysis needs to somehow skip execution paths that
 read more messages than the number of messages sent so far in the path, as
 such paths are infeasible at run time. Existing data flow analysis
 techniques do elide a subset of such infeasible paths, but have the
 restriction that they admit only finite abstract analysis domains. In this
 paper we propose a generalization of these approaches to admit infinite
 abstract analysis domains, as such domains are commonly used in practice
 to obtain high precision. We have implemented our approach, and have
 analyzed its performance on a set of 14 benchmarks. On these benchmarks
 our tool obtains significantly higher precision compared to a baseline
 approach that does not elide any infeasible paths and to another 
 baseline that elides infeasible paths but admits only finite abstract domains.

\keywords{Data Flow Analysis \and Message-passing systems.}
\end{abstract}

\section{Introduction}
\label{sec:intro}

Distributed software that communicates by asynchronous message passing is a
very important software paradigm in today's world. It is employed in
varied domains, such as distributed protocols and workflows, event-driven systems,
and UI-based systems.
Popular languages used in this domain include
Go ({\small \url{https://golang.org/}}),
Akka ({\small \url{https://akka.io/}}), and
P ({\small \url{https://github.com/p-org}}).

Analysis and verification of asynchronous systems is an important problem,
and poses a rich set of
challenges. The research community has focused historically on a variety of
approaches to tackle this overall problem, such as 
model checking and systematic concurrency
testing~\cite{holzmann1997model,deligiannis2015asynchronous}, formal
verification to check properties such as reachability or coverability of states~\cite{lynch1996distributed,abdulla1996verifying,abdulla1996general,geeraerts2006expand,finkel2001well,karp1969parallel,ganty2009verifying,abdulla1998fly}, and data flow
analysis~\cite{jhala2007interprocedural}.

Data flow analysis~\cite{kildall1973unified,kam1977monotone} is a specific
type of verification technique that propagates values from an
\emph{abstract domain} while accounting for all paths in a program. It can
hence be used to check whether a  property or assertion always holds. 
The existing verification and data flow analysis approaches mentioned earlier
have a major limitation, which is that they admit only finite abstract
domains.  This, in general, limits the classes of properties that can be
successfully verified. On the other hand, data flow analysis of sequential
programs using infinite abstract domains, e.g., \emph{constant
  propagation}~\cite{kildall1973unified}, \emph{interval
  analysis}~\cite{cousot1977abstract}, and
\emph{octagons}~\cite{mine2006octagon}, is a well developed area, and is
routinely employed in verification settings.  In this paper we seek to
bridge this fundamental gap, and develop a precise data flow analysis
framework for message-passing asynchronous systems that
admits infinite abstract domains.

\subsection{Motivating Example: Leader election}
\label{ssec:motivatingExample}

\begin{figure}
  \centering
  \input{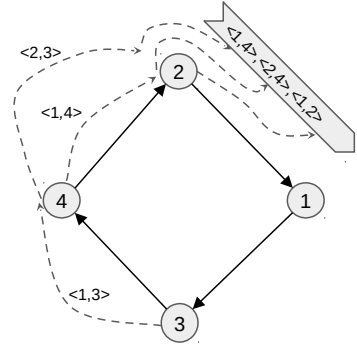}  
\caption{Pseudo-code of each process in leader election, and a partial run} 
\label{fig:introExample}
\end{figure}

To motivate our work we use a benchmark program\footnote{file
  \texttt{assertion.leader.prm} in
  \url{www.imm.dtu.dk/~albl/promela-models.zip}.} in the \emph{Promela}
language~\cite{holzmann1997model} that implements a \emph{leader election}
protocol~\cite{dolev1982n}. In the protocol there is a ring of processes,
and each process has a unique number. The objective is to discover the
``leader'', which is the process with the maximum number. The pseudo-code
of each process in the protocol is shown in the left side of
Figure~\ref{fig:introExample}. Each process has its own copy of local
variables \emph{max} and \emph{left}, whereas nr\_leaders is a global
variable that is common to all the processes (its initial value is
zero). Each process sends messages to the next process in the ring via an
unbounded FIFO channel. Each process becomes ``ready'' whenever a message
is available for it to receive, and at any step of the protocol any one
ready process (chosen non-deterministically) executes one iteration of its
``while'' loop. (We formalize these execution rules in a more general
fashion in Section~\ref{ssec:proc-to-vcfg}.) The messages are a 2-tuple $\langle x,i \rangle$, where $x$ can be 1 or 2, and $ 1 \leq i \leq \mathit{max}$. The right side of
Figure~\ref{fig:introExample} shows a snapshot at an intermediate point
during a run of the protocol. Each dashed arrow between two nodes
represents a send of a
message and a (completed) receipt of the same message. The block arrow
depicts the channel from Process~2 to Process~1, which happens to contain
three sent (but still unreceived) messages.

It is notable that in any run of the protocol, Lines~10-11 happen to get executed
only by the actual leader process, and that
too, exactly once. Hence, the assertion never fails. The argument for
this claim is not straightforward, and we refer the reader to the
paper~\cite{dolev1982n} for the details.

\subsection{Challenges in property checking}
\label{ssec:intro:challenges}

Data flow analysis could be used to verify the assertion in the example
above, e.g., using the \emph{Constant Propagation} (CP) abstract
domain. This analysis determines at each program point whether each
variable has a fixed value, and if yes, the value itself, across all runs
that reach the point.  In the example in Figure~\ref{fig:introExample}, all
actual runs of the system that happen to reach Line~10 come there with
value zero for the global variable nr\_leaders.

 A challenge for data flow
analysis on message-passing systems is that there may exist \emph{infeasible} paths in the
system. These are paths with more receives of a certain message than the
number of copies of this message that have been sent so far. For instance,
consider the path that consists of two back-to-back iterations of the
``while'' loop by the leader process, both times through
Lines~3,6,9-11. This path is not feasible, due to the impossibility of
having two copies of the message $\leaderMesg{1}{\mathit{max}}$ in the
input channel~\cite{dolev1982n}. The second iteration would bring the value 1 for nr\_leaders
at Line~10, thus inferring a non-constant value and hence declaring the
assertion as failing (which would be a false positive).

Hence, it is imperative in the interest of precision for any data flow
analysis or verification approach to track the channel contents as part of
the exploration of the state space. Tracking the contents of unbounded
channels precisely is known to be undecidable even when solving problems
such as reachability and coverability (which are simpler than data flow
analysis).  Hence, existing approaches either bound the channels (which in
general causes unsoundness), or use sound abstractions such as
\emph{unordered channels} (also known as the Petri Net or VASS abstraction)
or \emph{lossy channels}. Such abstractions suffice to elide a subset of
infeasible paths. In our running example, the unordered channel abstraction
happens to suffice to elide infeasible paths that could contribute to a
false positive at the point of the assertion. However, the analysis would
need to use an abstract domain such as CP to track the values of integer
variables. This is an infinite domain (due to the infinite number of
integers). The most closely related previous dataflow analysis approach for
distributed systems~\cite{jhala2007interprocedural} does use the unordered
channel abstraction, but does not admit infinite abstract domains,
and hence cannot verify assertions such as the one in the example above.

\subsection{Our Contributions}

 This paper is the first one to the best of our knowledge to propose an
 approach for data flow analysis for asynchronous message-passing systems
 that (a) admits infinite abstract domains, (b) uses a reasonably precise
 channel abstraction among the ones known in the literature (namely, the
 unordered channels abstraction), and (c) computes maximally precise
 results possible under the selected channel abstraction. Every other
 approach we are aware of exhibits a strict subset of the three attributes
 listed above. It is notable that previous approaches  do tackle the
 infinite state space induced by the unbounded channel contents. However,  they either do not
 reason about variable values at all, or only allow variables that are based on
 finite domains. 

 Our primary contribution is an approach that we call \emph{Backward
   DFAS}. This approach is maximally precise, and admits a class of infinite
 abstract domains. This class includes well-known examples such as Linear
 Constant Propagation (LCP)~\cite{sagiv1996precise} and Affine
 Relationships Analysis (ARA)~\cite{muller2004precise}, but does not
 include the full  (CP) analysis. We also propose
 another approach, which we call \emph{Forward DFAS}, which admits a
 broader class of abstract domains, but is not 
 guaranteed to be maximally precise on all programs.

 We describe a prototype implementation of both our approaches. On a set of
 14 real benchmarks, which are small but involve many complex idioms and
 paths, our tool verifies approximately 50\% more assertions than our
 implementation of the baseline approach~\cite{jhala2007interprocedural}.

The rest of the paper is structured as follows. Section~\ref{sec:prelim}
covers the background and notation that will be assumed throughout the
paper. We present the Backward DFAS approach in Section~\ref{sec:backward},
and the Forward DFAS approach in Section~\ref{sec:forward}.
Section~\ref{sec:impl-and-eval}
discusses our implementation and evaluation. Section~\ref{sec:relwork}
discusses related work, and Section~\ref{sec:concl} concludes the paper.

\section{Background and Terminology}
\label{sec:prelim}



Vector addition systems with states or VASS~\cite{hopcroft1979reachability}
are a  popular  modelling technique for  distributed
systems. We begin this section by defining an extension to VASS, which we call  a
\emph{VASS-Control Flow Graph} or \emph{VCFG}. 

\begin{definition}
	\label{def:vcfg}
A \emph{VASS-Control Flow Graph} or VCFG $\vcfggraph$ is a graph, and is described by the tuple $\langle Q, \delta, r, q_0, V, \pi, \theta \rangle$, where\\
$Q$ is a finite set of nodes,
$\delta \subseteq Q \times Q$ is a finite set of edges,\\
$r \in \natzero$, 
$q_0$ is the \emph{start} node, 
$V$ is a set of $variables$ or memory locations,\\
$\pi : \delta \rightarrow A$ maps each edge to an \emph{action}, where $A \equiv  ((V \rightarrow \mathbb{Z}) \rightarrow (V \rightarrow \mathbb{Z}))$,\\
$\theta : \delta \rightarrow \mathbb{Z}^r$ maps each edge to a vector in $\mathbb{Z}^r$. 
\end{definition}

For any edge $e = (q_1,q_2) \in \delta$, if $\pi(e)
= a$ and $\theta(e) = w$, then $a$ is called the \emph{action} of  $e$ and
$w$ is called the \emph{queuing vector} of $e$. This edge is depicted as
$q_1 \vcfgtrans{a}{w} q_2$. The variables and the \emph{actions} are the
only additional features of a VCFG over VASS.

A \emph{configuration} of a VCFG is a tuple $\langle q, c, \xi \rangle$,
where $q \in Q$, $c \in \natzero^r$ and $\xi \in (V \rightarrow
\mathbb{Z})$. The initial configuration of a VCFG is $\langle q_0, \vec{0},
\xi_0 \rangle$, where $\vec{0}$ denotes a vector with $r$ zeroes, and
$\xi_0$ is a given initial valuation for the variables. The VCFG can be said
to have $r$ \emph{counters}. The vector $c$ in each configuration
can be thought of as a valuation to the counters.  The transitions between
VCFG configurations are according to the rule below:\\[0.5em]

\noindent \begin{tabular}{c}
\begin{math}
e= (q_1, q_2), \hspace{4pt} e \in \delta, \hspace{4pt} \pi(e) =
a, \hspace{4pt} \theta(e) = w, \hspace{4pt} a(\xi_1) =
\xi_2, \hspace{4pt} c_1 + w = c_2, \hspace{4pt} c_2 \geq \vec{0}
\end{math}\rule[-0.25em]{0in}{0in}\\ \hline
\rule[0.25em]{0in}{0in}
\begin{math}
\langle q_1, c_1, \xi_1\rangle \vstatetrans_e \langle q_2, c_2, \xi_2 \rangle
\end{math}
\rule[-0.25em]{0in}{0in}
\end{tabular}




\subsection{Modeling of Asynchronous Message Passing Systems  as VCFGs}
\label{ssec:proc-to-vcfg}
Asynchronous systems are composed of finite number of independently executing processes that
communicate with each other by passing messages along FIFO channels.  The
processes may have local variables, and there may exist shared (or global)
variables as well. For simplicity of presentation we assume all variables
are global.

\begin{figure}
\centering
\begin{tabular}{lr}
\includegraphics[trim = 200 50 200 70, clip,  scale=0.3]{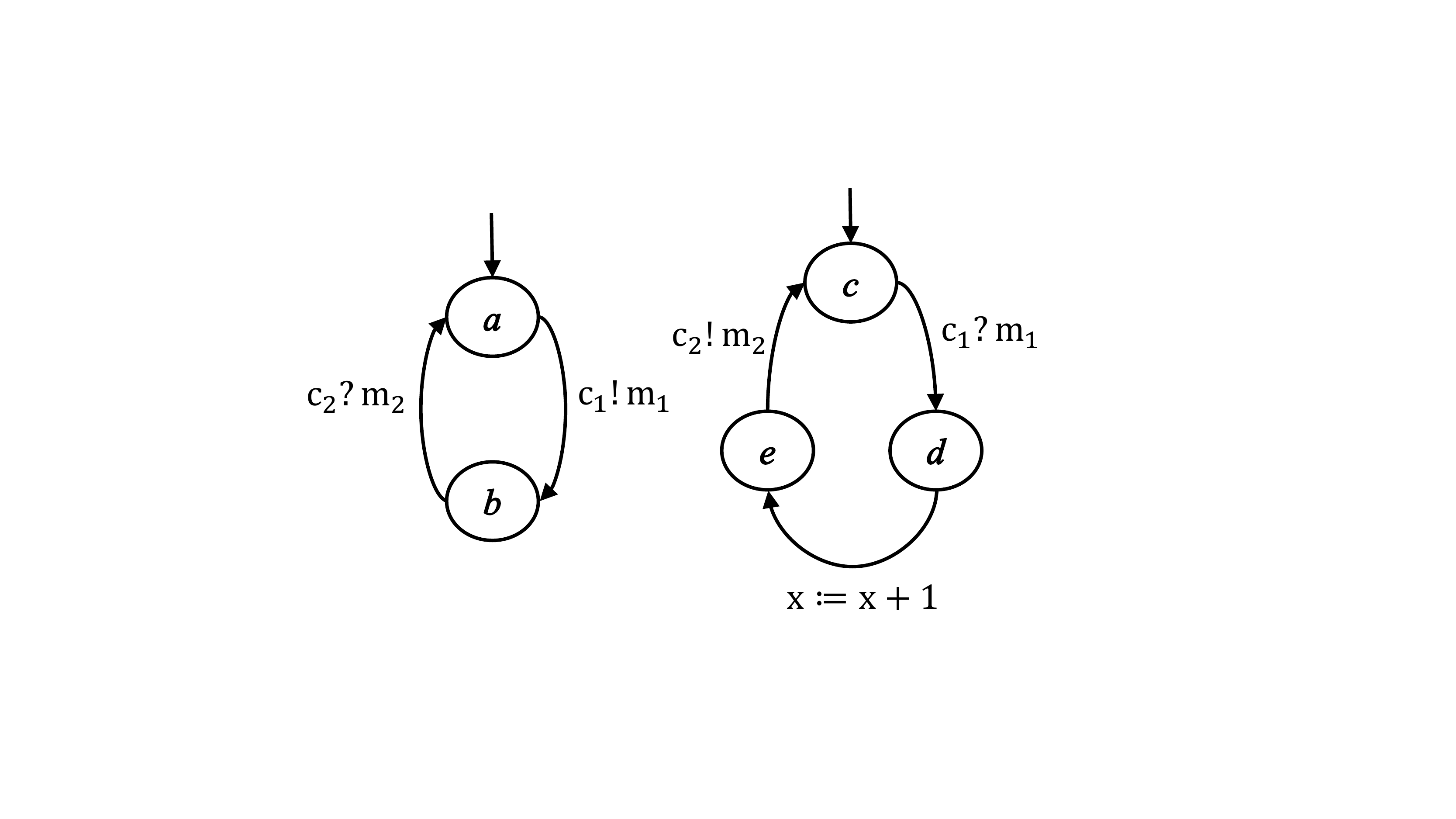} &
\includegraphics[trim = 200 50 200 50, clip, scale=0.3]{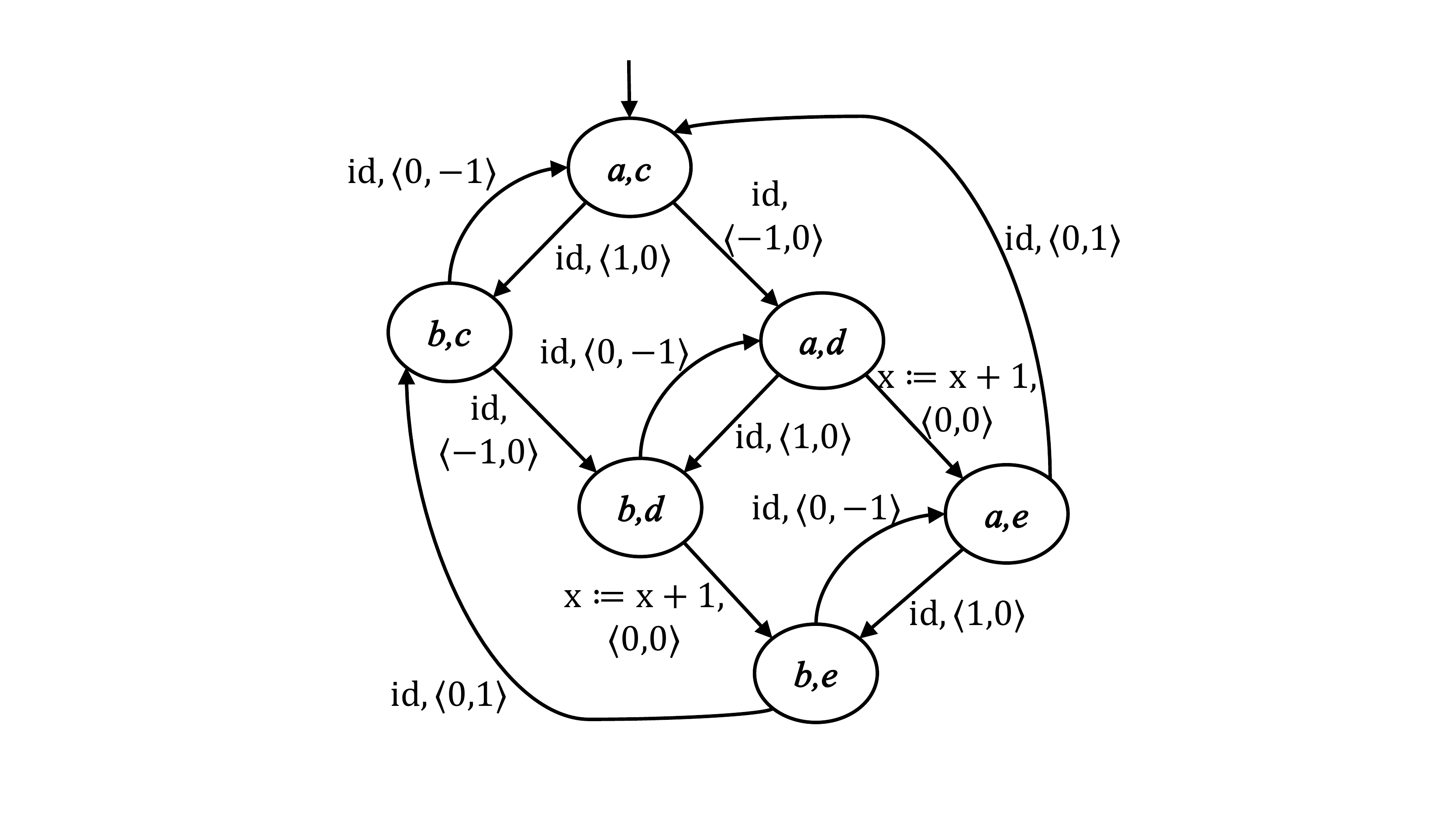} \\
\multicolumn{1}{c}{(a)} & \multicolumn{1}{c}{(b)}
\end{tabular}
\caption{(a) Asynchronous system with two processes, (b) its VCFG model}
\label{fig:notationAsyncSystem}
\end{figure}

Figure~\ref{fig:notationAsyncSystem}(a) shows a simple asynchronous system
with two processes. In this system there are two channels, $\mathrm{c}_1$
and $\mathrm{c}_2$, and a message alphabet consisting of two elements,
$\mathrm{m}_1$ and $\mathrm{m}_2$. The semantics we assume for
message-passing systems is the same as what is used by the tool
Spin~\cite{holzmann1997model}.  A configuration of the system consists of
the current control states of all the processes, the contents of all the
channels, and the values of all the variables. A single transition of the
system consists of a transition of one of the processes from its current
control-state to a successor control state, accompanied with the
corresponding queuing operation or variable-update action. A transition
labeled $\send{c}{m}$ can be taken unconditionally, and results in `m' being
appended to the tail of the channel `c'. A transition labeled $\recv{c}{m}$
can be taken only if an instance of `m' is available at the head of `c',
and results in this instance getting removed from `c'. (Note, based on the context, we
over-load the term ``message'' to mean either an element of the message
alphabet, or an instance of a message-alphabet element in a channel at
run-time.)

Asynchronous systems can be modeled as VCFGs, and our approach performs
data flow analysis on VCFGs.   We now illustrate how an asynchronous system can
be modeled as a VCFG. We assume a fixed number of
processes in the system. We do this illustration using the example VCFG in
Figure~\ref{fig:notationAsyncSystem}(b), which models the system in
Figure~\ref{fig:notationAsyncSystem}(a). Each node of the VCFG represents a
tuple of control-states of the processes, while each edge corresponds to a
transition of the system.  The action of a VCFG edge is identical to the
action that labels the corresponding process transition. (``id'' in
Figure~\ref{fig:notationAsyncSystem}(b) represents the \emph{identity}
action) The VCFG will have as many counters as the number of
unique pairs $(\mathrm{c_i}, \mathrm{m_j})$ such that the operation $\send{c_i}{m_j}$ is
performed by any process. If an edge $e$ in the VCFG corresponds to a send
transition $\send{c_i}{m_j}$ of the system, then $e$'s queuing vector would
have a +1 for the counter corresponding to $(\mathrm{c_i},\mathrm{m_j})$
and a zero for all the
other counters. Analogously, a receive operation gets modeled as -1 in the
queuing vector. In Figure~\ref{fig:notationAsyncSystem}(b), the first
counter is for (c$_1$,m$_1$) while the second counter is for (c$_2$,m$_2$).
Note that the +1 and -1 encoding (which are inherited from VASS's)
effectively cause FIFO channels to be treated as unordered channels.

When each process  can invoke procedures as part
of its execution, such systems can be modeled using \emph{inter-procedural}
VCFGs, or iVCFGs. These are extensions of VCFGs just as standard
inter-procedural control-flow graphs are extensions of control-flow
graphs. Constructing an iVCFG for a given system is straightforward,
under a restriction
that at most one of the processes in the system can be executing 
a procedure other than its main procedure at any
time. This restriction is also present in other related
work~\cite{jhala2007interprocedural,bouajjani2012analysis}.


\subsection{Data flow analysis over iVCFGs}
\label{ssec:prelim:dataflow}


Data flow analysis is based on a given \emph{complete lattice} $\dlattice$,
which serves as the abstract domain. As a pre-requisite step before we can
perform our data
flow analysis on iVCFGs, we first consider each edge $v \vcfgtrans{a}{w}
w$ in each procedure in the iVCFG, and replace the (concrete) action $a$
with an abstract action $f$, where $f: \dlattice \rightarrow \dlattice$ is
a given abstract transfer function that \emph{conservatively
  over-approximates}~\cite{cousot1977abstract} the behavior of the concrete
action $a$.


Let $\vpath{p}$ be a path in a iVCFG, let $p_0$ be the first node in the
path, and let $\xi_i$ be a valuation to the variables at the beginning of
$\vpath{p}$. The path \vpath{p} is said to be \emph{feasible} if, starting
from the configuration $\langle p_0, \vec{0}, \xi_i \rangle$, the
configuration $\langle q, d, \xi \rangle$ obtained at each successive point
in the path is such that $d \geq \vec{0}$, with successive configurations
along the path being generated as per the rule for transitions among VCFG
configurations that was given before Section~\ref{ssec:proc-to-vcfg}.
For any path $p = \vpath{e_1e_2 \ldots e_k}$ of an iVCFG, we define its
\emph{path transfer function} $\ptf{p}$ as $f_{e_k} \circ f_{e_{k-1}} \ldots
\circ f_{e_1}$, where $f_e$ is the abstract action associated with edge
$e$.


The standard data flow analysis problem for sequential programs is to
compute the join-over-all-paths (JOP) solution. Our problem statement is to
compute the join-over-all-feasible-paths (JOFP) solution for
iVCFGs. Formally stated, if $\startnode$ is the entry node of the ``main''
procedure of the iVCFG, given any node $\targetnode$ in any procedure of
the iVCFG, and an ``entry'' value $d_0 \in \dlattice$ at $\startnode$ such
that $d_0$ conservatively over-approximates $\xi_0$, we wish to compute the
JOFP value at
$\targetnode$ as defined by the following expression:
$$
\bigjoin_{\parbox{2.5in}{\centering $p$ is a feasible and interprocedurally valid
    path in the iVCFG from $\startnode$ to  ${\targetnode}$}} (\ptf{p})(d_0)
$$

Intuitively, due to the unordered channel abstraction, every run of the
system corresponds to a feasible path in the iVCFG, but not vice
versa. Hence, the JOFP solution above is guaranteed to conservatively
over-approximate the JOP solution on the \emph{runs} of the system (which
is not computable in general). 

\section{Backward DFAS Approach}
\label{sec:backward}

In this section we present our key contribution -- the \emph{Backward DFAS}
(Data Flow Analysis of Asynchronous Systems) algorithm -- an
interprocedural algorithm that computes the precise JOFP at any given node
of the iVCFG.


\begin{figure}[t]
	\centering
	\includegraphics[scale=0.27]{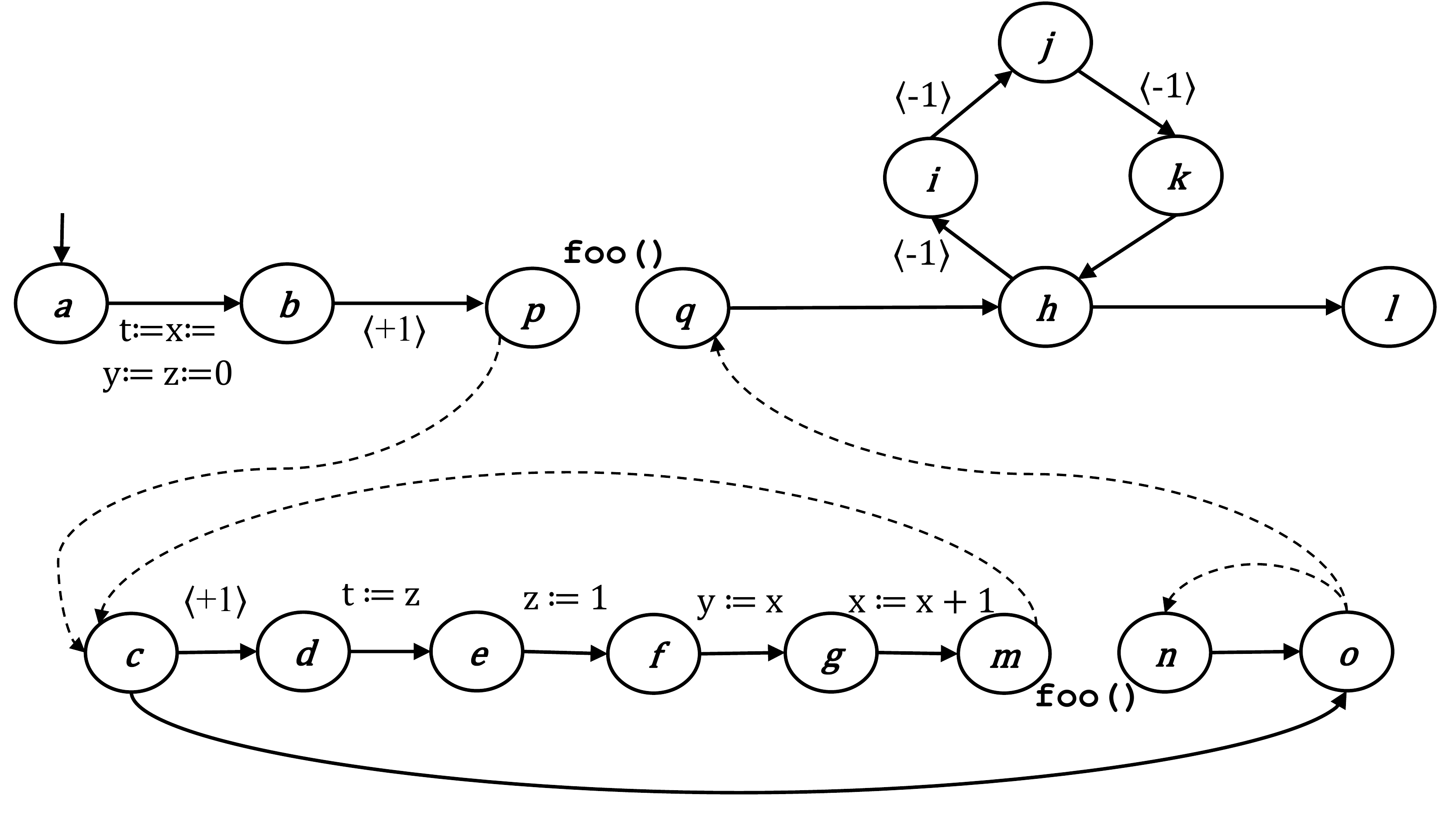}
	\caption{Example iVCFG}
	\label{fig:runningex}
\end{figure}

We begin by presenting a running example, which is the iVCFG with two
procedures depicted in Figure~\ref{fig:runningex}.  There is only one
channel and one message in the message alphabet in this example, and hence
the \emph{queuing vectors} associated with the edges are of size 1.  The
edges without the vectors are implicitly associated with zero vectors. The
\emph{actions} associated with edges are represented in the form of assignment
statements. The edges without assignment statements next to them have
\emph{identity} actions.
The upper part of the Figure~\ref{fig:runningex}, consisting of nodes $a, b, p, q, h, i, j, k, l$, is the VCFG of the ``main'' procedure. The remaining nodes
constitute the VCFG of the (tail) recursive procedure \texttt{foo}.
The solid edges are
intra-procedural edges, while dashed edges  are inter-procedural
edges.

Throughout this section we use \emph{Linear Constant Propagation}
(LCP)~\cite{sagiv1996precise} as our example data flow analysis. LCP, like
CP, aims to identify the variables that have constant values
at any given location in the system. LCP is based on the same infinite
domain as CP; i.e., each abstract domain element is a mapping from
variables to (integer) values. The ``$\dominates$'' relation for the LCP
lattice is also defined in the same way as for CP.
The encoding of the transfer functions in LCP is as
follows. Each edge (resp. path) maps the outgoing value of each variable
to \emph{either} a constant, \emph{or} to a
linear expression in the incoming value of at most one variable into the
edge (resp. path), \emph{or} to a special symbol $\top$ that indicates an
unknown outgoing value. For instance, for the edge $g \rightarrow m$ in
Figure~\ref{fig:runningex}, its transfer function can be represented
symbolically as (t'=t,x'=x+1,y'=y,z'=z), where the primed versions
represent outgoing values and unprimed versions represent incoming values.


Say we wish to compute the JOFP at node $k$. The only feasible paths that
reach node~$k$ are the ones that attain calling-depth of three or more in the
procedure \texttt{foo}, and hence encounter at least three \emph{send}
operations, which  are required to  clear the three \emph{receive} operations
encountered from node~$h$ to node~$k$.  All such paths happen to
bring the constant values (t = 1, z = 1) to the node $k$. Hence, (t = 1, z
= 1) is the precise JOFP result at node $k$. However, infeasible paths, if
not elided, can introduce imprecision. For instance, the path that directly
goes from node~$c$ to node~$o$ in the outermost call to the
Procedure~\texttt{foo} (this path is of calling-depth zero) brings values
of zero for all four variables, and would hence  prevent the precise
fact (t = 1, z = 1) from being inferred.


\subsection{Assumptions and Definitions}
\label{ssec:bdfas-assump}

The set of all $\dlattice \rightarrow \dlattice$ transfer functions clearly
forms a complete lattice based on the following ordering: $f_1 \sqsupseteq
f_2$ iff for all $d \in \dlattice$, $f_1(d) \sqsupseteq f_2(d)$.  Backward
DFAS makes a few assumptions on this lattice of transfer functions. The
first is that this lattice be of \emph{finite height}; i.e., all strictly
ascending chains of elements in this lattice are finite
(although no a priori bound on the sizes of these chains is required).  The
second is that a representation of transfer functions is available, as are
operators to compose, join, and compare transfer functions. Note, the two
assumptions above are also made by the classical ``functional''
inter-procedural approach of Sharir and
Pnueli~\cite{sharirPnueliInterproc}. Thirdly, we need distributivity, as
defined below: for any $f_1, f_2, f \in \dlattice \rightarrow \dlattice$,
$(f_1 \sqcup f_2) \circ f = (f_1 \circ f) \sqcup (f_2 \circ f)$. The
distributivity assumption is required only if the given system contains
recursive procedure calls.

Linear Constant Propagation (LCP)~\cite{sagiv1996precise} and Affine
Relationships Analysis (ARA)~\cite{muller2004precise} are well-known
examples of analyses based on infinite abstract domains that satisfy all of
the assumptions listed above. Note that the CP transfer-functions lattice
is not of finite height. Despite the LCP abstract domain being the same as
the CP abstract domain, the encoding chosen for LCP transfer functions
(which was mentioned above), ensures that LCP uses a strict, finite-height subset of the
full CP transfer-functions lattice that is closed
under join and function composition operations.  The
trade-off is that LCP transfer functions for assignment statements whose
RHS is not a linear expression and for conditionals are less
precise than the corresponding CP transfer functions.


Our final assumption is that procedures other than ``main'' may send
messages, but should not have any ``receive'' operations. Previous
approaches that have addressed data flow analysis or verification problems
for asynchronous systems with recursive procedures also have the same
restriction~\cite{sen2006model,jhala2007interprocedural,ganty2009verifying}.



We now introduce important terminology. The \textbf{\emph{demand}} of a
given path $p$ in the VCFG is a vector
of size $r$, and is defined as follows:

$\demandp{p} = \begin{cases} \mathit{max}(\vec{0} - w, \vec{0}),
 \hspace{45pt}\mathrm{if\ } p = (v \vcfgtrans{f}{w} z) \\ \mathit{max}(\demandp{p'} - w, \vec{0}), \mathrm{if\ } p = (e \concat p'),
 \mathrm{where\ } e \equiv (v   \vcfgtrans{f}{w} z) \end{cases}$\\

Intuitively, the demand of a path $p$ is the minimum required vector of
counter values in any starting configuration at the entry of the path for
there to exist a sequence of transitions among configurations that manages
to traverse the entire path  (following the rule given before
Section~\ref{ssec:proc-to-vcfg}).
It is easy to see that a path $p$ is
feasible \emph{iff} $\demandp{p} = \vec{0}$.

A set of paths $C$ is said to \textbf{\emph{cover}} a path $p$ iff: (a) all
paths in $C$ have the same start and end nodes (respectively) as $p$, (b)
for each $p' \in C$, $\demandp{p'} \leq \demandp{p}$, and (c) $(\join_{p'
  \in C} \ptf{p'}) \dominates \ptf{p}$. (Regarding (b), any binary vector
operation in this paper is defined as applying the same operation on every
pair of corresponding entries, i.e., point-wise.)

A \emph{path template} $(\vpath{p_1, p_2, \ldots, p_n)}$ of any procedure
$F_i$ is a sequence of paths in the VCFG of $F_i$ such that: (a) path $p_1$
begins at the entry node $\entrynode{F_i}$ of $F_i$ and path $p_n$ ends at
return node $\exitnode{F_i}$ of $F_i$, (b) for all $p_i, 1 \leq i < n$, $p_i$ ends at a
call-site node, and (c) for all $p_i, 1 < i \leq n$, $p_i$ begins at a
return-site node $v^i_r$, such that $v^i_r$ corresponds to the call-site
node $v^{i-1}_c$ at which $p_{i-1}$ ends.

\subsection{Properties of Demand and Covering}
\label{ssec:bdfas-properties}

At a high level, Backward DFAS works by growing paths in the
backward direction by a single edge at a time starting from the target node
(node $k$ in our example in Figure~\ref{fig:runningex}). Every time this
process results in a path reaching the \emph{start} node (node~$a$ in
our example), and the path is feasible, 
the approach simply transfers the entry value $d_0$ via this
path to the target node. The main challenge is that due to the presence of cycles and
recursion, there are an infinite number of feasible paths in general. In
this subsection we present a set of lemmas that embody our intuition on how a finite subset of the set of all
paths can be enumerated such that the join of the values brought by these
paths is equal to the JOFP. We then present our complete approach in
Section~\ref{ssec:backward-algo}.


\textit{\textbf{Demand Coverage Lemma:} Let $p_2$ and $p_2'$ be two paths from a node $v_i$ to a
node $v_j$ such that $\demandp{p_2'} \leq \demandp{p_2}$. If $p_1$ is any path ending at $v_i$, then $\demandp{p_1 \concat p_2'} \leq
\demandp{p_1 \concat p_2}$.} \hfill$\Box$

This lemma can be argued using induction on the length of path $p_1$. A
similar observation has been used to solve coverability of lossy channels
and well-structured transition systems in
general~\cite{abdulla1996verifying,finkel2001well,abdulla1996general}. An
important corollary of this lemma is that for any two paths $p_2'$ and $p_2$
from $v_i$ to $v_j$ such that $\demandp{p_2'} \leq \demandp{p_2}$, if there
exists a path $p_1$ ending at $v_i$ such that $p_1 \concat p_2$ is
feasible, then $p_1 \concat p_2'$ is also feasible.

\textit{\textbf{Function Coverage Lemma: } Let $p_2$ be a path from a node
  $v_i$ to a node $v_j$, and $P_2$ be a set of paths from $v_i$ to $v_j$
  such that $(\bigjoin_{p_2' \in P_2} \ptf{p_2'}) \dominates
  \ptf{p_2}$. Let $p_1$ be any path ending at $v_i$ and $p_3$ be any path
  beginning at $v_j$. Under the distributivity assumption stated in
  Section~\ref{ssec:bdfas-assump}, the following property holds:
  $(\bigjoin_{p_2' \in P_2} \ptf{p_1 \concat p_2' \concat p_3}) \dominates
  \ptf{p_1 \concat p_2 \concat p_3}$.} \hfill$\Box$


The following result follows from the Demand and Function Coverage
Lemmas and from monotonicity of the transfer functions:

\textit{\textbf{Corollary 1:} Let $p_2$ be a path from a node $v_i$ to a node
  $v_j$, and $P_2$ be a set of paths from $v_i$ to $v_j$ such that $P_2$ covers
  $p_2$. Let $p_1$ be any path ending at
  $v_i$. Then, the set of paths $\lbrace p_1.p_2' \mid p_2' \in P_2 \rbrace$
  covers the path $p_1.p_2$. } \hfill$\Box$



We now use the running example from Figure~\ref{fig:runningex} to
illustrate how we leverage Corollary~1 in our approach. When we grow paths
in backward direction from the target node $k$, two candidate paths that
would get enumerated (among others) are $p_i \equiv \vpath{hijk}$ and $p_j
\equiv \vpath{hijkhijk}$ (in that order). Now, $p_i$ covers
$p_j$. Therefore, by Corollary~1, any backward extension  $p_1 \concat
p_j$ of $p_j$ ($p_1$ is any path prefix) is guaranteed to be covered by the analogous backward extension $p_1
\concat p_i$ of $p_i$. By definition of covering, it follows that $p_1
\concat p_i$ brings in a data value that conservatively over-approximates
the value brought in by $p_1 \concat p_j$. Therefore, our approach discards
$p_j$ as soon as it gets enumerated. To summarize, our approach discards
any path as soon as it is enumerated if it is covered by some subset of the 
previously enumerated and retained paths.

Due to the finite height of the transfer functions lattice, and because
demand vectors cannot contain negative values, 
at some point
in the algorithm every new path that can be generated by backward extension at that point
would be discarded immediately.
At this point the approach would terminate, and soundness would be
guaranteed by definition of covering.

In the inter-procedural setting the situation is more complex. We first
present two lemmas that set the stage. The lemmas both crucially make use
of the assumption that recursive procedures are not allowed to have
``receive'' operations. For any path $p_a$ that contains no receive
operations, and for any demand vector $d$, we first define
$\supplypd{p_a}{d}$ as $\mathit{min}(s,d)$, where $s$ is the sum of the
queuing vectors of the edges of $p_a$.

\textit{ \textbf{Supply Limit Lemma: } Let $p_1, p_2$ be two paths from
  $v_i$ to $v_j$ such that there are no receive operations in $p_1$ and
  $p_2$. Let $p_b$ be any path beginning at $v_j$. If $\demandp{p_b} = d$, and
  if $\supplypd{p_1}{d} \geq \supplypd{p_2}{d}$, then $\demandp{p_1.p_b} \leq
  \demandp{p_2.p_b}$.} \hfill$\Box$

A set of paths $P$ is said to \textbf{\emph{$d$-supply-cover}} a path $p_a$ iff: (a) all
paths in $P$ have the same start node and same end node (respectively) as
$p_a$, (b)  $(\join_{p' \in P} \ptf{p'}) \dominates \ptf{p_a}$, and (c) for
each  $p' \in P$, $\supplypd{p'}{d} \geq \supplypd{p_a}{d}$.

\textit{ \textbf{Supply Coverage Lemma: } If $p_a \concat p_b$ is a path,
  and $\demandp{p_b} = d$, and if a set of paths $P$ $d$-supply-covers
  $p_a$, and $p_a$ as well as all paths in $P$ have no receive operations,
  then the set of paths $\{p' \concat p_b \, | \, p' \in P \}$
  covers the path $p_a \concat p_b$. }

\emph{Proof argument:} Since $P$ $d$-supply-covers $p_a$, by the Supply
Limit Lemma, we have (a): for all $p' \in P$, $\demandp{p' \concat p_b} \leq
\demandp{p_a \concat p_b}$. Since $P$ $d$-supply-covers $p_a$, we also have
$(\join_{p' \in P} \ptf{p'}) \dominates \ptf{p_a}$. From this, we use the
Function Coverage lemma to infer that (b): $(\join_{p' \in P} \ptf{p' \concat
  p_b}) \dominates \ptf{p_a \concat p_b}$. The result now follows from (a)
and (b). 
\hfill $\Box$

Consider path \vpath{hijk} in our example, which gets enumerated and
retained (as discussed earlier). This path gets extended back as
\vpath{qhijk}; let us denote this path as $p'$.  Let $d$ be the demand of
$p'$ (i.e., is equal to 3). Our plan now is to extend this path in the
backward direction all the way up to node $p$, by prepending
interprocedurally valid and complete (i.e., IVC) paths of procedure
\texttt{foo} in front of $p'$. An IVC path is one that begins at the entry
node of \texttt{foo}, ends at the return node of \texttt{foo}, is of
arbitrary calling depth, has balanced calls and returns, and has no pending
returns when it completes~\cite{reps1995precise}.
First, we enumerate the IVC path(s) with calling-depth zero (i.e.,
path \vpath{co} in the example), and prepend them
in front of $p'$. We then produce deeper IVC paths, in phases. In each
phase $i$, $i > 0$, we inline IVC paths of calling-depth $i-1$ that have
been enumerated and retained so far into the path templates of the
procedure to generate IVC paths of calling-depth $i$, and prepend these IVC
paths in front of $p'$. We terminate when each IVC path that is generated
in a particular phase $j$ is $d$-supply-covered by some subset $P$ of IVC
paths generated in previous phases.

The soundness of discarding the IVC paths of phase $j$ follows from the
Supply Coverage lemma ($p'$ would take the place of $p_b$ in the lemma's
statement, while the path generated in phase $j$ would take the place of
$p_a$ in the lemma statement).  The termination condition is guaranteed to be reached
eventually, because: (a) the supplies of all IVC paths generated are
limited to $d$,  and (b) the lattice of transfer functions is of finite
height. Intuitively, we could devise a sound termination condition even
though deeper and deeper IVC paths can increment counters more and more,
because a deeper IVC path that increments the counters beyond the
demand of $p'$ does not really result in lower overall demand when
prepended before $p'$ than a shallower IVC path that also happens to meet
the demand of $p'$ (Supply Limit lemma formalizes this).

In our running example, for the path \vpath{qhijk}, whose demand is
equal to three, prefix generation for it happens to terminate in the fifth
phase. The IVC paths that get generated in the five phases are,
respectively, $p_0 = \vpath{co}$, $p_1 = \vpath{cdefgmcono}$, $p_2 =
(\vpath{cdefgm})^2\vpath{co}(\vpath{no})^2$,  $p_3 =
(\vpath{cdefgm})^3\vpath{co}(\vpath{no})^3$, $p_4 =
(\vpath{cdefgm})^4\vpath{co}(\vpath{no})^4$, and $p_5 =
(\vpath{cdefgm})^5\vpath{co}(\vpath{no})^5$. $\supplypd{p_3}{3}$ =
$\supplypd{p_4}{3}$ = $\supplypd{p_5}{3}$ = 3. The LCP transfer functions
of the paths are  as follows. $\ptf{p_3}$ is (t’=1,
x’=x+3, y’=x+2, z’=1), $\ptf{p_4}$ is (t’=1, x’=x+4, y’=x+3, z’=1), while
$\ptf{p_5}$ is (t’=1, x’=x+5, y’=x+4, z’=1). $\{p_3, p_4\}$
$3$-supply-covers  $p_5$. 

We also need a result that when the IVC paths in the $j$th phase are
$d$-supply-covered by paths generated in preceding phases, then the IVC
paths that would be generated in the $(j+1)th$ would also be
$d$-supply-covered by paths generated in phases that preceded $j$. This can
be shown using a variant of the Supply Coverage Lemma, which we omit in the
interest of space. Once this is shown, it then follows inductively that
none of the phases after phase $j$ are required, which would imply that it
would be safe to terminate.

The arguments presented above were in a restricted setting, namely,
that there is only one call in each procedure, and that only recursive calls
are allowed. These restrictions were assumed only for simplicity, and are
not actually assumed in the algorithm to be presented below.


\subsection{Data Flow Analysis Algorithm}
\label{ssec:backward-algo}


\begin{algorithm}[t]
	\caption{Backward DFAS algorithm}
	\label{algo:main}
	\begin{algorithmic}[1]
		\Procedure{ComputeJOFP}{$\targetnode$}
	
		\Comment{\begin{minipage}{4in}Returns JOFP
				from $\startnode \in \nodes$ to  $\targetnode
				\in \nodes$, entry value $d_0 \in \dlattice$.\end{minipage}}
		
		\ForAll{$v \in \nodes$} \Comment{{\nodes} is the set of all nodes in the
			VCFG}
		
		\State \sPaths{v} = $\emptyset$
		
		\EndFor
		
		\State
		For each intra-proc {\small VCFG} edge $v \! \rightarrow \!
		\targetnode$, add this edge to $\worklist$ and to $\sPaths{v}$
		\label{step:intra:initw}
        
		\Repeat\label{step:intra:repeat}
		
		\State Remove any path $p$ from \worklist.\label{step:intra:removw}
		
		\State Let $v_1$ be the start node      of $p$.
		
		\If{$v_1$ is a return-site node, with incoming return edge from
			func.  $F_1$}\label{step:inter:retbeg}
		
		\State
		\begin{minipage}[t]{4in}
			Let $v_3$ be the call-site node corresponding to $v_1$,
			$e_1$ be the call-site-to-entry edge from $v_3$ to
			$\entrynode{F_1}$, and $r_1$ be the exit-to-return-site edge from
			$\exitnode{F_1}$ to $v_1$. 
		\end{minipage}
		
		\ForAll{$\vpath{p_1} \in \textsc{ComputeEndToEnd}(F_1, \demandp{p}) $}\label{step:inter:end2end}
		\State $p_2$ = $\vpath{e_1 \concat p_1 \concat r_1 \concat p}$\label{step:inter:growToCS}
		
		\If{\textsc{Covered}($\vpath{p_2}$,  $\sPaths{v_3}$) returns \fls} \label{step:inter:covret}
		\State Add $\vpath{p_2}$ to $\sPaths{v_3}$  and to $\worklist$.\label{step:inter:ivcext}					
		\EndIf				
		\EndFor \label{step:inter:retend}		
		
		\ElsIf{$v_1$  is the entry node of a func. $F_1$}\label{step:inter:enbeg}
		\ForAll{$v_3 \in \callsite{F_1}$}
		\State Let $e_1$ be the call edge from $v_3$ to $v_1$.
		\State $p_2$ = $\vpath{e_1 \concat p}$.
		\If{\textsc{Covered}($\vpath{p_2}$,  $\sPaths{v_3}$) returns \fls} \label{step:inter:covbeg}
		\State Add $\vpath{p_2}$ to $\sPaths{v_3}$  and to $\worklist$.					
		\EndIf	
		\EndFor \label{step:inter:enend}
		\Else 				
		\ForAll{intra-procedural edges $e$ = $v_3 \vcfgtrans{f}{w} v_1$ in the
			VCFG}\label{step:intra:inner-loop-begin} 
		
		\If{\textsc{Covered}($e \concat p$, $\sPaths{v_3}$) returns \fls}\label{step:intra:checkCov}
		
		\State Add the path ($e \concat p$) to \sPaths{v_3} and to
		\worklist. 
		
		\EndIf
		
		\EndFor\label{step:intra:inner-loop-end}
		
		\EndIf
		
		\Until{{\worklist} is empty}\label{step:intra:until}
		
		\State $P \ = \ \{p \ \, | \, \ p \in \sPaths{\startnode},
		\demandp{p} = \zerovector\}$\label{step:intra:P}
		
		\State \textbf{return} $\bigjoin_{p \in P} \,
		(\ptf{p})(d_0)$\label{step:intra:return} 
		
		\EndProcedure

		
		
		
		
	\end{algorithmic}
\end{algorithm}

\begin{algorithm}
	\caption{Routines invoked for inter-procedural processing in Backward
		DFAS algorithm}
	\label{algo:interfp}

	\begin{algorithmic}[1]
		\Procedure{ComputeEndToEnd}{$F$, $d$}
		
		\Comment{\begin{minipage}{4in} Returns a set of paths that
            $d$-supply-covers
				each IVC path of the procedure $F$.
		\end{minipage}}
		
		\ForAll{ $F_i \in \funcs$}
		
		\State Place all $0$-depth paths from $F_i$ in
		$\sIVCPaths{\mathit{F_i}}{d}$ \label{step:e2e:init} 
		
		\EndFor
		
		\Repeat\label{step:e2e:repeat}
		\State pathsAdded = \fls
		
		\ForAll{path template $(\vpath{p_1, p_2,  \ldots, p_n})$   in
			any function $F_i \in \funcs$}\label{step:e2e:inlbeg}
		
		\State \begin{minipage}[t]{4in}Let $F_1$ be the procedure called from the call-site at which
			$p_1$ ends, $F_2$ be the procedure called from the call-site at which
			$p_2$ ends, and so on.
		\end{minipage}
		
		\ForAll{$p_1' \in \sIVCPaths{F_1}{d}$, $p_2' \in \sIVCPaths{F_2}{d},
			\ldots$}\label{step:e2e:holecand}
		
		\State \begin{minipage}[t]{4in} Let $p' = \vpath{p_1 \concat e_1
				\concat p_1' \concat r_1 \concat p_2 \concat e_2 \concat p_2'
				\concat r_2 \concat \ldots p_n}$, where each $e_i$ is the call-edge
			that leaves the call-site node at which $p_i$ ends and $r_i$ is the
			return edge corresponding to $e_i$.
		\end{minipage}\label{step:e2e:fillhole}
		
		\If{\textsc{DSCovered}($p', d, \sIVCPaths{F_i}{d}$) returns \fls}\label{step:e2e:dscovcheck}
		\State Add the path $p'$ to $\sIVCPaths{F_i}{d}$.  pathsAdded = \tru. \label{step:e2e:retainpath}
		\EndIf	
		\EndFor		\label{step:e2e:endhole}
		
		\EndFor\label{step:e2e:inlend}
		
		\Until{pathsAdded is \fls\label{step:e2e:until}}	
		
		\State \textbf{return} $\sIVCPaths{F}{d}$
		
		\EndProcedure
		
	\end{algorithmic}
\end{algorithm}


Our approach is summarized in Algorithm~\ref{algo:main}. \textsc{ComputeJOFP} is the main routine. The algorithm works on a given iVCFG (which is an implicit parameter to the algorithm), and is given  a $\targetnode$ node at which the JOFP is to be computed. A key data structure in the algorithm is {\sPathsName}; for any node
$v$, $\sPaths{v}$ is the set of all paths that start from $v$ and end at
$\targetnode$ that the algorithm has generated and retained so far.  The
{\worklist} at any point stores a subset of the paths in {\sPathsName}, and
these are the paths of the iVCFG that need to be extended backward.

To begin with, all edges incident onto {\targetnode} are generated and
added to the sets $\sPathsName$ and $\worklist$
(Line~\ref{step:intra:initw} in Algorithm~\ref{algo:main}). In each step
the algorithm picks up a path $p$  from  {\worklist}
(Line~\ref{step:intra:removw}), and extends this path in the backward
direction. The backward extension has three cases based on the start node
of the path $p$. The simplest case is the intra-procedural case,
wherein the path is extended backwards in all possible ways by a single
edge
(Lines~\ref{step:intra:inner-loop-begin}-\ref{step:intra:inner-loop-end}). The
routine \textsc{Covered}, whose definition is not shown in the algorithm,
checks if its first argument (a path) is covered by its second argument (a
set of paths). Note, covered paths are not retained.

When the start node of $p$ is the entry node of a procedure $F_1$ (Lines~\ref{step:inter:enbeg}-\ref{step:inter:enend}), the path is extended backwards via all possible call-site-to-entry edges for procedure $F_1$. 


If the starting node of path $p$ is a return-site node $v_1$ (Lines~\ref{step:inter:retbeg}-\ref{step:inter:retend}) in a calling
procedure, we invoke a routine~\textsc{ComputeEndToEnd} (in
line~\ref{step:inter:end2end} of Algorithm~\ref{algo:main}).
This routine,  which
we explain later,  returns a set IVC paths of the called
procedure such that \emph{every} IVC path of the called procedure is
$d$-supply-covered by some subset of paths in the returned set, where $d$ denotes
$\demandp{p}$. These returned IVC paths are prepended before $p$
(Line~\ref{step:inter:growToCS}), with the call-edge $e_1$ and return edge
$r_1$ appropriately inserted. 


The final result returned by the algorithm (see
Lines~\ref{step:intra:P} and~\ref{step:intra:return} in
Algorithm~\ref{algo:main}) is the join of the values transferred by the
zero-demand paths (i.e., feasible paths) starting from the given entry value $d_0 \in \dlattice$.

\paragraph{Routine \textsc{ComputeEndToEnd}: } 

This routine is specified in
Algorithm~\ref{algo:interfp},  and  is basically a generalization of the approach that we
described in Section~\ref{ssec:bdfas-properties}, now handling multiple
call-sites in each procedure, mutual recursion, calls to non-recursive
procedures, etc.  We do assume for simplicity of presentation that there
are no cycles (i.e., loops) in the procedures, as this results in a fixed
number of path templates in each procedure. There is no loss of generality
here because we allow recursion. The routine incrementally populates a
group of sets -- there is a set named $\sIVCPaths{F_i}{d}$ for each
procedure $F_i$ in the system. The idea is that when the routine completes,
$\sIVCPaths{F_i}{d}$ will contain a set of IVC paths of $F_i$ that
$d$-supply-cover all IVC paths of $F_i$.  Note that we simultaneously
populate covering sets for all the procedures in the system in order to
handle mutual recursion.

The routine~\textsc{ComputeEndToEnd} first enumerates and saves all
zero-depth paths in all procedures (see Line~\ref{step:e2e:init} in
Algorithm~\ref{algo:interfp}). The routine then iteratively takes a
path template at a time, and fills in the ``holes'' between corresponding
(call-site, return-site) pairs of the form $v^{i-1}_c, v^i_r$ in the path
template with IVC paths of the procedure that is called from this pair of
nodes, thus generating a deeper IVC path (see the loop in
lines~\ref{step:e2e:inlbeg}-\ref{step:e2e:inlend}). A newly generated IVC
path $p'$ is retained only if it is not $d$-supply-covered by other IVC
paths already generated for the current procedure $F_i$
(Lines~\ref{step:e2e:dscovcheck}-\ref{step:e2e:retainpath}). The routine
terminates when no more IVC paths that can be retained are generated, and
returns the set $\sIVCPaths{F}{d}$. 


\subsection{Illustration}

We now illustrate our approach using the example in
Figure~\ref{fig:runningex}.  Algorithm~\ref{algo:main} would start from the
target node $k$, and would grow paths one edge at a time. After four steps
the path \vpath{hijk} would be added to $\sPaths{h}$ (the intermediate
steps would add suffixes of this path to $\sPaths{i}$, $\sPaths{j}$, and
$\sPaths{k}$). Next, path \vpath{khijk} would be generated and discarded, 
because it is covered by the ``root'' path \vpath{k}. Hence, further
iterations of the cycle are avoided. On the other hand, the path
\vpath{hijk} would get extended back to node $q$, resulting in path
\vpath{qhijk} being retained in $\sPaths{q}$.  This path would trigger a
call to routine~\textsc{ComputeEndToEnd}. As discussed in
Section~\ref{ssec:bdfas-properties}, this routine would return the
following set of paths: $p_0$ = \vpath{co}, and $p_i =
(\vpath{cdefgm})^i\vpath{co}(\vpath{no})^i$ for each $1 \leq i \leq
4$. (Recall, as discussed in Section~\ref{ssec:bdfas-properties}, that
$(\vpath{cdefgm})^5\vpath{co}(\vpath{no})^5$ and deeper IVC paths are
3-supply-covered by the paths $\{p_3, p_4\}$.)

Each of the paths returned above by the routine~\textsc{ComputeEndToEnd}
would be prepended in front of \vpath{qhijk}, with the corresponding call
and return edges inserted appropriately. These paths would then be extended
back to node $a$. Hence, the final set of paths in $\sPaths{a}$ would
be $\vpath{abpcoqhijk}$, $\vpath{abpcdefgmconoqhijk}$,
$\vpath{abp}(\vpath{cdefgm})^2\vpath{co}(\vpath{no})^2$,
$\vpath{abp}(\vpath{cdefgm})^3\vpath{co}(\vpath{no})^3$, and
$\vpath{abp}(\vpath{cdefgm})^4\vpath{co}(\vpath{no})^4$. Of these paths,
the first two are ignored, as they are not feasible. The initial data-flow
value (in which all variables are non-constant) is sent via the remaining
three paths. In all these three paths the final values of variables `t' and
`z' are one. Hence, these two constants are inferred at node $k$.

\subsection{Properties of the algorithm}
\label{ssec:intra-proof}

We provide argument sketches here about the key properties of Backward
DFAS.  Detailed proofs are available in the appendix.

\emph{Termination.} The argument is by contradiction.  For the algorithm to
not terminate, one of the following two scenarios must happen. The first is
that an infinite sequence of paths gets added to some set $\sPaths{v}$. By
Higman's lemma it follows that embedded within this infinite sequence there
is an infinite sequence $p_1, p_2, \ldots$, such that for all $i$,
$\demandp{p_i}$ $\leq$ $\demandp{p_{i+1}}$.  Because the algorithm never
adds covered paths, it follows that for all $i$: $ \bigjoin_{1 \leq k \leq
  i+1} \ptf{p_k} \sqsupset \bigjoin_{1 \leq k \leq i} \ptf{p_k} $. However,
this contradicts the assumption that the lattice of transfer functions is
of finite height. The second scenario is that an infinite sequence of IVC
paths gets added to some set $\sIVCPaths{F}{d}$ for some procedure $F$ and
some demand vector $d$ in some call to routine
\textsc{ComputeEndToEnd}. Because the ``supply'' values of the IVC paths
are bounded by $d$, it follows that embedded within the infinite sequence
just mentioned there must exist an infinite sequence of paths $p_1, p_2,
\ldots$, such that for all $i$, $\supplypd{p_i}{d}$ $\geq$
$\supplypd{p_{i+1}}{d}$. However, since $d$-supply-covered paths are never
added, it follows that for all $i$: $ \bigjoin_{1 \leq k \leq
  i+1} \ptf{p_k} \sqsupset \bigjoin_{1 \leq k \leq i} \ptf{p_k} $. However,
this contradicts the assumption that the lattice of transfer functions is
of finite height.

\emph{Soundness and Precision.}  We already argued informally in
Section~\ref{ssec:bdfas-properties} that the algorithm explores all
feasible paths in the system, omitting only paths that are covered by other
already-retained paths. By definition of covering, this is sufficient to
guarantee over-approximation of the JOFP.  The converse direction, namely,
under-approximation, is obvious to see as every path along which the data
flow value $d_0$ is sent at the end of the algorithm is a feasible
path. Together, these two results imply that the algorithm is guaranteed to
compute the precise JOFP.

\emph{Complexity.} We show the complexity of our approach in the
single-procedure  setting. Our analysis follows along the
lines of the analysis of the backwards algorithm for coverability in
VASS~\cite{bozzelli2011complexity}.  The overall idea, is to use the
technique of Rackoff~\cite{rackoff1978covering} to derive a bound on the
length of the paths that need to be considered. We derive a complexity bound
of $ O(\Delta.h^2.\lbound^{2r+1}.r.log(\lbound))$,  where $\Delta$ is
the total number of transitions in the VCFG,  $Q$ is the number of VCFG
nodes,  $h$ is the height of lattice of $\dlattice \rightarrow \dlattice$
functions, and $\lbound = (Q.(h+1).2)^{(3r)! + 1}$.

\section{Forward DFAS Approach}
\label{sec:forward}

The Backward DFAS approach, though precise, requires the transfer function
lattice to be of finite height. Due to this restriction, infinite-height
abstract domains like Octagons~\cite{mine2006octagon}, which need
\emph{widening}~\cite{cousot1977abstract}, are not accommodated by Backward
DFAS. To address this, we present the Forward DFAS approach, which admits
\emph{any} complete lattice as an abstract domain (if the lattice is of
infinite height then a widening operator should also be provided).
The trade-off is precision. Forward
DFAS elides only some of the infeasible paths in the VCFG, and hence, in
general, computes a conservative \emph{over-approximation} of the
JOFP. Forward DFAS is conceptually not as sophisticated as Backward
DFAS, but is still a novel proposal from the perspective of the
literature. 

The Forward DFAS approach is structured as an instantiation of Kildall's data flow analysis framework~\cite{kildall1973unified}.  This framework needs a given
complete lattice, the elements of which will be propagated around the VCFG
as part of the fix point computation. Let $\dlattice$ be the given underlying finite or infinite complete lattice. $\dlattice$ either needs to not
have any infinite \emph{ascending chains} (e.g., Constant Propagation), or $\dlattice$ needs to have an associated widening operator
``$\widen_\dlattice$''. The complete lattice $D$ that we use in our
instantiation of Kildall's framework is defined as  $ D \ \equiv
\ D_{r,\forwthresh} \rightarrow \dlattice$, where $\forwthresh \geq 0$ is a
user-given non-negative integer, and  $D_{r,\forwthresh}$ is the set of all
vectors of size $r$ (where $r$ is the
number of counters in the VCFG) such that all entries of the vectors are
integers in the range $[0,\forwthresh]$.  The ordering on this lattice is as follows: $(d_1 \in D) \sqsubseteq (d_2 \in D)$ iff $\forall c \in D_{r,\forwthresh}.\ d_1(c)
\sqsubseteq_\dlattice d_2(c)$. If a widening operator $\widen_\dlattice$
has been provided for $\dlattice$, we define a widening operator $\widen$
for $D$ as follows: $d_1 \widen d_2 \ \equiv \ \lambda c \in
D_{r,\forwthresh}. \: d_1(c) \, \widen_{\dlattice} \,d_2(c)$. 

We now need to define the abstract transfer functions with signature $D
\rightarrow D$ for the VCFG edges, to be used  within the
data flow analysis. As an intermediate step to this end, we define a
ternary relation $\boundedmovei$ as follows. Any triple of integers $(p, q, s) \in \boundedmovei$
iff
$$
\begin{array}{lr}
\ (0 \leq p \leq \forwthresh) \ \wedge \\
\ ( (q \geq 0 \wedge p + q \leq \forwthresh \wedge s = p + q)\: \vee
& \hspace*{2cm} \mathit{(a)}\\ 
\ \ \: (q \geq 0 \wedge p + q > \forwthresh \wedge s = \forwthresh)\:
\vee & \hspace*{2cm} \mathit{(b)}\\
\ \ \: (q < 0 \wedge p = \forwthresh \wedge 0 \leq s \leq \forwthresh
\wedge \forwthresh - s \leq -1*q) \: \vee& \hspace*{2cm} \mathit{(c)}\\
\ \ \: (q < 0 \wedge p < \forwthresh \wedge p + q \geq 0 \wedge s = p +
q)) & \hspace*{2cm} \mathit{(d)}
\end{array}
$$

We now define a ternary relation $\boundedmove$ on vectors. 
A triple of vectors $(c_1, c_2, c_3)$ belongs to 
relation $\boundedmove$ iff all three vectors are of the same size, and for
each index $i$, $(c_1[i], c_2[i], c_3[i]) \in \boundedmovei$.

We now define the $D \rightarrow D$ transfer function for the VCFG edge
$q_1 \vcfgtrans{f}{w} q_2$ as follows:
$$ \fun(l \in D) \ \equiv \ \lambda c_2 \in D_{r,\forwthresh}. \,
\left(\bigsqcup_{c_1 \mathrm{\ such\ that\ } (c_1,w,c_2) \in \boundedmove}
f(l(c_1))\right) $$


Finally, let $l_0$ denote following function: $\lambda c \in D_{r,\forwthresh}. \,
\mbox{if $c$ is $\vec{0}$ then $d_0$ else $\bot$}$, where $d_0 \in \dlattice$. We can now invoke Kildall's algorithm using the $\fun$ transfer functions
defined above at all VCFG edges, using $l_0$ as the fact at the ``entry''
to the ``main'' procedure.
After Kildall's algorithm has finished computing the fix point solution, if
$l^D_v \in D$ is the fix point solution at any node $v$, we return the
value $\left( \sqcup_{c \in D_{r,\forwthresh}} l^D_v(c) \right)$ as the final result at $v$.

The intuition behind the approach above is as follows. If $v$ is a vector
in the set $D_{r,\forwthresh}$, and if $(c,m)$ is a channel-message pair,
then the value in the $(c,m)$th slot of $v$ encodes the number of instances
of message $m$ in channel $c$ currently. An important note is that if this
value is $\forwthresh$, it actually indicates that there are $\forwthresh$
or more instances of message $m$ in channel $c$, whereas if the value is
less than $\forwthresh$ it represents itself. Hence, we can refer to
vectors in $D_{r,\forwthresh}$ as \emph{bounded queue configurations}. 
If $d \in D$ is a data flow
fact that holds at a node of the VCFG after data flow analysis terminates,
then for any $v \in D_{r,\forwthresh}$ 
if $d(v) = l$, it indicates that $l$ is a (conservative over-approximation)
of the join of the data flow facts brought by all feasible paths that reach
the node such that the counter values at the ends of these paths are as
indicated by $v$ (the notion of what counter values are indicated by a
vector $v \in D_{r,\forwthresh}$ was described earlier in this paragraph).

The relation $\boundedmove$ is responsible for blocking the propagation
along some of the infeasible paths. The intuition behind it is as follows.
Let us consider a VCFG edge $q_1 \vcfgtrans{f:\dlattice \rightarrow
  \dlattice}{w} q_2$. If $c_1$ is a bounded queue configuration at node
$q_1$, then, $c_1$ upon propagation via this edge will become a bounded
queue configuration $c_2$ at $q_2$ iff $(c_1, w, c_2) \in
\boundedmove$. Lines~(\emph{a}) and~(\emph{b}) in the definition of
$\boundedmovei$ correspond to sending a message; line~(\emph{b}) basically
throws away the precise count when the number of messages in the channel
goes above $\forwthresh$. Line~(\emph{c}) corresponds to receiving a
message when all we know is that the number of messages currently in the
channel is greater than or equal to $\forwthresh$.
Line~(\emph{d}) is key for
precision when the channel has less than $\forwthresh$ messages, as it
allows a receive operation to proceed only if the requisite number of
messages are present in the channel.

The formulation above extends naturally to inter-procedural VCFGs using
generic inter-procedural frameworks  such as the \emph{call strings}
approach~\cite{sharirPnueliInterproc}. We omit the details of this in the
interest of space. 

\paragraph{\textbf{Properties of the approach: }} Since Forward DFAS is an instantiation of Kildall's algorithm, it derives its properties from the same. As the set $D_{r,k}$ is a finite set, it is easy to see that the fix-point algorithm will terminate.

To argue the soundness of the algorithm, we consider the \emph{concrete
  lattice} $D_c \ \equiv\ D_r \rightarrow \dlattice$, and the following
``concrete'' transfer function for the VCFG edge $q_1 \vcfgtrans{f}{w}
q_2$: $ \funconc(l \in D_c) \ \equiv \ \lambda c_2 \in D_r. \,
\left(\bigsqcup_{c_1 \in D_r \mathrm{\ such\ that\ } c_1 + w = c_2}
f(l(c_1))\right)$, where $D_r$ is the set of all vectors of size $r$ of
natural numbers. We then argue that the abstract transfer function \emph{fun}
defined earlier is a \emph{consistent
  abstraction}~\cite{cousot1977abstract} of $\funconc$. This soundness
argument is given in detail in the appendix.


If we restrict our discussion to single-procedure systems, the complexity of
our approach is just the complexity of applying Kildall's algorithm. This
works out to $O(Q^2 \forwthresh^r h)$, where $Q$ is the  number of VCFG
nodes, and $h$ is either the height of the lattice $\dlattice$ or the
 maximum increasing sequence of values from $\dlattice$ that is
obtainable at any point  using the
lattice $\dlattice$ in conjunction with Kildall's algorithm,
using the given widening operation
$\widen_\dlattice$.

\paragraph{\textbf{Illustration:}}

\begin{figure}
  \centering
  \begin{scriptsize}
  \begin{tabular}{cccccc}
    \begin{tabular}{|l||l|l|l|l|}\hline
   $c$  & t & x & y & z \\\hline\hline
  1   & 0 & 0 & 0 & 0 \\\hline
  \end{tabular}~~ &

  \begin{tabular}{|l||l|l|l||l|}\hline
   $m$   & t & x & y & z \\\hline\hline
  2   & 0 & 1 & 0 & 1 \\\hline
  \end{tabular}~~ &

    \begin{tabular}{|l||l|l|l|l|}\hline
   $c$  & t & x & y & z \\\hline\hline
  1   & 0 & 0 & 0 & 0 \\\hline
  2   & 0 & 1 & 0 & 1 \\\hline
  \end{tabular}~~ &

  \begin{tabular}{|l||l|l|l||l|}\hline
   $m$   & t & x & y & z \\\hline\hline
  2   & 0 & 1 & 0 & 1 \\\hline
  3   & 1 & 2 & 1 & 1 \\\hline
  \end{tabular}~~ &

    \begin{tabular}{|l||l|l|l|l|}\hline
   $c$  & t & x & y & z \\\hline\hline
  1   & 0 & 0 & 0 & 0 \\\hline
  2   & 0 & 1 & 0 & 1 \\\hline
  3   & 1 & 2 & 1 & 1 \\\hline
  \end{tabular}~~ &

  \begin{tabular}{|l||l|l|l||l|}\hline
   $m$   & t & x & y & z \\\hline\hline
  2   & 0 & 1 & 0 & 1 \\\hline
  3   & 1 & $\top$ & $\top$ & 1 \\\hline
  \end{tabular} \\

  (1) & (2) & (3) & (4) & (5) & (6) \\

  \begin{tabular}{|l||l|l|l||l|}\hline
   $o$   & t & x & y & z \\\hline\hline
  1   & 0 & 0 & 0 & 0 \\\hline
  2   & 0 & 1 & 0 & 1 \\\hline
  3   & 1 & $\top$ & $\top$ & 1 \\\hline
  \end{tabular}~~ &

  \begin{tabular}{|l||l|l|l||l|}\hline
   $k$   & t & x & y & z \\\hline\hline
  1   & 1 & $\top$ & $\top$ & 1 \\\hline
  2   & 1 & $\top$ & $\top$ & 1 \\\hline
  3   & 1 & $\top$ & $\top$ & 1 \\\hline
  \end{tabular} \\

  (7) & (8)

  \end{tabular}
  \end{scriptsize}
  
	\caption{Data flow facts over a run of the algorithm}
	\label{fig:forw-illus}
\end{figure}

We illustrate Forward DFAS using the example in Figure~\ref{fig:runningex}.
Figure~\ref{fig:forw-illus} depicts the data flow values at four
selected nodes as they get updated over eight selected points of time
during the run of the algorithm.  In this illustration we assume a context
insensitive analysis for simplicity (it so happens that context sensitivity
does not matter in this specific example). We use the value $\forwthresh =
3$.  Each small table is a data flow fact, i.e., an element of $ D \ \equiv
\ D_{r,\forwthresh} \rightarrow \dlattice$. The top-left cell in the table shows the
node at which the fact arises. In each row the first column shows the
counter value, while the remaining columns depict the known constant value
of the variables ($\top$ indicates unknown). Here are some interesting
things to note. When any tuple of constant values transfers along the path
from node $c$ to node $m$, the constant values get updated due to the
assignment statements encountered, \emph{and} this tuple shifts from
counter $i$ to counter $i+1$ (if $i$ is not already equal to $\forwthresh$)
due to the ``send'' operation encountered. When we transition from Step~(5)
to Step~(6) in the figure, we get $\top$'s, as counter values~2 and~3 in
Step~(5) both map to counter value~3 in Step~(6) due to $\forwthresh$ being
3 (hence, the constant values get \emph{joined}). The value at node $o$ (in
Step~(7)) is the join of values from Steps~(5) and~(6). 
 Finally, when the value at node $o$ propagates to node $k$, the tuple
of constants associated with counter value~3 end up getting mapped to all
lower values as well due to the receive operations encountered. 

Note, the precision of our approach in general increases with the value of
$\forwthresh$ (the running time increases as well). For instance, if
$\forwthresh$ is set to 2 (rather than 3) in the example, some more
infeasible paths would be traversed. Only z = 1 would be inferred at node
$k$, instead of (t = 1, z = 1). 

\section{Implementation and Evaluation}
\label{sec:impl-and-eval}


We have implemented prototypes of both the Forward DFAS and Backward DFAS
approaches, in Java. Both the implementations have been parallelized, using
the ThreadPool library. With Backward DFAS the iterations of the outer
``repeat'' loop in Algorithm~\ref{algo:main} run in parallel, while with
Forward DFAS propagations of values from different nodes to their
respective successors happen in parallel. Our implementations currently
target systems without procedure calls, as none of our benchmarks had recursive procedure calls.

Our implementations accept a given system, and a ``target'' control state $q$
in one of the processes of the system at which the JOFP is desired. They then
construct the VCFG from the system (see Section~\ref{ssec:proc-to-vcfg}), and
identify the \emph{target set} of $q$, which is the set of VCFG nodes in
which $q$ is a constituent.  For instance, in
Figure~\ref{fig:notationAsyncSystem}, the target set for control state $e$
is $\{(a,e), (b,e)\}$. The JOFPs at the nodes in the target set are
then computed, and the join of these JOFPs is returned as the result for $q$.



Each variable reference in any transition leaving any control state is
called a ``use''. For instance, in Figure~\ref{fig:notationAsyncSystem},
the reference to variable x along the outgoing transition from state $d$ is one use.  In all our experiments, the objective is to find the
uses that are definitely constants by computing the JOFP at all uses. This
is a common objective in many research papers, as finding constants enables
optimizations such as constant folding, and also checking assertions in the
code.  We instantiate Forward DFAS with the Constant Propagation (CP)
analysis, and Backward DFAS with the LCP analysis (for the reason discussed
in Section~\ref{ssec:bdfas-assump}).  We use the bound $\forwthresh=2$ in
all runs of Forward DFAS, except with two benchmarks which are too large to
scale to this bound. We discuss this later in this section. All the
experiments were run on a machine with 128GB RAM and four AMD Opteron 6386
SE processors (64 cores total).

\subsection{Benchmarks and modeling}
\begin{table}
		\centering
	\caption{Information about the benchmarks. Abbreviations used: (a) prtcl
      = protocol, (b) comm = communication, (c) app = application}
	\label{tab:benchmarks}
	\begin{footnotesize}
		\begin{tabular}{|l|p{1.8in}|r|r|r|r|}
			\hline
			\textbf{Benchmark}& \textbf{Description} & \textbf{\#Proc} & \textbf{\#Var} & \textbf{$r$} & \textbf{\#VCFG} \\
			\textbf{(1)} & \textbf{(2)} & \textbf{(3)} & \textbf{(4)} & \textbf{(5)} & \textbf{nodes (6)} \\ \hline
			mutex & mutual exclusion example & 3 & 1 & 6 & 4536 \\ \hline
			bartlett & Bartlett's alternating-bit prtcl &  3 & 3 & 7 & 17864 \\ \hline
			leader & leader election prtcl &  2 & 11 & 12 & 16002 \\ \hline
			lynch & distorted channel comm prtcl &  3 & 5 & 27 & 168912 \\ \hline
			peterson & Peterson's mutual exclusion prtcl &  3 & 4 & 4 & 6864 \\ \hline
			boundedAsync  & illustrative example  &  3 & 5 & 10 & 14375 \\ \hline
			receive1 & illustrative example &  2 & 5 & 13 & 1160 \\ \hline
			server &  actor-based client server app &  3 & 3 & 6 & 1232 \\ \hline
			chameneos  & Chameneos concurrency game &  3 & 9 & 10 & 45584 \\ \hline
			replicatingStorage  & replicating storage system &  4 & 4 & 8 & 47952 \\ \hline
			event\_bus\_test  & publish-subscribe system &  2 & 2 & 5 & 160 \\ \hline
			jobqueue\_test  & concurrent job queue system &  4 & 1 & 10 & 28800 \\ \hline
			bookCollectionStore  & REST app &  2 & 2 & 12 & 2162 \\ \hline			
			nursery\_test  & structured concurrency app &  3 & 2 & 4 & 1260 \\ \hline			
		\end{tabular}
	\end{footnotesize}
    %
    %
\end{table}


We use 14 benchmarks for our evaluations. These are described in
the first two columns  of
Table~\ref{tab:benchmarks}. Four benchmarks -- bartlett, leader,
lynch, and peterson -- are Promela models for the
Spin model-checker.
Three benchmarks -- boundedAsync, receive1, and replicatingStorage -- are
from the P language repository ({\small \url{www.github.com/p-org}}). Two
benchmarks -- server and chameneos -- are  from the Basset
repository
({\small \url{www.github.com/SoftwareEngineeringToolDemos/FSE-2010-Basset}}). Four benchmarks --
event\_bus\_test, jobqueue\_test, nursery\_test, and bookCollectionStore -- are
real world Go programs.
There is one toy example ``mutex'', for ensuring mutual exclusion, via blocking receive messages, that we have made
ourselves.  We provide precise links to the benchmarks
in the appendix.



Our DFAS implementations expect the asynchronous system to be specified in
an XML format. We have developed a custom XML schema for this, closely based
on the Promela modeling language used in Spin~\cite{holzmann2004spin}. We
followed this direction in order to be able to evaluate our approach on
examples from different languages. We manually translated each benchmark
into an XML file, which we call a \emph{model}.  As the input XML schema is
close to Promela, the Spin models were easily translated. Other benchmarks
had to be translated to our XML schema by understanding their semantics.

Note that both our approaches are  expensive in the worst-case (exponential
or worse in the number of counters $r$). Therefore, we have chosen
benchmarks that are moderate in their complexity metrics. Still, these
benchmarks  are real and contain
 complex logic (e.g., the leader election example from Promela, which
was discussed in detail in Section~\ref{ssec:motivatingExample}). We have
also performed some manual simplifications to the benchmarks to aid
scalability (discussed below). Our evaluation is aimed towards
understanding the impact on precision due to infeasible paths in real
benchmarks, and not necessarily to evaluate applicability of our approach to large systems.

We now list some of the simplifications referred to above.
Language-specific idioms  that were irrelevant to the core logic of the benchmark were  removed.
The number of instances of identical processes in some of the
models were reduced in a behavior-preserving manner according to our best judgment.  In many of the benchmarks, messages carry \emph{payload}. Usually the payload
is one byte. We would have needed 256 counters
just to encode the payload of one 1-byte
message. Therefore,  in the interest of keeping the analysis time manageable, the payload size was reduced
to 1 bit or 2 bits. The reduction was done while preserving key behavioral
aspects according to our best judgment. Finally, procedure calls were
inlined (there was no use of recursion in the benchmarks).

In the rest of this section, whenever we say ``benchmark'', we actually
mean the model we created corresponding to the benchmark.
Table~\ref{tab:benchmarks} also shows various metrics of our benchmarks
(based on the XML models).  Column~3-6 depict, respectively, the number of
processes, the total number of variables, the number of ``counters'' $r$,
and the total number of nodes in the VCFG.
We provide our XML models of all our benchmarks, as well as full output
files from the runs of our approach, as a downloadable
folder~(\url{https://drive.google.com/drive/folders/181DloNfm6_UHFyz7qni8rZjwCp-a8oCV}).

\subsection{Data flow analysis results}

\begin{table}
	\caption{Data flow analysis results}
	\label{tab:dataresults}
	\begin{minipage}{\columnwidth}
			\centering
		\begin{footnotesize}
			\begin{tabular}{|l|r|r|r|r|r|r||r|r|r|r|}
				\hline 
				& & & \multicolumn{4}{|c||}{DFAS Approach} & \multicolumn{4}{|c|}{Baseline Approaches} \\ \cline{4-11} 
				Benchmark & \multicolumn{1}{|c|}{\#Var.} & \#Asserts &  \multicolumn{2}{|c|}{\#Consts.} & \multicolumn{2}{|c||}{\#Verified}&\multicolumn{2}{|c|}{\#Consts.} & \multicolumn{2}{|c|}{\#Verified} \\ 
				(1) & \multicolumn{1}{|c|}{uses (2)} & \multicolumn{1}{|c|}{ (3)} &  \multicolumn{2}{|c|}{(4)} & \multicolumn{2}{|c||}{(5)}&\multicolumn{2}{|c|}{(6)} & \multicolumn{2}{|c|}{(7)} \\ \cline{4-11}
				&  &  & Forw. & Back. & Forw. & Back. & JOP & CCP & JOP & CCP \\ \hline
				mutex & 6 & 2 & 6 & 6 & 2 & 2 & 0 & 0 & 0 & 0 \\ \hline           
				bartlett & 9 & 1 & 0 & 0 & 0 & 0 & 0 & 0 & 0 & 0  \\ \hline           
				leader & 54 & 4 & 20 & 6 & 4 & 0 & 6 & 6 & 2 & 0 \\ \hline           
				lynch & 6 & 2 & 4 & 3 & 0 & 0 & 4 & 3 & 0 & 0 \\ \hline           
				peterson & 14 & 2 & 0 & 0 & 0 & 0 & 0 & 0 & 0 & 0 \\ \hline           
				boundedAsync & 24 & 8 & 8 & 8 & 0 & 0 & 8 & 8 & 0 & 0  \\ \hline           
				receive1 & 9 & 5 & 8 & 8 & 4 & 4 & 2 & 8 & 2 & 4 \\ \hline           
				server & 4 & 1 & 0 & 0 & 0 & 0 & 0 & 0 & 0 & 0 \\ \hline           
				chameneos & 35 & 2 & 2 & 2 & 0 & 0 & 2 & 2 & 0 & 0 \\ \hline           
				replicatingStorage & 8 & 1 & 2 & 0 & 1 & 0 & 0 & 0 & 0 & 0  \\ \hline           
				event\_bus\_test & 5 & 3 & 3 & 3 & 3 & 3 & 0 & 2 & 0 & 2  \\ \hline           
				jobqueue\_test & 3 & 1 & 0 & 1 & 0 & 1 & 0 & 0 & 0 & 0 \\ \hline           
				bookCollectionStore & 10 & 8 & 8 & 10 & 6 & 8 & 0 & 8 & 0 & 6 \\ \hline 
				nursery\_test & 2 & 2 & 2 & 2 & 2 & 2 & 0 & 2 & 0 & 2  \\ \hline\hline    
				\textbf{Total} & 189 & 42 & 63 & 49 & 22 & 20 & 22 & 39 & 4 & 14 \\ \hline      
			\end{tabular}
			
		\end{footnotesize}
	\end{minipage}

\end{table}


We structure our evaluation as a set  of research questions (RQs) below. 
Table~\ref{tab:dataresults} summarizes results for the first three RQs, while Table~\ref{tab:timeresults} summarizes results for RQ 4.

\paragraph{\textbf{RQ 1: } How many constants are identified by the Forward and Backward DFAS approaches?} 



Column (2) in Table~\ref{tab:dataresults} shows the number of \emph{uses} in each
benchmark. Columns~(4)-Forw and (4)-Back show the number of uses identified
as constants by the Forward and Backward DFAS approaches, respectively. In
total across all benchmarks Forward DFAS identifies 63 constants whereas
Backward DFAS identifies 49 constants.

Although in aggregate Backward DFAS appears weaker than Forward DFAS,
Backward DFAS infers more constants than Forward DFAS in two benchmarks --
jobqueue\_test and bookCollectionStore. Therefore, the two approaches are
actually incomparable. The advantage of Forward DFAS is that it can use
relatively more precise analyses like CP that do not satisfy the assumptions of
Backward DFAS, while the advantage of Backward DFAS is that it always computes the precise JOFP. 


\paragraph{\textbf{RQ 2: } How many assertions are verified by the approaches?}


Verifying assertions that occur in code is a useful activity as it gives
confidence to developers. All but one of our benchmarks had assertions (in
the original code itself, before modeling). We carried over these
assertions into our models.  For instance, for the benchmark \emph{leader}, the assertion appears in Line~11 in Figure~\ref{fig:introExample}. In some benchmarks, like jobqueue\_test, the
assertions were part of test cases. It makes sense to verify these
assertions as well, as unlike in testing, our technique 
considers all possible interleavings of the processes. As ``bookCollectionStore'' did not come with any assertions, a
graduate  student who was unfamiliar with
our work studied the benchmark and suggested  assertions.



Column~(3) in Table~\ref{tab:dataresults} shows the number of assertions
present in each benchmark. Columns~(5)-Forw and (5)-Back in
Table~\ref{tab:dataresults} show the number of assertions declared as safe
(i.e., verified) by the Forward and Backward DFAS approaches,
respectively. An assertion is considered verified iff constants (as opposed
to ``$\top$'') are inferred for all the variables used in the assertion, and
if these constants satisfy the assertion.  As can be seen from the last row
in Table~\ref{tab:dataresults}, both approaches verify a substantial
percentage of all the assertions -- 52\% by Forward DFAS and 48\% by
Backward DFAS.
We believe these results are surprisingly useful, given that our technique
needs no loop invariants or usage of theorem provers.

\paragraph{\textbf{RQ 3: } Are the DFAS approaches more precise than baseline approaches?}

We compare the DFAS results with two baseline approaches. The first
baseline is a Join-Over-all-Paths (JOP) analysis, which basically performs CP
analysis on the VCFG without eliding any infeasible paths.  Columns~(6)-JOP
and (7)-JOP in Table~\ref{tab:dataresults} show the number of constants
inferred and the number of assertions verified by the JOP baseline. It can
be seen that Backward DFAS identifies 2.2 times the number of constants as
JOP, while Forward DFAS identifies 2.9 times the number of constants as
JOP (see columns~(4)-Forw, (4)-Back, and (6)-JOP in the \textbf{Total} row
in Table~\ref{tab:dataresults}).
In terms of assertions, each of them verifies almost 5 times as many
assertions as JOP  (see  columns~(5)-Forw, (5)-Back, and (7)-JOP in
\textbf{Total} row in
Table~\ref{tab:dataresults}.) It is clear from the results that eliding
infeasible paths is extremely important for precision.

The second baseline is Copy Constant Propagation
(CCP)~\cite{reps1995precise}. This is another variant of constant propagation
that is even less precise than LCP. However, it is based on a finite lattice,
specifically, an IFDS~\cite{reps1995precise} lattice. Hence this baseline
represents the capability of the closest related work to
ours~\cite{jhala2007interprocedural}, which elides infeasible paths but
supports only IFDS lattices, which are a sub-class of finite
lattices. (Their implementation also used a finite lattice of predicates,
but we are not aware of a  predicate-identification tool that would work
on our benchmarks out of the box.)
We implemented the CCP baseline within our Backward DFAS framework. This
baseline hence computes the JOFP using CCP (i.e., it elides infeasible
paths).

Columns (6)-CCP and (7)-CCP in Table~\ref{tab:dataresults} show the number
of constants inferred and the number of assertions verified by the CCP
baseline. From the \textbf{Total}
row in Table~\ref{tab:dataresults} it can be seen that Forward DFAS
finds 62\% more constants than CCP, while Backward DFAS finds 26\% more
constants than CCP.  With respect to number of assertions verified, the
respective gains are 57\% and 43\%. In other words, infinite domains such
as CP or LCP can give significantly more precision than closely related
finite domains such as CCP.

\begin{table}[t]
  \caption{Execution time in seconds}
  \label{tab:timeresults}
  \centering
  \begin{footnotesize}
    \begin{tabular}{|l|r|r|r|r|r|r|r|r|r|r|r|r|r|r|}\hline
         & mut. & bar. & lea. & lyn. & pet. & bou. & rec. & ser. & cha. & rep. & eve. & job. & boo. & nur. \\\hline
      Forw & 1.2 & 14.0 & 1.3 & 8.0 & 1.2 & 21.0 & 1.2 & 1.2 & 18.0 & 2.4 & 1.2 & 1.2 & 1.2 & 1.2 \\\hline
      Back & 5.0 & 11.0 & 284.0 & 118.0 & 13.0 & 21.0 & 8.0 & 3.0 & 220.0 & 21.0 & 3.0 & 140.0 & 16.0 & 1.0 \\\hline
      JOP & 1.2 & 1.3 & 1.6 & 8.0 & 1.2 & 1.4 & 1.3 & 1.2 & 3.1 & 3.0 & 1.1 & 1.4 & 1.2 & 1.2 \\\hline
      CCP & 5.0 & 12.0 & 226.0 & 116.0 & 12.0 & 14.0 & 8.0 & 3.0 & 156.0 & 24.0 & 3.0 & 51.0 & 30.0 & 1.0 \\\hline
    \end{tabular}
  \end{footnotesize}
\end{table}

\paragraph{\textbf{RQ 4:} How does the execution cost of DFAS approaches compare to the cost of the JOP baseline?}

 The columns in Table~\ref{tab:timeresults}
 correspond to the benchmarks (only first three letters of each benchmark's name
 are shown in the interest of space). The rows show the running times for
 Forward DFAS, Backward DFAS, JOP baseline, and CCP baseline, respectively.

The JOP baseline was quite fast on almost all benchmarks (except lynch). This is
because it maintains just a single data flow fact per VCFG node, in contrast
to our approaches.
Forward DFAS was generally quite efficient, except on chameneos and
lynch. On these two benchmarks, it scaled only with $\forwthresh=1$ and
$\forwthresh=0$, respectively, encountering memory-related crashes at
higher values of $\forwthresh$ (we used $\forwthresh=2$ for all other benchmarks).
These two benchmarks have large number of nodes and a high value of $r$, which increases the size of the data flow facts.

The running time of Backward DFAS is substantially higher than the JOP
baseline. One reason for this is that being a demand-driven approach, the
approach is invoked separately for each \emph{use} (Table~\ref{tab:dataresults}, Col.~2), and the cumulative time
across all these invocations is reported in the table. In fact, the mean time per query for Backward DFAS is less than the total time for Forward DFAS on 9 out of 14 benchmarks, in some cases by a factor of 20x. Also, unlike Forward DFAS, Backward DFAS visits a small portion of the VCFG in each invocation. Therefore, Backward DFAS is more memory efficient and scales to all our benchmarks. Every invocation of Backward DFAS 
consumed less than 32GB of memory, whereas with Forward DFAS, three
benchmarks (leader, replicatingStorage, and jobqueue\_test) required more
than 32GB, and two (lynch and chameneos) needed more than the 128 GB that was available in the machine. On the
whole, the time requirement of Backward DFAS is still acceptable
considering the large precision gain over the JOP baseline.



\subsection{Limitations and Threats to Validity}

The results of the evaluation using our prototype implementation are very
encouraging, in terms of both usefulness and efficiency. The evaluation does
however pose some threats to the validity of our results. The benchmark
set, though extracted from a wide set of sources, may not be exhaustive in
its idioms. Also, while modeling, we had to simplify some of the features of
the benchmarks in
order to let the approaches scale. Therefore, applicability of our approach
directly on real systems with all their language-level complexities, use of
libraries, etc., is not yet established, and would be a very interesting
line of future work.

\section{Related Work}
\label{sec:relwork}

The modeling and analysis of \emph{parallel systems}, which include
asynchronous systems, multi-threaded systems, distributed systems,
event-driven systems, etc., has been the focus of a large body of work, for
a very long time. We discuss some of the more closely related previous
work, by dividing the work into four broad categories.


\paragraph{Data Flow Analysis:} The work of Jhala et al.~\cite{jhala2007interprocedural} is the closest work that addresses
similar challenges as our work. They combine the Expand, Enlarge and Check
(EEC) algorithm~\cite{geeraerts2006expand} that answers control state
reachability in WSTS~\cite{finkel2001well}, with  the unordered channel abstraction, and  the IFDS~\cite{reps1995precise} algorithm for data flow analysis, to
compute the JOFP solution for all nodes. They admit only IDFS abstract
domains, which are finite by definition. Some recent work has extended this
approach for analyzing JavaScript ~\cite{yee2019precise} and
Android~\cite{mishra2016asynchrony} programs. Both our approaches are
dissimilar to theirs, and we admit  infinite lattices (like CP and LCP).
On the other hand, their approach is able to handle parameter
passing between procedures, which we do not.

Bronevetsky et al.~\cite{bronevetsky2009communication} address generalized
data flow analysis of a very restricted class of systems, where any receive
operation must receive messages from a specific process, and channel
contents are not allowed to cause non-determinism in control flow.  Other
work has addressed analysis of asynchrony in 
web applications~\cite{jensen2011modeling,madsen2015static}. These approaches
are  efficient, but over-approximate the JOFP by eliding only certain
specific types of infeasible paths.


\paragraph{Formal Modeling and Verification:} Verification of asynchronous
systems has received a lot of attention over a long
time. VASS~\cite{karp1969parallel} and Petri nets~\cite{reisig2012petri}
(which both support unordered channel abstraction)
have been used widely to model parallel and asynchronous
processes~\cite{karp1969parallel,lautenbach74petrinet,sen2006model,jhala2007interprocedural,ganty2009verifying,bouajjani2012analysis}. Different analysis problems based on these models
have been studied, such as  reachability of configurations~\cite{brand1983communicating,mayr1981complexity,kosaraju1982decidability,lambert1992structure},
coverability and boundedness~\cite{karp1969parallel,abdulla1996verifying,abdulla1996general,finkel2001well,geeraerts2006expand,bozzelli2011complexity},
and coverability in the presence of stacks or other data structures~\cite{Torre2008ContextBoundedAO,bouajjani2012analysis,cai2013well,chadha2007decidability,leroux2014hyper}. 

The \emph{coverability} problem mentioned above is considered equivalent to
control state reachability, and has received wide
attention~\cite{abdulla1998fly,delzanno2002towards,jhala2007interprocedural,ganty2009verifying,sen2006model,geeraerts2015verification,kochems2013safety,bouajjani2012analysis,stievenart2017mailbox}. Abdulla
et al.~\cite{abdulla1996verifying} were the first to provide a backward
algorithm to answer coverability.  Our Backward DFAS approach is
structurally similar to their approach, but is a strict generalization, as we incorporate data flow analysis using infinite abstract
domains. (It is noteworthy that when the abstract domain is finite, then data
flow analysis can be reduced to coverability.)
One difference is that we use the unordered
channel abstraction, while they use the lossy channel abstraction. It is
possible to modify our approach to use lossy channels as well (when there
are no procedure calls, which they also do not allow); we omit the
formalization of this due to lack of space.

Bouajjani and Emmi~\cite{bouajjani2012analysis} generalize over previous
coverability results by solving the coverability problem for a  class of
multi-procedure systems called recursively parallel programs. Their class
of systems is somewhat broader than ours, as they allow a caller to receive
the messages sent by its callees. Our \textsc{ComputeEndToEnd} routine in
Algorithm~\ref{algo:interfp} is structurally similar to their
approach. They admit finite abstract  domains only.
It would be interesting future work to extend the Backward DFAS approach to their class of systems.

Our approaches explore all interleavings between the processes, following
the Spin semantics. Whereas, the closest previous
approaches~\cite{jhala2007interprocedural,bouajjani2012analysis}
only address ``event-based'' systems, wherein a set of processes execute
sequentially without interleaving at the statement level, but over an
unbounded schedule (i.e., each process executes from start to finish
whenever it is scheduled).

\paragraph{Other forms of  verification:} Proof-based techniques have been
explored for verifying asynchronous and distributed
systems~\cite{hawblitzel2015ironfleet,wilcox2015verdi,padon2016ivy,v2019pretend}. These
techniques need inductive variants and are not as user-friendly as data
flow analysis techniques. Behavioral types have been used to tackle
specific analysis problems such as deadlock detection and correct usage of
channels~\cite{lange2017fencing,lange2018static,scalas2019verifying}.

\paragraph{Testing and Model Checking:} Languages and tools such as Spin
and Promela~\cite{holzmann2004spin}, P~\cite{desai2013p},
P\#~\cite{deligiannis2015asynchronous}, and JPF-Actor~\cite{jpfActor} have
been used widely to model-check asynchronous systems.
A lot of work has been done in testing of asynchronous systems~\cite{desai2015systematic,deligiannis2015asynchronous,sen2006automated,guo2011practical,yang2009modist} as well. Such techniques are bounded in nature and cannot provide the strong verification guarantees that data flow analysis provides.

\section{Conclusions and Future Work}
\label{sec:concl}

In spite of the substantial body of work on analysis and verification of distributed
systems, there is no existing approach  that  performs precise
data flow analysis of such systems using infinite abstract domains, which
are otherwise very commonly used with sequential programs.  We propose two
data flow analysis approaches that solve this problem -- one computes the
precise JOFP solution always, while the other one admits a fully general
class of infinite abstract domains.  We have implemented
our approaches, analyzed 14 benchmarks using the implementation, and have
observed substantially higher precision from our approach over
two different baseline approaches.

Our approach can be extended in many ways. One interesting extension would
be to make  Backward DFAS  work with infinite height lattices, using
widening. Another possible extension could be the handling of parameters in
procedure calls.  There is significant scope for improving
the scalability using better engineering, especially for Forward DFAS. One
could explore the integration of partial-order
reduction~\cite{clarke1999state} into both our approaches.  Finally, we
would like to build tools based on our approach that apply directly to
programs written in commonly-used languages for distributed programming.


\section{Appendix}

\subsection{Benchmark Sources}

Following are the links to the sources of the benchmarks used in the paper.

\begin{enumerate}
	\item mutex : self-made
	\item bartlett : \url{www.imm.dtu.dk/~albl/promela-models.zip}, file assertion.barlett.prm 
	\item leader :  \url{www.imm.dtu.dk/~albl/promela-models.zip}, file assertion.leader.prm 
	\item lynch : \url{www.imm.dtu.dk/~albl/promela-models.zip}, file assertion.lynch.prm 
	\item peterson : \url{www.github.com/nimble-code/Spin/blob/master/Examples/peterson.pml}
	\item boundedAsync : \url{www.github.com/p-org/PSharp/tree/master/Samples/Framework/BoundedAsync}
	\item receive1 : \url{www.github.com/p-org/P/blob/master/Tst/RegressionTests/Feature2Stmts/Correct/receive1/receive1.p}
	\item server : \url{www.github.com/SoftwareEngineeringToolDemos/FSE-2010-Basset/tree/master/jpf/jpf-actor/src/examples/server}
	\item chameneos : \url{www.github.com/SoftwareEngineeringToolDemos/FSE-2010-Basset/tree/master/jpf/jpf-actor/src/examples/chameneos}
	\item replicatingStorage :  \url{www.github.com/p-org/PSharp/tree/master/Samples/Framework/ReplicatingStorage}
	\item event\_bus\_test : \url{www.github.com/asaskevich/EventBus/blob/master/event_bus_test.go}  
	\item jobqueue\_test : \url{www.github.com/dirkaholic/kyoo/blob/master/jobqueue_test.go}  
	\item bookCollectionStore : \url{www.github.com/jirenius/go-res/tree/master/examples/04-book-collection-store}
	\item nursery\_test : \url{www.github.com/arunsworld/nursery/blob/master/nursery_test.go} 
	
\end{enumerate}

\subsection{Complete Proofs for Backward DFAS}

In this section, we formally prove the termination and correctness of our algorithm \textsc{ComputeJOFP}. The proofs in the appendix are self-contained and only refer to Algorithm~1 and Algorithm~2 in the paper. Before presenting the proofs, we revise the important definitions.

\begin{definition}[Demand]
	\label{def:demand}
	For a path $p$, and vector $d \in {\natzero}^r$
	$$
	\begin{array}{lcll}
	\demandpd{p}{d} & = & \mu(d - w), & \mathrm{if\ } p = (v \vcfgtrans{f}{w} v') \\
	& = & \mu(\demandpd{p'}{d} - w), 
	& \mathrm{if\ } p = (e \concat p'), \mathrm{where\ } e \equiv (v
	\vcfgtrans{f}{w} v')
	\end{array}
	$$
\end{definition}

Here, $\mu \equiv \lambda z \in \mathbb{Z}.(max(0,z))$. $\mu$ can be applied to vectors of integers in the natural manner, that is, by applying $\mu$ to each component of the vector. This definition is equivalent to the definition presented in the paper (if $d$ is replaced by $\zerovector$).

\begin{definition}[Covering]
	\label{def:cover}	
	A set of paths $C$ is said to \emph{cover} a path $p$ iff
	
	\begin{itemize}
		\item all paths in $C$ have the same start and end nodes
		(respectively) as $p$, and
		\item for each path $p' \in C$, $\demandp{p'} \leq \demandp{p}$, and
		\item 
		the join of the path transfer functions of all these paths dominates the
		path transfer function of $p$.
	\end{itemize}
\end{definition}

\begin{definition}[Path Template]
	\label{def:pathtemplate}
	
	A \emph{path template} $(\vpath{p_1, p_2, \ldots, p_n)}$ of a
	procedure $F \in \funcs$ is a sequence of paths in the VCFG of $F$ such
	that
	
	\begin{itemize}
		\item $n \geq 2$
		\item path $p_1$ begins at node $\entrynode{F}$ (the entry node of the VCFG of procedure $F$) and path $p_n$ ends at node $\exitnode{F}$ (the designated exit node of the VCFG of procedure $F$)
		\item for all $p_i, 1 \leq i < n$, $p_i$ ends at a
		call-site node, and
		\item for all $p_i, 1 < i \leq n$, $p_i$ begins
		at the return-site  node corresponding to the call-site node at which $p_{i-1}$
		ends. 
	\end{itemize}
\end{definition}

\begin{definition}[D-Covering]
	\label{def:dcover}
	A path $p$ is \emph{d-covered} by a set of paths $S$, for given demand $d$ \textit{iff},
	\begin{enumerate}
		\item if $p$ begins in vertex $v_i$ and ends at vertex $v_j$, all paths in $S$ start at $v_i$ and end at $v_j$
		\item for all paths $p’ \in S$, $(\demandpd{p'}{d} \leq \demandpd{p}{d})$
		\item $\underset{p' \in S}{\bigjoin} \ptf{p'} \dominates \ptf{p}$
	\end{enumerate}	
\end{definition}

Note, when $d = \zerovector$, then $\zerovector-$covering is simply equivalent to covering as defined in Defintion~\ref{def:cover}.

\subsubsection{Termination}
\label{sec:inter:proof:termination}

\begin{theorem}[Termination]
	\label{thm:algo2:term}
	The algorithm terminates if the lattice of transfer procedures $(\dlattice \rightarrow \dlattice)$ is of finite height.
\end{theorem}

\paragraph{Proof: } We prove the theorem in two parts. First, we prove that
each invocation of the form \textsc{ComputeEndToEnd(\textit{F}, \textit{d})}, where $F \in \funcs$ and $d \in \natzero^r$, necessarily terminates. 

First, it is clear that the loop at lines 2-3 in \textsc{ComputeEndToEnd} terminates, as there are only finite number of 0-depth paths.

Now we reason about the other loop in the routine from lines~4-12. Let each visit to Line 11 in the routine \textsc{ComputeEndToEnd} in algorithm  (where a path is added to a set $\sIVCPaths{\textit{F}_i}{d}$) during the current invocation be considered as an ``event". Each event is fully described by a triple of parameters, namely:

(the procedure $F_i$ being currently visited,

\hspace{3pt} the path IVC path $p'$ in procedure $F_i$ that is currently being added to
$\sIVCPaths{F_i}{d}$,

\hspace{3pt} $\demandpd{p'}{d}$)

Therefore, the entire invocation corresponds to a sequence of events of the
kind mentioned above. Let this sequence be called $S$. Clearly, the invocation is non-terminating \textit{iff}  $S$ is infinitely long. 

Since the procedures contain only send operations, for any path $p$ that
is fully within the procedures, by Definition~\ref{def:demand}, $\demandpd{p}{d} \leq d$. Therefore, there is a
finite number of values possible in the third components of the triples
mentioned above. Also, the number of procedures in \textit{Funcs} is
finite. Therefore, if $S$ is infinite, there must exist a procedure  $F_1$ and a
vector $d'$ such that an infinite number of events in $S$ have $F_1$ in their
first component and $d'$ in their third component. Let $S'$ be the subsequence
of $S$ consisting of all  events with $F_1$ in their first component and $d'$ in
their third component; thus, $S'$ is infinite.

For any $j$, let $S'[j].\textit{path}$ denotes the second component of the event at
$S'[j]$. 

By the definition of \textsc{CoveredWithDemand}, and from Line 23 in the procedure \textsc{ComputeEndToEnd}, it follows that for every index $i \geq 1$:

$$\underset{0 \leq j < i}{\bigjoin} \ptf{S'[j].\textit{path}}$$

does not dominate $\ptf{S'[i].\textit{path}}$.

From this, it follows that for every $i \geq 1$,

$$\underset{0 \leq j \leq i}{\bigjoin} \ptf{S'[j].\textit{path}}$$ 

strictly dominates

$$\underset{0 \leq j < i}{\bigjoin} \ptf{S'[j].\textit{path}}$$

This implies that the following infinite sequence is a strictly increasing
sequence as per the $\dominates$ ordering in the lattice of transfer procedures:

$$\underset{0 \leq j \leq 0}{\bigjoin} \ptf{S'[j].\textit{path}}$$

$$\underset{0 \leq j \leq 1}{\bigjoin} \ptf{S'[j].\textit{path}}$$

$$\underset{0 \leq j \leq 2}{\bigjoin} \ptf{S'[j].\textit{path}}$$

$$...$$

However, this contradicts our assumption that the transfer procedure lattice
has no infinitely increasing chains (refer assumptions section).

Therefore, $S'$ cannot be infinite, and hence $S$ cannot be infinite. Therefore, we have proved that every invocation \textsc{ComputeEndToEnd(\textit{F}, \textit{d})} must necessarily terminate.

Next we prove that \textsc{ComputeJOFP} in Algorithm~1  always terminates. If every call to \textsc{ComputeEndToEnd} is terminating, then the
only way \textsc{ComputeJOFP} will go into non-termination is if for some node $v$, an infinite number of paths get inserted into $\sPaths{v}$ at Lines 13, 19, and 23 in \textsc{ComputeJOFP}. Here, $v$ can be any node in the VCFG of any procedure. Let

$$ S_1    =   p_1,p_2, \ldots$$

be the infinite sequence of paths inserted into $\sPaths{v}$. Since
the set of all demand vectors form a well-quasi ordering w.r.t. the $\leq$
comparison on demand vectors, there must exist an infinite subsequence $S_1'$ of $S_1$ such
that for all $i \geq 1$, $\demandp{S_1'[i]} \geq \demandp{S_1'[i-1]}$. From this, Lines 12, 18, and 22 in \textsc{ComputeJOFP} algorithm, and the definition of \textsc{covered}, it follows
that for all $i \geq 1$:

$$\underset{0 \leq j < i}{\bigjoin} \ptf{{S_1'[j]}}$$

does not dominate $\ptf{S_1'[i]}$. This implies that the following infinite sequence is strictly increasing as per the $\dominates$ ordering in the lattice of transfer procedures:

$$\underset{0 \leq j \leq 0}{\bigjoin} \ptf{S_1'[j]}$$

$$\underset{0 \leq j \leq 1}{\bigjoin} \ptf{S_1'[j]}$$

$$\underset{0 \leq j \leq 2}{\bigjoin} \ptf{S_1'[j]}$$

$$\ldots$$

However, this again contradicts our assumption that the transfer procedure lattice
has no infinite increasing chains.

Therefore, we have contradicted our initial assumption that the sequence $S_1'$
is infinite. Therefore, $S$ cannot have an infinite subsequence $S_1'$. Hence $S$ will be finite. Therefore, procedure \textsc{ComputeJOFP} always terminates. 

\emph{Hence Proved.}

\subsubsection{Soundness}

Soundness of the algorithm is characterized by the following theorem:

\begin{theorem}[Soundness]
	\label{thm:soundness}
For any node $v$, let $d \in \dlattice$ be the JOFP value computed by \textsc{ComputeJOFP} in algorithm  for $v$, treating $d_0 \in \dlattice$ as the initial value at the $\startnode$ node, and $\startnode$ as the initial node. Then, 

$$ d \dominates \underset{\substack{\textit{p is a feasible, interprocedurally valid} \\ \textit{path from $\startnode$ to $v$}}}{\bigsqcup} (\ptf{p})(d_0)$$

\end{theorem}

The proof of this theorem requires a set of lemmas and intermediate theorems. Therefore, we will first present the necessary lemmas, then the intermediate theorems, and then the final proof of correctness.

\paragraph{Important Lemmas}
\label{sssec:inter:proof:correct:lemma}
We require the following lemmas.


\begin{lemma}
	\label{lemma:1}
	If $d \in \natzero^r$ is a vector, $r \geq 1$, $p_1$ and $p_2$ are paths from $v_i$ to $v_j$ such that $\demandpd{p_1}{d} \leq \demandpd{p_2}{d} $, and  $p_0$ is any path ending at $v_i$, then $\demandpd{p_0.p_1}{d} \leq  \demandpd{p_0.p_2}{d}$.
\end{lemma}

\paragraph{Proof:} This is the \emph{Demand Supply Lemma} presented in the paper. We prove the lemma using induction on the length of the path $p_0$.

We first consider the base case when $p_0 = v_1 \vcfgtrans{f}{w_0} v_i$, i.e., it is a single edge. Let $d_1 = \demandpd{p_1}{d}$, $d_2 = \demandpd{p_2}{d}$.

We are  given, $d_1 \leq d_2$

Subtracting $w_0$ from both sides we get,

$d_1 - w_0 \leq d_2 - w_0$ $\stmtno{1}$

We now prove that the application of $\mu$ preserves the ordering in the inequation $(1)$, or in other words, $\mu$ applied to vectors is a monotone function. 

Recall that by Definition~\ref{def:demand}, $\demandpd{p_0.p_1}{d} = \mu(d_1 - w_0)$, and $\demandpd{p_0.p_2}{d} = \mu(d_2 - w_0)$, i.e., $\mu$ applied on the LHS and RHS of $(1)$ gives the demands of $\vpath{p_0.p_1}$ and $\vpath{p_0.p_2}$. 

As $\mu$ is applied component-wise on a vector and addition/subtraction of vectors is also component-wise, it suffices to show that for any $i \in [1 \ldots r],$ $ (\mu(d_1[i] - w_0[i]) \leq \mu(d_2[i] - w_0[i])$. Based on the values of $d_1[i], d_2[i]$ and $w_0[i]$, we have the following possibilities:

\begin{enumerate}
	\item \textbf{Case 1 : }$d_1[i] - w_0[i] > 0$ and $d_2[i] - w_0[i] > 0$. In this case, by the definition of $\mu$,
	
	$\mu(d_1[i] - w_0[i]) = d_1[i] - w_0[i] \stmtno{2}$
	
	and
	
	$\mu(d_2[i] - w_0[i]) = d_2[i] - w_0[i]\stmtno{3}$
	
	From $(1), (2)$ and $(3)$ we can infer that 
	
	$\mu(d_1[i] - w_0[i]) \leq \mu(d_2[i] - w_0[i]) \stmtno{4}$
	
	\item \textbf{Case 2 : }$d_1[i] - w_0[i] \leq 0$ and $d_2[i] - w_0[i] \leq 0$. In this case, by the definition of $\mu$ we have,
	
	$\mu(d_1[i] - w_0[i]) =0 \stmtno{5}$
	
	and
	
	$\mu(d_2[i] - w_0[i]) = 0 \stmtno{6}$
	
	From $(5)$ and $(6)$ we can infer that 
	
	$\mu(d_1[i] - w_0[i]) \leq \mu(d_2[i] - w_0[i]) \stmtno{7}$
	
	\item \textbf{Case 3 : }$d_1[i] - w_0[i] \leq 0$ and $d_2[i] - w_0[i] > 0$. In this case by the definition of $\mu$ we have,
	
	$\mu(d_1[i] - w_0[i]) =0 \stmtno{8}$
	
	and
	
	$\mu(d_2[i] - w_0[i]) > 0 \stmtno{9}$
	
	From $(8)$ and $(9)$ we can infer that 
	
	$\mu(d_1[i] - w_0[i]) \leq \mu(d_2[i] - w_0[i]) \stmtno{10}$	
	
\end{enumerate}

Due to $(1)$, the fourth case $(d_1[i] - w_0[i] > 0 \wedge d_2[i] - w_0[i] \leq 0)$ cannot occur. From $(4), (7)$ and $(10)$ we can infer that for all $i \in [1 \ldots r]$, $ (\mu(d_1[i] - w_0[i]) \leq \mu(d_2[i] - w_0[i])$, irrepective of the values of $d_1[i], d_2[i]$ and $w_0[i]$.

Therefore, it can be seen that $\mu(d_1 - w_0) \leq \mu(d_2 - w_0)$ $\stmtno{11}$ 

Now, using the definition of demand for $\vpath{p_0.p_1}$ and $\vpath{p_0.p_2}$, and from $(11)$ it follows that,

$\demandpd{p_0.p_1}{d} \leq \demandpd{p_0.p_2}{d}$

This proves the base case. Now, for the inductive case, let the length of $p_0$ be $n+1$. Let $p_0 = (v_1 \vcfgtrans{f}{w} v_2). p_0'$, where we have from the hypothesis that, 

$\demandpd{p_0'.p_1}{d} \leq \demandpd{p_0'.p_2}{d}$.

The inductive case can be proved the same way as the base case, by replacing $p_1$ by $p_0'.p_1$ , and $p_2$ by $p_0'.p_2$ in the base case proof.

\emph{Hence proved.}

\setcounter{equation}{0}

\begin{lemma}
	\label{lemma:2}
	Let $p_1$ be a path from $v_i$ to $v_j$ and $S_2$ be a set of paths from $v_i$ to $v_j$ such that $ \underset{p_2 \in S_2}{\bigjoin} \ptf{p_2} \dominates \ptf{p_1}$ . Let $p_0$ be any path ending at $v_i$. Then, $\underset{p_2 \in S_2}{\bigjoin} \ptf{p_0.p_2} \dominates \ptf{p_0.p_1}$.
\end{lemma}

\paragraph{Proof: } Let $S_2 = \lbrace p_{21}, p_{22}, \ldots, p_{2n} \rbrace$,

We are given,

$$ \ptf{p_{21}} \join \ptf{p_{22}} \ldots \join \ptf{p_{2n}} \dominates \ptf{p_{1}} \stmtno{1}$$, 

By composing on the left-side of the LHS and RHS in $(1)$ using $\ptf{p_0}$ and the monotonicity of composition operation, we obtain,

$$ \ptf{p_0} \circ ( \ptf{p_{21}} \join \ptf{p_{22}} \ldots \join \ptf{p_{2n}}) \dominates \ptf{p_0} \circ \ptf{p_1} \stmtno{2}$$

It is given that the path transfer functions form a complete lattice. As a consequence, the path transfer function composition left-distributes over function join, i.e., $f \circ (f_1 \join f_2 \join \ldots \join f_n) = (f \circ f_1) \join (f \circ f_2) \join \ldots \join (f \circ f_n)$. Therefore, expanding $(2)$ using left-distributivity we obtain,

$$ ( \ptf{p_0} \circ \ptf{p_{21}} ) \join (\ptf{p_0} \circ \ptf{p_{22}}) \ldots \join (\ptf{p_0} \circ \ptf{p_{2n}}) \dominates (\ptf{p_0} \circ \ptf{p_1}) \stmtno{3}$$

The path transfer function for a path $p =\vpath{p_a.p_b}$  is given by $\ptf{p_a.p_b} = \ptf{p_a} \circ \ptf{p_b}$. Therefore,  by rewriting $(3)$ in terms of path transfer functions, we get

$$\ptf{p_0.p_{21}} \join \ptf{p_0.p_{22}} \join \ldots \join \ptf{p_0.p_{2n}} \dominates \ptf{p_0.p_{1}} \stmtno{4}$$

As $S_2 = \lbrace p_{21}, p_{22}, \ldots, p_{2n} \rbrace$, we can condense $(4)$ to 

$$ \underset{p_2 \in S_2}{\bigjoin}\ptf{p_0.p_2} \dominates \ptf{p_0.p_1}$$ 

\emph{Hence proved.}

\begin{lemma}
\label{lemma:demEquiv}
Let $p_1 = \vpath{e_1 \concat e_2 \ldots e_n}$ be a path, such that for $ i \in [1 \ldots n]$, $w_{e_i} = \theta(e_i)$. Let $p_2$ be any path that starts at the end node of $p_1$, and let $p = p_1 \concat p_2$. Let $d$ be any element of $\natzero^r$. If $p_1$ does not receive any messages,  then $\demandpd{p}{d} = \mu(\demandpd{p_2}{d} - w_{p_1})$, where $w_{p_1} = w_{e_1} + w_{e_2} + \ldots w_{e_n}$.
\end{lemma}

\paragraph{Proof: } We prove the lemma by induction on the length of $p_1$, and taking any arbitrary $p_2$. 

We first prove the base case. The base case is when $p_1$ is of length 1, i.e., it is an edge $e$. Let $w_e$ be the queuing vector of $e$, and therefore $w_{p_1} = w_e$. 

By Definition~\ref{def:demand}, we have,

$\demandpd{p}{d} = \mu(\demandpd{p_2}{d} - w_{e})$

Replacing the value of $w_{e}$ using $w_{p_1} = w_{e}$, 

$\demandpd{p}{d} = \mu(\demandpd{p_2}{d} - w_{p_1})$

This proves the base case.

We now proceed to the inductive case. Let $p =\vpath{ e \concat p_1' \concat p_2}$, where $\vpath{e \concat p_1'}$ is of length $n+1$. As $p_1'$ is of length $n$, the inductive hypothesis holds for $p_1'$ and path $p' = \vpath{ p_1' \concat p_2}$. Therefore, by the hypothesis we have,

$\demandpd{p'}{d} = \mu(\demandpd{p_2}{d} - w_{p_1'}) \stmtno{1}$

From Definition~\ref{def:demand} we have for path $p$,

$\demandpd{p}{d} = \mu(\demandpd{p'}{d} - w_{e}) \stmtno{2}$

After replacing the value of $\demandpd{p'}{d}$ from $(1)$ in $(2)$ we get,

$\demandpd{p}{d} = \mu(\mu(\demandpd{p_2}{d} - w_{p_1'}) - w_{e}) \stmtno{3}$

Now we argue that $\demandpd{p}{d} = \mu(\demandpd{p_2}{d} - w_{p_1'} - w_{e})$; i.e., the inner application of $\mu$ from $(3)$ can be dropped. In order to prove this, we will argue that for any $i \in [1 \ldots r], \mu(\mu(\demandpd{p_2}{d}[i] - w_{p_1'}[i]) - w_{e}[i]) = \mu(\demandpd{p_2}{d}[i] - w_{p_1'}[i] - w_{e}[i])$. Proving this suffices as the two operations involved, $\mu$ and vector addition/subtraction are component-wise. Therefore, proving the required result for all components, will prove it for the full vector.

Thus, we proceed to the proof. Let $x = \demandpd{p_2}{d} - w_{p_1'}$. Based on the value of $x[i]$ for any $i \in [1 \ldots r]$, we have the following possible scenarios:

\begin{enumerate}
\item $x[i] > 0 $ : In this case $\mu(x[i]) = x[i]$. 

Therefore by replacing the value of $\mu(x[i])$ we have,  $\mu(\mu(x[i]) - w_{e}[i]) = \mu(x[i] - w_{e}[i]) \stmtno{4}$. 

\item $x[i] \leq 0$ :  We are given by the lemma assumption $w_e[i] \geq 0$, i.e. $e_i$ cannot be a receive operation. Therefore subtracting $w_e[i]$ from $x[i]$ will further reduce the value, that is,

 $x[i] - w_e[i] \leq 0 \stmtno{5}$

From $(5)$, and the definition of $\mu$, we have

 $\mu(x[i] - w_e[i]) = 0 \stmtno{6}$.

Again, because $x[i] \leq 0$, therefore $\mu(x[i]) = 0$. Hence subtracting $w_e[i]$ from $\mu(x[i])$ will result in zero or lower value, that is,

 $\mu(x[i]) - w_e[i] \leq 0 \stmtno{7}$
 
From $(7)$, and the definition of $\mu$, 

we have $\mu(\mu(x[i]) - w_e[i] ) = 0 \stmtno{8}$
 
 As the RHS are equal in $(6)$ and $(8)$, we can infer that in this case,
 
$\mu(\mu(x[i]) - w_e[i]) = \mu(x[i] - w_{e}[i]) \stmtno{9}$
\end{enumerate}

Therefore, from $(4)$ and $(9)$, we have proved that for any $i \in [1 \ldots r], \hspace{3pt} (\mu(\mu(x[i]) - w_e[i]) = \mu(x[i] - w_{e}[i]))$. Therefore, we can rewrite $(3)$ to,

$\demandpd{p}{d} = \mu(\demandpd{p_2}{d} - w_{p_1'} - w_{e})$

As the paths $p_1$ and $p_1'$ do not receive any messages $w_{p_1} = w_{e.p_1'} = w_{p_1'} + w_{e}$. Thus, replacing $(- w_{p_1'} - w_{e})$ in the above equation by $ - w_{p_1}$,

$\demandpd{p}{d} = \mu(\demandpd{p_2}{d} - w_{p_1})$

Therefore, the inductive case holds as well.

\emph{Hence proved.}

As a consequence of Lemma~\ref{lemma:demEquiv}, we have the following corollary.

\textit{\textbf{Corollary 2:} Let $p = \vpath{e_1 \concat e_2 \ldots e_n}$ be a path, such that for $ i \in [1 \ldots n]$, $w_{e_i} = \theta(e_i)$. Let $d$ be any element of $\natzero^r$. If $p$ does not receive any messages,  then $\demandpd{p}{d} = \mu(d - w_{p})$, where $w_{p_1} = w_{e_1} + w_{e_2} + \ldots w_{e_n}$.}

\paragraph{Proof : } The proof of the corollary is the same as that of Lemma~\ref{lemma:demEquiv}, and can be obtained by simply replacing $\demandpd{p_2}{d}$ by $d$.

\vspace{10pt}

\begin{lemma}
\label{lemma:2a}
Let $p_1$ be a path from $v_i$ to $v_j$ and $S_2$ be a set of paths from $v_i$ to $v_j$ such that $ \underset{p_2 \in S_2}{\bigjoin} \ptf{p_2} \dominates \ptf{p_1}$ . Let $p_0$ be any path beginning at $v_j$. Let procedure  composition be right-distributive over join for all the path transfer procedure s. Then, $\underset{p_2 \in S_2}{\bigjoin} \ptf{p_2.p_0} \dominates \ptf{p_1.p_0}$.
\end{lemma}

\paragraph{Proof: } Let $S_2 = \lbrace p_{21}, p_{22}, ..., p_{2n} \rbrace$. We are given,

$$(\ptf{p_{21}} \join \ptf{p_{22}} \ldots \join \ptf{p_{2n}}) \dominates \ptf{p_{1}} \stmtno{1}$$

Composing the LHS and RHS of inequation $(1)$ from the right-side by $\ptf{p_0}$ and due to the monotonicity of $\ptf{p_0}$ and procedure  composition, we get,

$$(\ptf{p_{21}} \join \ptf{p_{22}} \ldots \join \ptf{p_{2n}}) \circ \ptf{p_0}\dominates \ptf{p_{1}} \circ \ptf{p_0} \stmtno{2}$$

As procedure  composition is right-distributive over procedure  join, i.e. for any $f, f_1, \ldots f_n \in \dlattice  \rightarrow \dlattice$, $(f_1 \join  \ldots \join f_n) \circ f = (f_1 \circ f) \join \ldots \join (f_n \circ f)$, we can expand $(2)$ as,

$$(\ptf{p_{21}} \circ \ptf{p_0}) \join (\ptf{p_{22}} \circ \ptf{p_0}) \ldots \join (\ptf{p_{2n}} \circ \ptf{p_0}) \dominates \ptf{p_{1}} \circ \ptf{p_0} \stmtno{3}$$

The path transfer procedure  for a path $p =\vpath{p_a.p_b}$  is given by $\ptf{p_a.p_b} = \ptf{p_a} \circ \ptf{p_b}$. Therefore,  by rewriting $(3)$ in terms of path transfer procedure s, we get

$$(\ptf{p_{21} \concat p_0} \join \ptf{p_{22}  \concat p_0} \ldots \join \ptf{p_{2n}  \concat p_0}) \dominates \ptf{p_{1}  \concat p_0} \stmtno{4}$$

As $S_2 = \lbrace p_{21}, p_{22}, \ldots, p_{2n} \rbrace$, we can condense $(4)$ to 

$$\underset{p_2 \in S_2}{\bigjoin} \ptf{p_2.p_0} \dominates \ptf{p_1.p_0}$$

\emph{Hence proved.}

\setcounter{equation}{0}
\begin{lemma}
	\label{lemma:3}
	Let $d \in \natzero^r$ be a vector,$v_i$ and $v_j$ be any two nodes $p_1$ and $p_2$ be paths from $v_i$ to $v_j$ such that $\demandpd{p_1}{d} \leq \demandpd{p_2}{d} $. Let $p_0$ be any path ending at $v_i$ and $p_3$ be any path beginning from $v_j$. Let the paths $p_1, p_2, p_0$ and $p_3$ be such that they do not receive any messages. Then, $\demandpd{p_0.p_1.p_3}{d} \leq  \demandpd{p_0.p_2.p_3}{d}$. 
\end{lemma}

\paragraph{Proof: } The proof is in two parts. We first prove that  $\demandpd{p_1.p_3}{d} \leq \demandpd{p_2.p_3}{d}$. The second part proves $\demandpd{p_0.p_1.p_3}{d} \leq  \demandpd{p_0.p_2.p_3}{d}$.


To prove the first part, we use Corollary~2. As the paths $p_1$, $p_2$ do not receive any messages, Corollary~2 is applicable on $p_1$ and $p_2$. Let $w_{p_1}$ and $w_{p_2}$ be the sums of the queuing vectors of edges in paths $p_1$ and $p_2$ respectively. Therefore, by mapping $p_1$ and $p_2$ to $p$ in Corollary~2, we have,

$\demandpd{p_1}{d} = \mu(d -  w_{p_1}) \stmtno{1}$

$\demandpd{p_2}{d} = \mu(d - w_{p_2}) \stmtno{2}$

We are given,

$\demandpd{p_1}{d} \leq \demandpd{p_2}{d} \stmtno{3}$

Substituting the values from Equations $(1)$ and $(2)$ into $(3)$, we obtain

$\mu(d -  w_{p_1}) \leq \mu(d -  w_{p_2}) \stmtno{4}$

Equation $(4)$ will hold true \textit{iff}  for all $i \in [1 \ldots r]$, it holds that :

\begin{itemize}
	\item \textbf{Condition 1: } either $(w_{p_1}[i] < d[i] \vee w_{p_2}[i] < d[i] ) \wedge (w_{p_1}[i] \geq w_{p_2}[i]) $, or
	\item  \textbf{Condition 2: }  $w_{p_1}[i] \geq d[i] \wedge w_{p_2}[i] \geq d[i]$
\end{itemize}

We now argue that irrespective of which of the two conditions above holds for any given $1 \leq i \leq r$, $(\demandpd{p_1.p_3}{d}[i] \leq \demandpd{p_2.p_3}{d}[i])$.

Let $\demandpd{p_3}{d} = d_3$. The first case is when Condition 1 holds for some $i$. In this case,

$w_{p_1}[i] \geq w_{p_2}[i] \stmtno{5}$

Negating both sides of Equation $(5)$ and adding $d_3[i]$ to both sides, we obtain,

$d_3[i] - w_{p_1}[i] \leq d_3[i] - w_{p_2}[i] \stmtno{6}$

By Definition~\ref{def:demand} and Lemma~\ref{lemma:demEquiv}, we have

$\demandpd{p_1.p_3}{d} = \mu(d_3 - w_{p_1})$ $\stmtno{7}$

and 

$\demandpd{p_2.p_3}{d} = \mu(d_3 - w_{p_2})$ $\stmtno{8}$

That is, to obtain the demand, $\mu$ will be applied on both sides. 

We proved in Lemma~\ref{lemma:1} that $\mu$ is a monotone procedure , i.e., if $x \leq y$ then $\mu(x) \leq \mu(y)$. Thus, the application of $\mu$ on both sides of Equation $(6)$ preserves the ordering of the inequation, and we obtain

$$\mu(d_3[i] - w_{p_1}[i]) \leq \mu( d_3[i] - w_{p_2}[i]) \stmtno{9}$$

Using the definitions of demands of paths $\vpath{p_1 \concat p_3}$ and $\vpath{p_2 \concat p_3}$ from Equations $(7)$ and $(8)$, in conjunction with Equation $(9)$ it follows that,

$$\demandpd{p_1.p_3}{d}[i] \leq \demandpd{p_2.p_3}{d}[i] \stmtno{10}$$

The second case is when condition 2 holds for $i$. In this case we have,

$w_{p_1}[i] \geq d[i] \stmtno{11} $

$w_{p_2}[i] \geq d[i] \stmtno{12}$

Negating both sides of $(11)$ and $(12)$ and adding $d[i]$ to both sides in both equations, we obtain

$d[i] - w_{p_1}[i] \leq 0$ and $d[i] - w_{p_2}[i] \leq 0 \stmtno{13}$.

As $p_3$ does not receive any messages, therefore by the definition of demand,

$\demandpd{p_3}{d} \leq d \stmtno{14}$

As $d_3$ is lower than or equal to $d$, then from Equations $(13)$ and $(14)$ we get,

$d_3[i] - w_{p_1}[i] \leq 0$ and $d_3[i] - w_{p_2}[i] \leq 0 \stmtno{15}$

From $(15)$ and the definition of $\mu$, we have

$\mu(d_3[i] - w_{p_1}[i]) = 0$ and $\mu(d_3[i] - w_{p_2}[i]) = 0 \stmtno{16}$ 

By Lemma~\ref{lemma:demEquiv} we have $\demandpd{p_1.p_3}{d} = \mu(d_3 - w_{p_1})$ and $\demandpd{p_2.p_3}{d} = \mu(d_3 - w_{p_2})$. 

Using Equation $(16)$, and the above definitions, we can infer

$$\demandpd{p_1.p_3}{d}[i] \leq \demandpd{p_2.p_3}{d}[i] \stmtno{17}$$

Since $(10)$ and $(17)$ hold for all $i$, we get:

$$\demandpd{p_1.p_3}{d} \leq \demandpd{p_2.p_3}{d} \stmtno{18}$$

Now we prove the second part, i.e., if $\demandpd{p_1}{d} \leq \demandpd{p_2}{d}$ then $\demandpd{p_0.p_1.p_3}{d} \leq \demandpd{p_0.p_2.p_3}{d}$.

Applying Lemma~\ref{lemma:1} by mapping  $p_1.p_3$ to $p_1$ in the lemma statement, $p_2.p_3$ to $p_2$ in the lemma statement, and $p_0$ to $p_0$ in the lemma statement, and using $(18)$, we have 

$$	\demandpd{p_0.p_1.p_3}{d} \leq \demandpd{p_0.p_2.p_3}{d}$$

\emph{Hence proved.}

\setcounter{equation}{0}
\begin{lemma}
	\label{lemma:4}
	Let $v_i$ and $v_j$ be any two nodes. Let $p_1$ be a path from $v_i$ to $v_j$ and $S_2$ be a set of paths from $v_i$ to $v_j$ such that $\underset{p_2 \in S_2}{\bigjoin} \ptf{p_2} \dominates \ptf{p_1}$. Let procedure  composition be both left- and right- distributive over procedure  join for all the path transfer procedure s. Let $p_0$ be any path ending at $v_i$ and $p_3$ be any path starting from $v_j$. Then $\underset{p_2 \in S_2}{\bigjoin} \ptf{p_0.p_2.p_3} \dominates \ptf{p_0.p_1.p_3}$.
\end{lemma}

\paragraph{Proof: } This is the \emph{Function Coverage Lemma}, as presented in the paper. To prove lemma we use the results of Lemma~\ref{lemma:2} and \ref{lemma:2a}. From Lemma~\ref{lemma:2}, we can infer that,

$$\left( \underset{p_2 \in S_2}{\bigjoin} \ptf{p_0.p_2} \right) \dominates \ptf{p_0.p_1} \stmtno{1}$$

Let $S_1 = \lbrace p_0.p_2 \mid  p_2 \in S_2 \rbrace$. Therefore, we can rewrite $(1)$ as

$$\left( \underset{p \in S_1}{\bigjoin} \ptf{p} \right) \dominates \ptf{p_0.p_1} \stmtno{2}$$

Now, from Lemma~\ref{lemma:2a}, we can infer,

$$\left( \underset{p \in S_1}{\bigjoin} \ptf{p.p_3} \right)\dominates \ptf{p_0.p_1.p_3} \stmtno{3}$$

Rewriting $(3)$ in terms of $S_2$ using the definition of $S_1$, we obtain

$$ \left( \underset{p_2 \in S_2}{\bigjoin} \ptf{p_0.p_2.p_3} \right) \dominates \ptf{p_0.p_1.p_3} $$

\emph{Hence proved.}

\setcounter{equation}{0}
\begin{lemma}
\label{lemma:5}
Let $d \in \natzero^r$ be any vector. Say a set of paths $S$ $d$-covers a path $p$. For any path $p'$, let $S_{p'}$ denote any set of paths that $d$-covers $p'$. Then, the set of paths $\underset{p' \in S}{\bigcup} S_{p'} $ $d$-covers $p$.
\end{lemma}

\paragraph{Proof: } Let $S_1 = \underset{p' \in S}{\bigcup} S_{p'}$. Let $p$ begin at vertex $v_i$ and end at $v_j$.  As $S$ $d$-covers $p$, we have the following facts from Definition~\ref{def:cover}:

\begin{enumerate}
	\item all paths in $S$ start at $v_i$ and end at $v_j$
	\item for all paths $p’ \in S, \hspace{3pt} (\demandpd{p'}{d} \leq \demandpd{p}{d})$
	\item $\underset{p' \in S}{\bigjoin} \ptf{p'} \dominates \ptf{p}$
\end{enumerate}	

Similarly, we have the following facts for any $p' \in S$ the set $S_{p'}$ that $d$-covers $p'$. 

\begin{enumerate}
	\item as $p'$ starts at $v_i$ and ends at $v_j$, all paths in $S_{p'}$ start at $v_i$ and end at $v_j$
	\item for all paths $p'' \in S_{p'}, \hspace{3pt} (\demandpd{p''}{d} \leq \demandpd{p'}{d})$
	\item $\underset{p'' \in S_{p'}}{\bigjoin} \ptf{p''} \dominates \ptf{p'}$
\end{enumerate}	

From the facts above and the definition of $S_1$, it can be directly inferred that every path in $S_1$ begins at $v_i$ and ends at $v_j$. $\stmtno{1}$

Now we prove that for any path $p_1 \in S_1$, $\demandpd{p_1}{d} \leq \demandpd{p}{d}$. 

Let $p' \in S$. For any path $p_1 \in S_{p'}$,

$ \demandpd{p_1}{d} \leq \demandpd{p'}{d}$. $\stmtno{2}$

We are given that $(\demandpd{p'}{d} \leq \demandpd{p}{d})$. Therefore from $(2)$, we can infer that 

$(\demandpd{p_1}{d} \leq \demandpd{p}{d})$. $\stmtno{3}$

Now we prove that the join of the transfer procedure s of the paths in $S_1$ dominates the path transfer procedure  of $p$. Because $S$ covers $p$, we have

$$\underset{p' \in S}{\bigjoin} \ptf{p'} \dominates \ptf{p}$$

Say $S = \lbrace p_1', p_2', \ldots p_n' \rbrace$, then by expanding the above equation we get

$$\ptf{p_1'} \join \ptf{p_2'} \join \ldots \join \ptf{p_n'} \dominates \ptf{p} \stmtno{4}$$

We are given for all $p' \in S$ that $\underset{p_1 \in S_{p'}}{\bigjoin} \ptf{p_1} \dominates \ptf{p'}$. By the property of join operation we have,

$$\underset{p_1 \in S_{p'}}{\bigjoin} \ptf{p_1} \join \underset{p_1 \in S_{p_2'}}{\bigjoin} \ptf{p_1} \join \ldots \join  \underset{p_1 \in S_{p_n'}}{\bigjoin} \ptf{p_1}   \dominates \ptf{p'} \join \ptf{p_2'} \join \ldots \join \ptf{p_n'} \stmtno{5} $$

where $S_{p_1'}, S_{p_2'}, \ldots S_{p_n'}$ are the sets $d$-covering paths $p_1', p_2',  \ldots p_n'$ respectively. From $(4)$ and $(5)$, it can be seen that

$$\underset{p_1 \in S_{p_1'}}{\bigjoin} \ptf{p_1} \join \underset{p_1 \in S_{p_2'}}{\bigjoin} \ptf{p_1} \join \ldots \join \underset{p_1 \in S_{p_n'}}{\bigjoin} \ptf{p_1} \dominates \ptf{p} \stmtno{6} $$

By the property of join $(6)$ can be rewritten as,

$$\underset{p_1 \in (S_{p_1'} \cup S_{p_2'} \cup \ldots \cup S_{p_n'}) }{\bigjoin} \ptf{p_1}  \dominates \ptf{p} \stmtno{7}$$ 

As $S_1 = {S_{p_1'}} \cup {S_{p_2'}} \ldots \cup S_{p_n'}$, therefore $(7)$ can be rewritten as

$$\underset{p_1 \in S_1 }{\bigjoin} \ptf{p_1}  \dominates \ptf{p}  \stmtno{8}$$

From $(1), (3)$ and $(8)$, $S_1$ $d$-covers $p$.

\emph{Hence proved}.

Our algorithm can be seen as generating paths iteratively, and storing each generated path if it is not covered by other paths. A path is generated at every visit to Lines 11, 17, and 21 of routine \textsc{ComputeJOFP}. For a given path $p$, we say that  it is generated by our algorithm  if it is generated during any visit to any of the lines mentioned above. Over the course  of a run of the algorithm, note that for any path $p$ that is generated, the path necessarily ends at the node \targetnode, although it could begin at any node. Also note that when any path $p$ is generated, if $p$ begins at a node $v_i$, then $p$ is stored in $\sPaths{v_i}$ unless $\sPaths{v_i}$ already stores a previously generated set of paths that cover $p$.

\begin{lemma}
	\label{lemma:6}
	If $p$ is a path from a node $v_i$ in the VCFG of any procedure to the $\targetnode$ node, and if the algorithm generates a set of paths $S$ that cover $p$, then when \textsc{ComputeJOFP} terminates there is guaranteed to be a set of paths in $\sPaths{v_i}$ that cover $p$.
\end{lemma}

\paragraph{Proof: } Note, after any path $p' \in S$ is generated, the algorithm would invoke the routine \textsc{Covered}($p'$). The two following outcomes can result from this invocation.

\begin{enumerate}
	\item The routine \textsc{Covered} returns \fls\:. In this case $p'$ is added to $\sPaths{v_i}$, That is, $p'$ is retained.
	
	\item The routine \textsc{Covered} returns \tru\:. In this case, $p'$ is not added to $\sPaths{v_i}$, as there exists a set of paths $\cover{p'} \subseteq \sPaths{v_i}$ such that $\cover{p'}$ covers $p'$.
\end{enumerate} 

We now prove that irrespective of the outcome above of \textsc{Covered}, there exists a set of paths in $\sPaths{v_i}$ that cover $p$. 

Let $S_1 = \lbrace p' \in S \mid  \textsc{Covered}(p') \textit{ returned } \fls\rbrace $

Let $S_2 = \lbrace p'' \in \cover{p'} \mid p'\in S \wedge \textsc{Covered}(p') \textit{ returns } \tru \rbrace$

Clearly, $S_1 \cup S_2 \subseteq \sPaths{v_i}$, and they begin at $v_i$ and end at $v$. $\stmtno{1}$

As S covers $p$, for all paths $p' \in S_1$, $\demandp{p'} \leq \demandp{p}$. $\stmtno{2}$

Similarly, as S covers $p$, and $S_2$ contains paths due to sets that cover paths $p'\in S$, therefore for all paths $p''\in S_2. \hspace{1pt} (\demandp{p''} \leq \demandp{p})$. $\stmtno{3}$

From $(2)$ and $(3)$, we can infer that for all paths $p' \in S_1 \cup S_2$,

$\demandp{p'} \leq \demandp{p}) \stmtno{4}$

Now we prove the relation between the path transfer procedure s of $p$ and paths in $\sPaths{v_i}$. We are given,

$\underset{p' \in S}{\bigjoin} \ptf{p'} \dominates \ptf{p}$

Splitting the above using the definition of $S_1$ we obtain,

$\underset{p' \in S_1}{\bigjoin} \ptf{p'} \join \underset{p'' \in S - S_1}{\bigjoin} \ptf{p''}  \dominates \ptf{p} \stmtno{5}$

We can expand the set $S- S_1 = \lbrace p_1'', p_2'', \ldots p_n''\rbrace$ and write $(5)$ as,

$\underset{p' \in S_1}{\bigjoin} \ptf{p'} \join \ptf{p_1''} \join \ptf{p_2''} \ldots \join \ptf{p_n''}  \dominates \ptf{p} \stmtno{6}$

From the two outputs of \textsc{Covered}, we know that  $S_1 \subseteq \sPaths{v_i}$. Also, for each $p'' \in S - S_1$ there exists a set $\cover{p''} \subseteq \sPaths{v_i}$ such that $\cover{p''}$ covers $p''$. 

Thus replacing the path transfer procedure s of all $p'' \in S - S_1$ by $\underset{p_1 \in \cover{p''}}{\bigjoin} \ptf{p_1}$ in $(6)$, and the property that if $a \dominates b$ then $c \join a \dominates c \join b$,  we get

$$\underset{p' \in S_1}{\bigjoin} \ptf{p'} \join \underset{p_1 \in \cover{p_1''}}{\bigjoin} \ptf{p_1} \join  \ldots \join \underset{p_1 \in \cover{p_n''}}{\bigjoin} \ptf{p_1}  \dominates \ptf{p} \stmtno{7}$$

Let $S'= S_1 \cup \cover{p_1''} \cup \ldots \cup \cover{p_n''}$. Rewriting $(7)$ in terms of $S'$ we get

$$ \underset{p' \in S'}{\bigjoin} \ptf{p'} \dominates \ptf{p} \stmtno{8}$$

From the definition of $S'$ we know that $S'= S_1 \cup S_2$ and  $S' \subseteq \sPaths{v_i}$. 

Therefore, from $(1), (4)$ and $(8)$, we can infer that $S' \subseteq \sPaths{v_i}$ covers $p$.

\emph{Hence proved.}

\vspace{10pt}

Similar to the generation of paths discussed above, the routine \textsc{ComputeEndToEnd} also generates paths iteratively in each invocation, and stores a generated path if it is not covered by other paths. The routine \textsc{ComputeEndToEnd} is invoked with a procedure $F$ and a vector $d$ (which is the demand of a path). In each invocation, the routine \textsc{ComputeEndToEnd} generates an interprocedurally valid and complete path, using path templates, at every visit to Line~9. For a given path $p$, we say that it is generated by \textsc{ComputeEndToEnd} if it is generated during any visit to Line~9. Any path $p$ generated by \textsc{ComputeEndToEnd} begins at the entry node of the VCFG of some procedure $F_i$ and ends at the designated exit node of $F_i$. If $p$ is generated and it begins at the entry node of $F_i$, then it is stored in $\sIVCPaths{F_i}{d}$ unless $\sIVCPaths{F_i}{d}$ already stores a previously generated set of paths that $d$-cover $p$.

\begin{lemma}
\label{lemma:6a}
Let $d \in \natzero^r$ be a given vector such that \textsc{ComputeEndToEnd} is invoked with procedure  $F_i \in \funcs$ and $d$ as arguments. If $p$ is an interprocedurally valid and complete path from the entry node $\entrynode{F}$ of any procedure  $F \in \funcs$ to the exit node $\exitnode{F}$, and if \textsc{ComputeEndToEnd} generates (at Line 11) a set of paths $S$ that $d$-cover $p$ for the given vector $d$, then when the above-mentioned invocation to  \textsc{ComputeEndToEnd} terminates there is guaranteed to be a set of paths in $\sIVCPaths{F}{d}$ that $d$-cover $p$.

\end{lemma}

\paragraph{Proof : } The proof of this lemma is similar to the proof of Lemma~\ref{lemma:6}. In this case also, for each $p'\in S$, either $p'$ is retained in $\sIVCPaths{F}{d}$, or a set of $d$-covering paths is already present. Therefore, this lemma holds.

\begin{lemma}
	\label{lemma:7}
	Let $p_0$, $p_1$, $p_2$ be paths where $p_1$ and $p_2$ end at $v_i$ and $p_0$ begins at $v_i$. If $\demandpd{p_1}{\demandp{p_0}} \leq \demandpd{p_2}{\demandp{p_0}} $, then $\demandp{p_1.p_0} \leq \demandp{p_2.p_0}$
\end{lemma}
\paragraph{Proof: } To prove the lemma, we first prove an intermediate result, i.e., for any paths $p,q$, such that end node of $p$ is the same as the start node of $q$, $\demandpd{p \concat q}{d} = \demandpd{p}{\demandpd{q}{d}}$. We prove this using induction on the length of $p$.

The base case is when $p$ is a single edge $e$ with queuing vector $w_e$. Then by Definition~\ref{def:demand},

$$\demandpd{e.q}{d} = \mu(\demandpd{q}{d} - w_e) \stmtno{1}$$

Again by Definition~\ref{def:demand} $\demandpd{e}{\demandpd{q}{d}} = \mu(\demandpd{q}{d} - w_e) \stmtno{2}$.

From $(1)$ and $(2)$, we have $\demandpd{e.q}{d}= \demandpd{e}{\demandpd{q}{d}}$.

We now prove the inductive case. Let $p=e.p'$, where $p$ is of length $n+1$ and $p'$ is of length $n$. By the induction hypothesis we have, $\demandpd{p'.q}{d}= \demandpd{p'}{\demandpd{q}{d}}$.

By Definition~\ref{def:demand}, we have

$$\demandpd{p \concat q }{d} = \mu(\demandpd{p'.q}{d} - w_e)$$

Replacing the value in the RHS using induction hypothesis,

$$\demandpd{p \concat q }{d} = \mu(\demandpd{p'}{\demandpd{q}{d}} - w_e) \stmtno{3}$$

Also, since $p = e.p'$, using the definition of demand, we have

$$\demandpd{p}{\demandpd{q}{d}} = \mu(\demandpd{p'}{\demandpd{q}{d}} - w_e) \stmtno{4}$$

From $(3)$ and $(4)$, we can infer that $\demandpd{p \concat q }{d} =\demandpd{p}{\demandpd{q}{d}} $ for the inductive case as well. Therefore we have proved that for any paths $p$ and $q$, $\demandpd{p \concat q }{d} = \demandpd{p}{\demandpd{q}{d}}$.

By taking $d  = \zerovector$ in the result above, we get

$\demandp{p_1.p_0} = \demandpd{p_1}{\demandp{p_0}} \stmtno{5}$

$\demandp{p_2.p_0} = \demandpd{p_2}{\demandp{p_0}} \stmtno{6}$.

We can conclude from $(5)$ and $6$ that

$\demandpd{p_1}{\demandp{p_0}} \leq \demandpd{p_2}{\demandp{p_0}} \Rightarrow \demandp{p_1.p_0} \leq \demandp{p_2.p_0}$.

\emph{Hence proved.}

\begin{lemma}
\label{lemma:IV}
Let v be any node in the VCFG of any procedure, such that for a given node \targetnode\, the algorithm computes the set $\sPaths{v}$ for v on termination. Any path $p \in \sPaths{v}$ is an interprocedurally valid path. 
\end{lemma}

\paragraph{Proof: } According to the definition of an interprocedurally valid path, a path $p$ will not be interprocedurally valid if at least once during the traversal of the path, the symbol popped from the top of the stack on encountering a return-site node is not the same as the corresponding call-node. The above scenario can result only if the path has a return edge that does not have a `matching' call-edge.

We prove by induction on the length of the path $p$, that whenever a path is added to $\sPaths{v}$, it is interprocedurally valid (i.e., there are no unbalanced return edges).

For the base case, the length of $p$ is 1. A path of length 1 will be interprocedurally invalid if it requires that an empty stack should be popped along the traversal of the path. Paths of length 1 are added only at Line~4 in \textsc{ComputeJOFP}, and all the added paths (rather edges) are intra-procedural edges. Therefore, $p$ does not contain an unbalanced return edge that will cause the stack to pop from an empty stack. 

Hence all paths of length 1 added to $\sPaths{v}$ are interprocedurally valid.

Moving on to the inductive case, let $p$ be a path of length $n+1$. Let the inductive hypothesis hold for all paths of length upto $n$. There are 3 points in the algorithm where paths are added to $\sPaths{v}$ -- lines~13, 19, and 23  in \textsc{ComputeJOFP}. Therefore, based on the location in algorithm where $p$ was added, we have the following cases. 

\begin{itemize}
\item \textbf{Case 1 :} $p = e.p'$ and $p$ was added to $\sPaths{v}$ at line~23 in \textsc{ComputeJOFP}. In this case, $e$ is an intra-procedural edge, and the path $p'$ is an interprocedurally valid path by the induction hypothesis.

Therefore, the concatenation of $e$ and $p'$ does not introduce any un-balanced return edges, and hence $p$ is also an interprocedurally valid path. $\stmtno{1}$

\item \textbf{Case 2 :} $p = c.p'$ and $p$ was added to $\sPaths{v}$ at line~19 in \textsc{InterProcExt}. In this case, $c$ is a call edge. Traversal of $c$ does not pop the stack, and from the hypothesis we know that the traversal of $p'$ is also interprocedurally valid. 

Therefore, $p$ is interprocedurally valid in this case. $\stmtno{2}$

\item \textbf{Case 3:} $p=p_1 \concat p_2$, where $p$ was added to $\sPaths{v}$ at line~13 in \textsc{InterProcExt}, $p_1$ begins at $v$ and ends at a return-site node $v_r$ and $p_2$ begins at $v_r$, and $p_1$ is an interprocedurally valid and complete path. Both $p_1$ and $p_2$ are of length $\leq n$ such that their combined length is equal to $n$. Therefore, the induction hypothesis holds for both $p_1$ and $p_2$.

As $p_1$ is an IVC path, at the end of the traversal of $p_1$ (i.e., at node $v_r$) , the stack will be empty. From the induction hypothesis we know that the traversal of $p_2$ from $v_r$ is interprocedurally valid. 

Therefore, the path $p = p_1.p_2$ is also interprocedurally valid. $\stmtno{3}$
\end{itemize}

From $(1)$, $(2)$ and $(3)$, it follows that $p$ is an interprocedurally valid path.

\emph{Hence Proved}.

\paragraph{Intermediate Theorems}
\label{sssec:inter:proof:correct:main}

We first prove that the set of paths returned by the routine~\textsc{ComputeEndToEnd} for a given demand $d$, $d$-covers all IVC paths of any procedure $F \in \funcs$.

\begin{theorem}[ComputeEndToEnd Cover]
	\label{thm:algo2ComputeE2Ecover}
	Let $d \in \natzero^r$ be a given vector. Let $F_i$ be a procedure in $\funcs$ and let \textsc{ComputeEndToEnd} be invoked with aguments $F_i$ and $d$. Let $\sIVCPaths{F_i}{d}$ be the set of interprocedurally valid and complete (IVC) paths computed and returned by \textsc{ComputeEndtoEnd}. Let $F$ be any procedure in $\funcs$. Let $p$ be any IVC path from start of $F$ to end of $F$.  Then there exists a set of paths $\cover{p} \subseteq \sIVCPaths{F}{d} $ such that $\cover{p}$ $d$-covers $p$.
\end{theorem}

\paragraph{Proof: }

We prove the theorem using induction on the depth of path $p$. The \emph{depth} of path $p$ is the maximum number of (non-sequential) calls made from within the path $p$.

The base case is when $p$ is of depth $0$. For all methods $F_1 \in \funcs$ and given demand $d$, all the paths of depth $0$ in $F_1$ are added to $\sIVCPaths{F_1}{d}$ at Line 3 in \textsc{ComputeEndToEnd}. Hence, $p \in \sIVCPaths{F}{d}$, and $d$-covers itself.

\begin{figure}[h]	
	\centering
	\includegraphics[trim = 0 60 0 60, clip, scale=0.45]{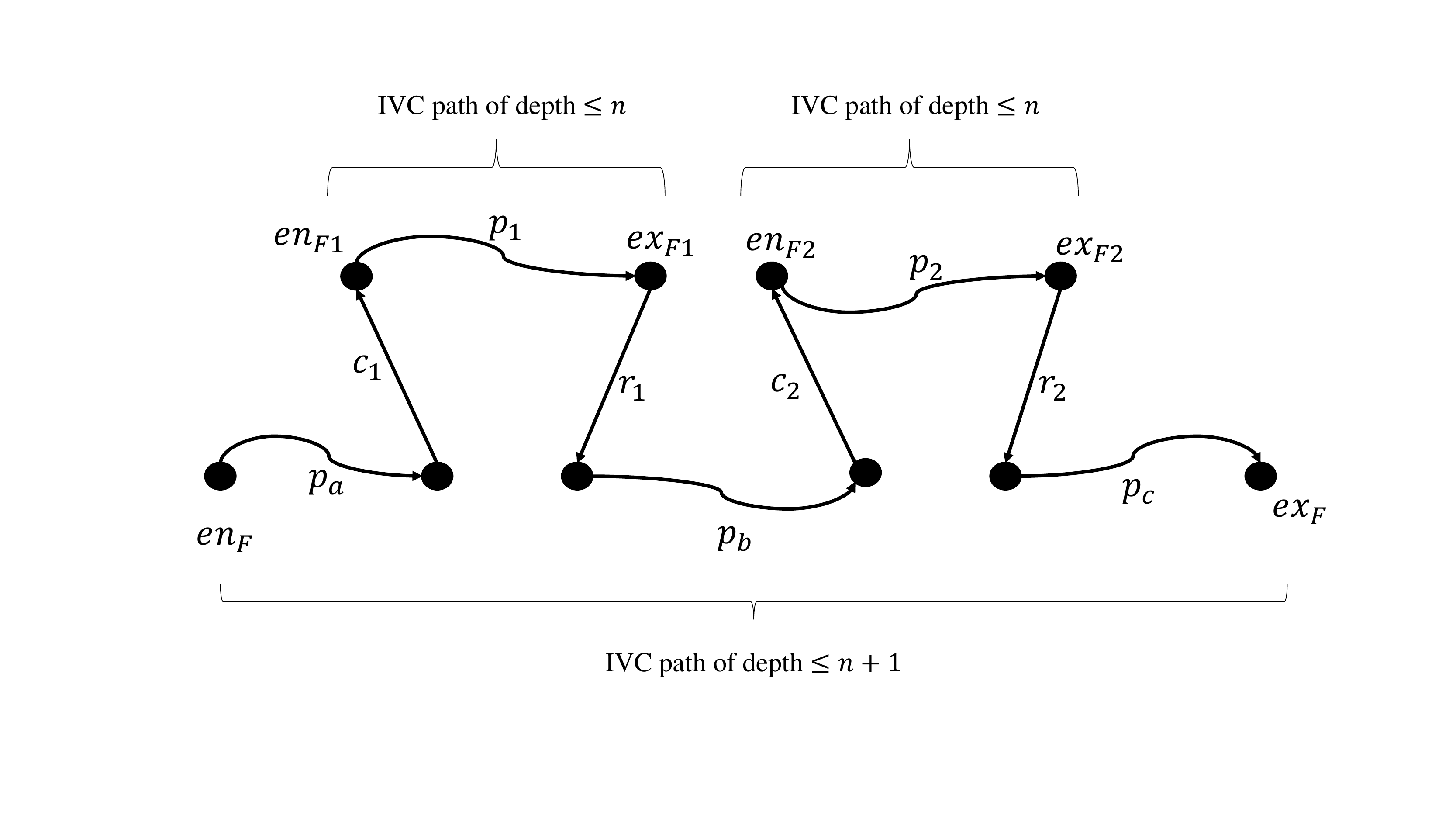}
	\caption{IVC path \vpath{p} of depth $n+1$ }
	\label{fig:computeetoe}
\end{figure}

We now prove the inductive case. We assume that the hypothesis holds for all paths of depth up to $n$, i.e., for a given $d$, all paths of depth up to $n$ are $d$-covered by their respective $\sIVCPathsName$ sets. 

Let $p$ be an IVC path from $\entrynode{F}$ to $\exitnode{F}$ of depth $n+1$.

Any IVC path of $F$ depth $n+1$ will have the following structure. It will start at the method entry $\entrynode{F}$, then reach a call-site ,traverse an IVC path of depth upto $n$, then return to $F$, then reach another call-site, traverse another $\leq n$ depth IVC path, and so on, until it reaches the exit node $\exitnode{F}$.

For the simplicity of discussion, for now we assume that $p$ contains just two outermost level calls. Figure~\ref{fig:computeetoe} shows a schematic structure of such a path. At the end of the proof, we will discuss how to extend the proof for paths with arbitrary number of calls. 

In Figure~\ref{fig:computeetoe}, as $p$ is of depth $n+1$, $\vpath{p_1}$ and $\vpath{p_2}$ are paths of depth $n$, and we have from the hypothesis that $\vpath{p_2}$ is $d$-covered by $\cover{p_2} \in \sIVCPaths{F2}{d}$, and $\vpath{p_1}$ is $d$-covered by $\cover{p_1} \in \sIVCPaths{F1}{d}$. In the figure, the paths $p_1, p_2, p_a, p_b$ and $p_c$ are all in non-\main procedures and hence do not receive any messages. The edges $c_1, c_2, r_1,$ and $r_2$ are call and return edges.

We now obtain a set of paths that $d$-covers $\vpath{p_b \concat c_2 \concat p_2 \concat r_2 \concat p_c}$

Let $S_2 = \lbrace \vpath{p_b \concat c_2 \concat p_2'\concat r_2 \concat p_c \mid p_2' \in \cover{p_2}}\rbrace$

By the inductive hypothesis, $\demandpd{p_2'}{d} \leq \demandp{p_2}{d}$ for each $p_2' \in \cover{p_2}$. 

Therefore, according to Lemma~\ref{lemma:3},  each path $p_j \in S_2$ is such that $\demandpd{p_j}{d} \leq  \demandpd{\vpath{p_b \concat c_2 \concat p_2 \concat r_2 \concat p_c }}{d}$.

Again by the hypothesis, and by Lemma~\ref{lemma:4}, the join of the path transfer procedures of the paths in $S_2$ dominates the path transfer procedure of $\vpath{p_b \concat c_2 \concat p_2 \concat r_2 \concat p_c}$. Therefore, $S_2$ $d$-covers the path $p_b \concat c_2 \concat p_2 \concat r_2 \concat p_c$. $\stmtno{1}$

Let $S_{12} = \lbrace \vpath{p_a \concat c_1 \concat p_1 \concat r_1 \concat p_i \mid  p_i \in S_2 } \rbrace$.

By Lemma~\ref{lemma:1}, we have for all $p'\in S_{12}$

$\demandpd{p'}{d} \leq \demandpd{p}{d}$

Applying Lemma~\ref{lemma:2}, we can infer that the join of  path transfer procedures of paths in $S_{12}$ dominates $\ptf{p}$. Thus, $S_{12}$ $d$-covers $p$. $\stmtno{2}$

Now for each $p_i \in S_2$, let $S_{i2}$ be the set 

$S_{i2} = \lbrace \vpath{p_a \concat c_1 \concat p_1' \concat r_1 \concat p_i} \mid p_1' \in \cover{p_1} \rbrace$

From Lemma~\ref{lemma:3} and \ref{lemma:4}, by taking $p_0 = \vpath{p_a \concat c_1}$, $p_3 = \vpath{r_1 \concat p_i}$, and $\cover{p_1}$ to be $S_2$, it can be seen that $S_{i2}$ $d$-covers $p'= \vpath{p_a \concat c_1 \concat p_1 \concat r_1 \concat p_i}$. $\stmtno{3}$

From the definition of $S_{12}$ it is clear that $p' \in S_{12}$. $\stmtno{4}$

Let $S = \underset{p_i \in S_2}{\bigcup} S_{i2}$.

From $(2), (3), (4)$, and  Lemma~\ref{lemma:5}, the set $S$ $d$-covers $p$. $\stmtno{5}$

As $\cover{p_1} \subseteq \sIVCPaths{F1}{d}$, and $\cover{p_2} \subseteq \sIVCPaths{F2}{d}$, by Lines 8-9 in procedure \textsc{ComputeEndToEnd} in algorithm, we know that each path in the set $S$ is generated by the algorithm. Therefore, by $(5)$ and Lemma~\ref{lemma:6a}, the set $\sIVCPaths{F}{d}$ $d$-covers $p$.

For arbitrary structure, the above reasoning can be repeated the required number of times, i.e., for every call made in the path $p$,  the proof will find the $d$-covering paths for the path suffix starting from the end of the current call to the exit node, and extend it backwards with the $d$-covering paths of the IVC path of the current call, as done in this proof. Therefore, an inductive proof with induction on the number of calls in a path will be able to prove it using the same arguments.
\emph{Hence proved.}

We use the concept of \emph{segment} in the subsequent proofs, which can be defined  as follows. 

\begin{definition}[Segment]
	\label{def:segment}
	Any interprocedurally valid path between any two nodes, where each node can bes a node in the VCFG of any procedure in the system, can be considered as a sequence of segments. Each \emph{segment} is either a single intra-procedural edge, a call edge, or a path consisting of:
	
	$\langle $ a call edge from a call-site $v_i$ to a procedure $F$,\\
	\hspace*{3pt} followed by an IVC path from the start of $F$ to the end of $F$,\\
	\hspace*{3pt} followed by the return edge from the exit of $F$ to the return site
	corresponding to $v_i$ $\rangle$
	
\end{definition}

Similar to the proof of Lemma~\ref{lemma:IV}, it is easy to see that this definition is valid, that is, the three kinds of segments are sufficient. For instance, a return edge cannot be a segment, as then it would allow a  single return edge to be deemed an interprocedurally valid path, which is not correct.

The next theorem ensures that the paths that do not go through \main and end at the node \targetnode\, in the VCFG of any $F \in \funcs$ are covered by the paths generated and stored by the algorithm.

\begin{theorem}[Covering in non-main procedures]
	\label{thm:algo2coverNonMain}
	Let $\vpath{p}$ be any interprocedurally valid path from a node $v_1$ in the VCFG of some procedure $F_1$, and say the \targetnode\, node is in some procedure $F$(i.e. \targetnode\, is not in \texttt{main}). When the algorithm terminates, there exists a set of paths  $\cover{p} \subseteq \sPaths{v_1}$ such that $\cover{p}$ covers $\vpath{p}$.
\end{theorem}

\paragraph{Proof: } We prove the theorem using induction on the number of segments in the given path $p$, where segments are as defined in Definition~\ref{def:segment}. Without loss of generality, we assume that the \targetnode\, node cannot be a return-site node or an entry node of any procedure. In order to compute the JOFP for these nodes, one can always introduce dummy successor nodes from these nodes.

The base case is when the path has only one segment. Due to the assumptions on $\targetnode\,$ stated above, $p$ must be of the form $v_i \vcfgtrans{f}{w} \targetnode$, i.e., a path consisting of only a single intra-procedural edge (the other segments end at either a return-site node or an entry node). All such edges are added to $\sPaths{v_i}$ at line 4 of routine \textsc{ComputeJOFP} in the algorithm. Therefore, $p \in \sPaths{v_i}$ and each as path covers itself by definition, the base case holds.

Now we prove the inductive case. From the inductive hypothesis we have that all paths $p_1$ having $n$ segments are covered.

Based on the types of segments, the inductive case has 3 cases. The first case is when $p$ is of the form $p = (v_i \vcfgtrans{f}{w} v_j) \concat p_1$, where $p_1$ is a
path from $v_j$ to $\targetnode\,$ having $n$ segments, and $p_1$ is covered by $\cover{p_1} \subseteq \sPaths{v_j}$. 

Consider the following set of paths:

$$S_1   =   \lbrace (v_i \rightarrow v_j).p_j \mid p_j \in \cover{p_1} \rbrace$$

By mapping $d$ in Lemma~\ref{lemma:1} to be the zero vector, $(v_1 \rightarrow v_i)$ to $p_0$, and each path $p_j \in \cover{p_1}$ to $p_2$, we have for every path $p_i$ in $S_1$ is such that

\begin{equation}
\label{eqn:thm2e1}
\demandp{p_i} \leq \demandp{p}
\end{equation}

By the inductive hypothesis, the join of the path transfer procedures of the
paths in $\cover{p_1}$ dominates the path transfer procedure of $p_1$. Therefore,
by Lemma~\ref{lemma:2}, we have,

\begin{equation}
\label{eqn:thm2e2}
\underset{p_i \in S_1}{\bigjoin} \ptf{p_i} \dominates \ptf{p}
\end{equation}

From Equations~\ref{eqn:thm2e1} and~\ref{eqn:thm2e2} it follows that $S_1$ covers $p$.

Since every path in $\cover{p_1}$ is present in $\sPaths{v_j}$ (inductive
hypothesis), the algorithm would have generated every path in $S_1$ (at Lines 21 and 22). Therefore, by applying Lemma~\ref{lemma:6} and using the fact that $S_1$ covers $p$ we can infer that a set paths that covers $p$ is present in $\sPaths{v_i}$ when \textsc{ComputeJOFP} terminates.

Case 2 is when the first segment of $p$ is a call-return path as described
in Definition~\ref{def:segment}. Let $\vpath{p} = \vpath{p_1 \concat p_2}$, where $\vpath{p_1}$ is the call-return path, and $\vpath{p_2}$ is the remainder of $\vpath{p}$. Let $v_j$ be the end node of $\vpath{p_1}$ (i.e., a return-site node) and the start node of $\vpath{p_2}$ as well. Let $v_i$ be the start node of $\vpath{p_1}$ (i.e., the call-site node corresponding to $v_j$), and let $F$ be the procedure that is called from $v_i$.

By the inductive hypothesis, there exists a set of paths  $\cover{\vpath{p_2}}$  in $\sPaths{v_j}$  that cover $\vpath{p_2}$. $\stmtno{3}$ 

Consider the following set of paths:

$$S_1 = \lbrace \vpath{p_1.p_i} \mid  \vpath{p_i}  \in \cover{\vpath{p_2}} \rbrace$$

Due to the assumptions on the VCFG, the \main procedure is not called by any other procedure. Therefore, $v_i$ and $\targetnode\,$ are both not in \texttt{main}, the paths considered in the theorem do not go through the \main procedure. As a result the demands of all paths are $\zerovector$. Therefore, each path $\vpath{p_j}$ in $S_1$ is such that $\demandp{\vpath{p_j}} \leq \demandp{\vpath{p}}$. Again by the inductive hypothesis, and by Lemma~\ref{lemma:2}, the join of the path transfer procedures of the paths in $S_1$ dominate the path transfer procedure of $\vpath{p}$.

Therefore, $S_1$ covers $\vpath{p}$ $\stmtno{4}$ 

Let $c$ be the call edge from $v_i$ to $\entrynode{F}$ and $r$ be the corresponding return edge from $\exitnode{F}$ to $v_j$. Consider any path $\vpath{p_i}$ in $\cover{\vpath{p_2}}$ and the following set:

\begin{math}
S_{\vpath{p_i}} = \lbrace \vpath{c \concat p_j \concat r \concat p_i} \mid  \vpath{p_j} \in  \textsc{computeEndToEnd}(F, \zerovector) \rbrace
\end{math}

From Theorem~\ref{thm:algo2ComputeE2Ecover}, we know that the set of paths $\sIVCPaths{F}{\zerovector}$ $d$-covers $p_1$ where $d = \zerovector$. Therefore, using Lemma~\ref{lemma:3}, by mapping $c$ to $p_0$, $r$ to $p_3$, $p_1$ to $p_1$ and each path $p_j \in  \textsc{computeEndToEnd}(F, \zerovector)$ to $p_2$, it follows that for every path $\vpath{p_k} \in S_{\vpath{p_i}}$,

$ \demandp{\vpath{p_k}} \leq \demandp{\vpath{p_1 \concat p_i}} $  $ \stmtno{5}$ 

Also, by Lemma~\ref{lemma:4} by mapping $\textsc{computeEndToEnd}(F, \zerovector)$ to $S_2$, it follows that the join of the path
transfer procedures of the paths in $S_{\vpath{p_i}}$ dominates the path transfer
procedure of $\vpath{p_1.p_i}$.         $\stmtno{6}$

From $(5)$ and $(6)$,  it follows that the set $S_{\vpath{p_i}}$ covers the path $\vpath{p_1 \concat p_i}$.  $\stmtno{7}$

The path $\vpath{p_i}$ was generated by the algorithm (as per definition of  $\cover{\vpath{p_2}}$). Thus, from lines 10-11 in the pseudocode of \textsc{ComputeJOFP}, it is clear that algorithm  generates all paths in $S_{\vpath{p_i}}$.   $\stmtno{8}$

Consider the following set:

\begin{math}
S_2   =   \underset{\vpath{p_i} \in \cover{p_2}}{\bigcup} S_{\vpath{p_i}}
\end{math}

From the definition of $S_1$, from statements $(4)$ and $(8)$, from Lemma~\ref{lemma:5}, and from the definition of $S_2$, it follows that:

$S_2$ covers $\vpath{p}$  $\hspace{15pt} - (7)$

From Statement $(8)$, it is clear that the algorithm generates every path in
$S_2$. From this, and from Statement $(7)$ and Lemma~\ref{lemma:6}, we infer that
when algorithm  terminates, $\sPaths{v_i}$ will contain a set of
paths that cover $p$.

The third case is when $p$ is of the form $p = \vpath{e_c \concat p_1}$, where $e_c$ is a call-site-to-entry-node edge for method $F_2$, and $p_1$ is covered by $\cover{p_1} \subseteq \sPaths{\entrynode{F_2}}$. The proof in this case is again similar to the first case.

\emph{Hence proved.}

\vspace{10pt}

Next we need to prove that the algorithm generates and stores paths that cover all the paths in the system. In order to do this, we first prove the following important lemma.

\begin{lemma}
	\label{lemma:8}
	Let $\vpath{p_j}$ be any inter-procedurally valid path from the entry of a procedure
	$F_i$ to the target node $v$ that is inside some procedure in $\funcs$ such that the algorithm has
	added $\vpath{p_j}$ to $\sPaths{\entrynode{F_i}}$. If $\vpath{p_1}$ is any interprocedurally valid path such that $\vpath{p_1}$ begins in some vertex $v_i$ in \main and ends at a call-site node $v_k$ in \main from which there is a call-edge $c$ to $\entrynode{F_i}$, then it can be shown that when the algorithm terminates there exist a set of paths in $\sPaths{v_i}$ such
	that this set of paths covers the path $p =\vpath{p_1 \concat c \concat p_j}$.	
\end{lemma}

\paragraph{Proof: }

By Lines 14-16 in the procedure \textsc{ComputeJOFP} in algorithm , and from our assumption that $\vpath{p_j}$ is in $\sPaths{\entrynode{F_i}}$, it follows that  $\sPaths{v_k}$ will contain a
set of paths, denoted as $\cover{\vpath{c \concat p_j}}$, that cover the path $\vpath{c \concat p_j}$.   $\stmtno{1}$

The proof of this lemma is by induction on the number of segments, as defined in Definition~\ref{def:segment}, in $\vpath{p_1}$. As $\vpath{p_1}$ is an IVC path in \main (if $p_1$ is not IVC then the edge $c$ will not be from \main to $\entrynode{F_i}$, and hence will not satisfy the requirements of the lemma), there can be only two kinds of segments in $\vpath{p_1}$ - intra-procedural edge in \main, or an IVC call return path.

The base case is when $\vpath{p_1}$ has a single segment. This segment has to be of the form $v_i \rightarrow v_k$, where $v_i \rightarrow v_k$ is an edge in \main (the other kind of segment ends at a return-site node, not at a call-site node).  Consider the
following set of paths:

$$S_1   =   \lbrace (v_i \rightarrow v_k) \vpath{\concat p_k}\mid  \vpath{p_k} \in \cover{\vpath{c \concat p_j}}\rbrace$$

Since there are no receive operations inside the procedures, the $\demandp{\vpath{c \concat p_j}} = \bar{0}$. Therefore, from Statement $(1)$, since every path $\vpath{p_j}$ in $\cover{\vpath{c \concat p_j}}$ covers $\vpath{c \concat p_j}$, every path $\vpath{p_k}$ in $\cover{\vpath{c \concat p_j}}$ is such that $\demandp{\vpath{p_k}} = 0$.  Therefore, for every path $\vpath{p_l} \in S_1, \demandp{\vpath{p_l}} \leq \demandp{p}$.  $\stmtno{2}$

From Statement $(1)$, since $\cover{\vpath{c \concat p_j}}$ covers $\vpath{c \concat p_j}$, the join of the path transfer procedures of the paths in $\cover{\vpath{c \concat p_j}}$  dominates the path transfer procedure of $\vpath{c \concat p_j}$. Therefore, by Lemma~\ref{lemma:2}, the join of the path transfer procedures of the paths in $S_1$ dominates the path transfer procedure of $\vpath{p}$.  $\stmtno{3}$

From Statements $(2)$ and $(3)$, it follows that $S_1$ covers the path $\vpath{p}$.  $\stmtno{4}$

Since every path in $\cover{\vpath{c \concat p_j}}$ is present in $\sPaths{v_k}$
(Statement $(1)$ above), the algorithm would have generated every path in $S_1$, and
would have checked whether to add this path to $\sPaths{v_i}$ or whether this path is already covered by paths in $\sPaths{v_i}$ (Line 21-22 in the pseudocode for \textsc{ComputeJOFP}). This, in conjunction with Statement $(4)$ above and Lemma~\ref{lemma:6} lets us infer that a set of paths that covers \vpath{p} is present in $\sPaths{v_i}$ when the algorithm terminates.

We now move onto the inductive case. We assume that the lemma is true
whenever the path ending at $v_k$ is of length at most $n$ segments. Let $\vpath{p_1}$
consists of $n+1$ segments. Based on types of segments, the argument proceeds under two cases.

The first case is that $\vpath{p_1}$ is of the form $(v_i \rightarrow v_j).p_2$, where $v_i \rightarrow v_j$ is an edge in \main, and $v_j$ is the first vertex in the suffix path $\vpath{p_2}$.

Since $\vpath{p_2}$ has at most $n$ segments, the inductive hypothesis is applicable on
the path $\vpath{p_2 \concat c \concat p_j}$. The remainder of the argument is identical to the same inductive case in the proof of Theorem~\ref{thm:algo2coverNonMain}.

The second case is that $\vpath{p_1}$ is of the form $\vpath{p_3 \concat p_2}$, where $\vpath{p_3}$ is an IVC call-return path ($\vpath{p_3}$ is the first segment of $\vpath{p_1}$), and $\vpath{p_2}$ is the remainder of $\vpath{p_1}$. Since $\vpath{p_2}$ has at most $n$ segments, the inductive hypothesis is applicable on the path $\vpath{p_2 \concat c \concat p_j}$. 

Let $v_j$ be the end node of $p_3$ (i.e., a return-site node) and start node of $p_2$ as well. Let $v_i$ be the start node of $p_3$ (i.e., the call-site node corresponding to $v_j$), and let $F$ be the procedure that is called from $v_i$.

Let $p' = p_2.c.p_j$. By the inductive hypothesis, there exists a set of paths  $\cover{\vpath{p'}}$  in $\sPaths{v_j}$  that cover $\vpath{p'}$. $\hspace{15pt} - (5)$ 

Consider the following set of paths:

$$S_1 = \lbrace \vpath{p_3.p_i} \mid  \vpath{p_i}  \in \cover{\vpath{p'}} \rbrace$$

By the inductive hypothesis, $\demandp{\vpath{p_i}} \leq \demandp{\vpath{p'}}$ for each $\vpath{p_i}$ in $\cover{\vpath{p'}}$. Therefore, according to Lemma~\ref{lemma:1} (taking $d$ in that lemma to be the zero vector), each path $\vpath{p_j}$ in $S_1$ is such that $\demandp{\vpath{p_j}} \leq \demandp{\vpath{p}}$. Again by the inductive hypothesis, and by Lemma~\ref{lemma:2}, the join of the path transfer procedures of the paths in $S_1$ dominate the path transfer procedure of $\vpath{p}$.

Therefore, $S_1$ covers $\vpath{p}$ $\hspace{15pt} - (6)$

Consider any path $\vpath{p_i}$ in $\cover{\vpath{p'}}$. Let $c_1$ be the call edge from $v_i$ to $\entrynode{F}$ and $r_1$ be the corresponding return edge from $\exitnode{F}$ to $v_j$. Consider the following sets:

\begin{math}
T_{\vpath{p_i}} = \lbrace \vpath{p_j} \mid  \vpath{p_j} \in \textsc{computeEndToEnd}(F, \demandp{\vpath{p_i}}), \demandpd{\vpath{p_j}}{\demandp{\vpath{p_i}}} \leq \demandpd{\vpath{p_3}}{\demandp{\vpath{p_i}}} \rbrace
\end{math}

\begin{math}
S_{\vpath{p_i}} = \lbrace \vpath{c_1 \concat p_j \concat r_1 \concat p_i} \mid  \vpath{p_j} \in T_{\vpath{p_i}} \rbrace
\end{math}

From Theorem~\ref{thm:algo2ComputeE2Ecover}, it follows that the set $T_{p_i}$ $d$-covers the path fragment from $\entrynode{F}$ to $\exitnode{F}$ in $p_3$, where $d = \demandp{p_i}$.

Therefore, from Theorem~\ref{thm:algo2ComputeE2Ecover}, Lemma~\ref{lemma:7} by taking $p_0$ to be $p_i$ in Lemma~\ref{lemma:7}, and $Lemma~\ref{lemma:3}$, it follows that for every path $\vpath{p_k} \in S_{\vpath{p_i}}$,

$ \demandp{\vpath{p_k}} \leq \demandp{\vpath{p_3 \concat p_i}} $  $ \hspace{15pt} - (7)$ 

From Theorem~\ref{thm:algo2ComputeE2Ecover}, it also follows that the
join of the path transfer procedures of the paths in $T_{\vpath{p_i}}$ dominates the
path transfer procedure of the path fragment from $\entrynode{F}$ to $\exitnode{F}$ in $p_3$. Therefore, by Lemma~\ref{lemma:4}, the join of the path
transfer procedures of the paths in $S_{\vpath{p_i}}$ dominates the path transfer
procedure of $\vpath{p_3.p_i}$.         $\hspace{15pt} - (8)$

From $(7)$ and $(8)$,  it follows that the set $S_{\vpath{p_i}}$ covers the path $\vpath{p_3 \concat p_i}$.  $\hspace{15pt} - (9)$

$\vpath{p_i}$ was generated by the algorithm (as per definition of  $\cover{\vpath{p'}}$). Thus, from lines 10-11 in the pseudocode of \textsc{ComputeJOFP}, it is clear that algorithm generates all paths in $S_{\vpath{p_i}}$.   $\hspace{15pt}-  (10)$

Consider the following set:

\begin{math}
S_2   =   \underset{\vpath{p_i} \in \cover{p'}}{\bigcup} S_{\vpath{p_i}}
\end{math}

From the definition of $S_1$, from statement $(6)$ and $(9)$, from Lemma~\ref{lemma:5}, and
from the definition of $S_2$, it follows that:

$S_2$ covers $\vpath{p}$  $\hspace{15pt} - (11)$

From Statement $(10)$, it is clear that the algorithm generates every path in
$S_2$. From this, and from Statement $(11)$ and from Lemma~\ref{lemma:6}, we infer that
when algorithm  terminates, $\sPaths{v_i}$ will contain a set of
paths that cover $p$.

\emph{Hence proved.}

\begin{theorem}
	\label{thm:algo2cover}
	If \vpath{p} is an interprocedurally valid path from a node $v_i$ in \main to a target node $v$ such that $v$ is in any procedure (including \texttt{main}), then when algorithm  terminates,  there exists a set of paths $\cover{p} \subseteq \sPaths{v_i}$ such that $\cover{p}$ covers \vpath{p}.
	
\end{theorem}
\paragraph{Proof: } Based on the structure of $p$, there can be two possible cases.

First is when the target $v$ is in \texttt{main}, and therefore the structure of $p$ is that it starts from $v_i$ in \texttt{main}, goes via vertices in \texttt{main}, enters procedures whose calls it encounters in \main and returns to \texttt{main}, again goes via vertices in \texttt{main}, and so on, and then ends at $v$ in \texttt{main}. In this case, the proof is similar to the proof of Lemma~\ref{lemma:8}, by taking the interprocedurally valid path suffix $c.p_j$ to be empty, and the induction is on the number of segments in $p$.

The other case is when the structure of $p$ is that it starts from vertex $v_i$ in  \texttt{main}, goes via vertices in \texttt{main}, enters procedures whose calls it encounters in \main
and returns to \texttt{main}, again goes via vertices in \texttt{main}, and so on, until it
makes a final entry into a procedure $F_i$ such that after this entry it
eventually reaches the target vertex $v$ (which may be in $F_i$ or a transitive
callee of $F_i$) without returning from $F_i$.

Therefore, \vpath{p} is of the form $\vpath{p_1 \concat c \concat p_2}$, where $ \vpath{p_2}$ is the suffix of  \vpath{p} from the $\entrynode{F_i}$ to $v$ (without returning from $F_i$), $\vpath{p_1}$ is a path from $v_i$ to a call-site node in \texttt{main}, and $c$ is a call-edge from this call-site node to $\entrynode{F_i}$. 

According to Theorem~\ref{thm:algo2coverNonMain}, taking $v_1$ to be $\entrynode{F_i}$, when algorithm terminates, there exists a set of inter-procedurally valid paths $\cover{\vpath{p_2}}$ in $\sPaths{\entrynode{F_i}}$ such that all these paths are from $\entrynode{F_i}$ to $v$, and the join of the path transfer procedures of these paths dominates $\ptf{p_2}$. $\stmtno{1}$

Consider the following set of paths:

$$S_1  =  \lbrace \vpath{p_1 \concat c \concat p_j} \mid  p_j \in \cover{p_2} \rbrace$$

From $(1)$, and from Lemmas~\ref{lemma:1} and~\ref{lemma:2}, it follows that $S_1$ covers $p$. $\stmtno{2}$

For any path $p_j$ in $\cover{p_2}$, according to Lemma~\ref{lemma:8}, when the algorithm
terminates, a set of paths that cover $\vpath{p_1 \concat c \concat p_j}$ exist in $\sPaths{v_i}$. Therefore, it follows from the definition of $S_1$ that a set of paths
exist in $\sPaths{v_i}$ that cover $S_1$.

The statement above, together with Statement $(2)$ and Lemma~\ref{lemma:5}, implies that
a set of paths exist in  $\sPaths{v_i}$ that cover $p$.

\emph{Hence proved.}

\vspace{10pt}

\paragraph{Proof of Theorem~\ref{thm:soundness} }

Now we are ready to prove the soundness of the algorithm.

As proved in Theorem~\ref{thm:algo2cover} any interprocedurally valid path $p$ from \startnode\, node to $\targetnode$ node  is covered by $\cover{p} \subseteq \sPaths{\startnode}$. From Lemma~\ref{lemma:IV}, we know that all paths in $\sPaths{\startnode}$ are interprocedurally valid. As the set of all feasible paths is subset of all possible paths, Theorem~\ref{thm:algo2cover} holds for all feasible paths as well.

All feasible paths from $\startnode$ to $v$ have a demand of $\zerovector$ (else they have more receives than sends in some prefix of the path). Therefore, by the definition of covering, all paths in the set $\cover{p}$ will have a demand $\zerovector$.  

Let $S  = \lbrace p_i \mid p_i \in \sPaths{\startnode} \wedge \demandp{p} = \zerovector \rbrace $.

Clearly, $S \supseteq \cover{p} $, thus $S$ covers $p$.

From Lines 25-26 in \textsc{ComputeJOFP} we have,

$$d = \underset{p_i \in S}{\bigjoin} \ptf{p_i} (d_0)$$ 

where $d$ is the value returned by the algorithm. As $S$ covers $p$, therefore

$$\underset{p_i \in S}{\bigjoin} \ptf{p_i} \dominates \ptf{p}$$

Hence, $d \dominates \ptf{p}(d_0)$, for any $p$. 

As $d \dominates \ptf{p}(d_0)$ for any feasible $p$, and by the property $(a \dominates b \wedge a \dominates c) \Rightarrow (a \dominates b \join c) $, for the set of all feasible paths $P = \lbrace p_1, p_2, \ldots \rbrace$ that begin at \startnode, and end at $v$,

$$d \dominates \ptf{p_1}(d_0) \join \ptf{p_2}(d_0) \join \ldots $$

By condensing the above inequation, we get

$$ d \dominates \underset{\substack{\textit{p is a feasible, interprocedurally valid} \\ \textit{path from $\startnode$ to $\targetnode$}}}{\bigsqcup} (\ptf{p})(d_0)$$

\emph{Hence proved.}

\subsubsection{Precision}
\label{ssec:inter:proof:precision}

Our algorithm claims to compute the precise JOFP of multi-procedure VCFGs, \emph{under the unordered channel abstraction}. That is, we do not include values due to paths which are infeasible under this abstraction, but may include values due to paths which are infeasible when the order of messages in the channel is maintained. The following theorem proves the precision of our algorithm.

\begin{theorem}[Precision]
	\label{thm:precision}
For any node $\targetnode$, let $d \in \dlattice$ be the JOFP value computed by \textsc{ComputeJOFP} in algorithm  for $\targetnode$, treating $d_0 \in \dlattice$ as the initial value at the $\startnode$ node, and $\startnode$ as the initial node. Then, 

$$ d = \underset{\substack{\textit{p is a feasible, interprocedurally valid} \\ \textit{path from $\startnode$ to $\targetnode$}}}{\bigsqcup} (\ptf{p})(d_0)$$
\end{theorem}

\paragraph{Proof: } From the Soundness theorem for algorithm , we know that,

$$ d \dominates \underset{\substack{\textit{p is a feasible, interprocedurally valid} \\ \textit{path from $\startnode$ to $\targetnode$}}}{\bigsqcup} (\ptf{p})(d_0)$$

We know that for any $p_i \in \sPaths{\startnode}$, where $p_i$ has a demand of $\zerovector$, $p_i$ is a feasible path.

Therefore, the set $S = \lbrace p_i \mid p_i \in \sPaths{\startnode} \wedge \demandp{p_i} = \zerovector \rbrace$ is a subset of the set of feasible paths reaching $v$. Also, we know from Lemma~\ref{lemma:IV} that the all the paths in $\sPaths{\startnode}$ are interprocedurally valid. Thus,

$$\underset{\substack{\textit{p is a feasible, interprocedurally valid} \\ \textit{path from $\startnode$ to $\targetnode$}}}{\bigsqcup} (\ptf{p}) \dominates \underset{p_i \in S}{\bigjoin}\ptf{p_i} \stmtno{2}$$

According to Lines 17-18 of method \textsc{ComputeJOFP} in the algorithm ,

$$d = \underset{p_i \in S}{\bigjoin} \ptf{p_i} (d_0) \stmtno{3} $$

From $(2)$ and $(3)$, we have 

$$\underset{\substack{\textit{p is a feasible, interprocedurally valid} \\ \textit{path from $\startnode$ to $\targetnode$}}}{\bigsqcup} (\ptf{p}) \dominates d  \stmtno{4}$$

From $(1)$ and $(4)$, we can infer,

$$\underset{\substack{\textit{p is a feasible, interprocedurally valid} \\ \textit{path from $\startnode$ to $\targetnode$}}}{\bigsqcup} (\ptf{p})= d  \stmtno{4}$$

\emph{Hence proved.}

\subsubsection{Complexity Analysis of Backward DFAS Algorithm}
\label{sec:complexity}

We present here the complexity derivation of the single procedure case in the Backward DFAS algorithm. The derivation additionally assumes that the transfer functions are right-distributive.

Let $Q$ be the number of locations,  $K$ be the number of counters and
let $h$ be the height of the transfer function lattice. Wlog. we assume
that each transition changes the value of any counter by at most $1$.
For this section we assume that the composition operator is distributive
on the lattice of transfer functions.

Let $\pi$ be a run from $(p,\ol{u})$ to $(q,\ol{v})$, written
$(p,\ol{u}) \Path{\pi} (q,\ol{v})$. We shall write $f_\pi$ to denote
the transfer function defined by composing those associated with the transitions
along the run.
We say that a set $P$ of runs covers $\pi$, if 
for each $\rho \in P$,  $(p,\ol{u}) \Path{\rho}
(q,\ol{w})$ with $\ol{v} \below \ol{w}$ and $f_\pi \below \sqcup_{\rho \in P} f_\rho$. Note that this notion of covering is defined over runs and as we shall
see it is related to the notion defined for paths earlier.
We say that $P$ strictly covers $\pi$ if for each $\rho \in P$, the final configuration is identical to $(q,\ol{v})$. We shall write $f_P$ to denote the function $\sqcup_{\rho \in P} f_\rho$.  The following are easy to see.

\paragraph{Fact 1:} If the set of runs $P$ covers (resp. strictly covers)  $\pi$,
and for each $\rho \in P$, the set $P_\rho$ covers (resp. strictly covers)  $\rho$ then 
$\bigcup_{\rho \in P} P_\rho$ covers (resp. strictly covers) $\pi$. 

\paragraph{Fact 2:} Suppose $P$  covers (resp. strictly covers)  $\pi$ from $(p,\ol{u})$ to $(q,\ol{v})$.  Suppose $(q,\ol{v}) \Move{\delta} (r,\ol{w})$. Then, $\{(p,\ol{u})\Path{\rho}(q,\ol{v})\Move{\delta} (r,\ol{w}) ~|~ \rho \in P\}$ strictly covers $(p,\ol{u})\Path{\rho}(q,\ol{v})\Move{\delta}(r,\ol{w})$.

We shall often write $\rho.\delta$ to refer to $(p,\ol{u})\Path{\rho}(q,\ol{v})\Move{\delta}(r,\ol{w})$.

\paragraph{Fact 3:} If $P_1$ is a set of runs that strictly cover
$\pi_1$ and suppose $P_2$ is a set of runs that cover $\pi_2$. Then,
$P_1.P_2$ covers $\pi_1.\pi_2$.
\smallskip

We write $\length(P)$ for the maximum of the lengths of the
runs in a finite set  $P$.
Let $\pi$ be a run from $(q,\ol{u})$ to $(t,\ol{v})$ for some $v$.
Then, we let $\elength(\pi)$ (\emph{effective length})  to be 
$$ Minimum ~ \{ \length(P) ~|~ P \mbox{ covers } \pi\}$$

First we consider the case when all configurations along the run are bounded
by a value $B$, i.e., the value of each counter in each configuration along
the run (including the initial and final configurations) is bounded by $B$. 
We say that such a run is $B$-bounded.

\begin{lemma}\label{lem:BBound}
	For a $B$-bounded run $(p,\ol{u}) \Path{\pi} (q,\ol{v})$,  we
	have a finite set of runs  $P$, with $\length(P) \leq  Q.(h+1).B^K$ that strictly covers $\pi$, where $K$ is the number of counters.
\end{lemma}
\emph{Proof.}
	We prove this by induction on the
	length of $\pi$.
	If the length of $\pi$ is less than $Q.(h+1).B^K$ then  we may
	take $P = \{\pi\}$.  Otherwise, since the number of $B$ bounded configurations is bounded
	by $Q.B^K$,  we may break up the run as:
	$$(p,\ol{u}) \Path{\pi_0} (r,\ol{x}) \Path{\pi_1} (r,\ol{x}) \Path{\pi_2} \ldots (r,\ol{x})\Path{\pi_h} (r,\ol{x})\Path{\pi'} (q,\ol{v})$$
	where each $\pi_i$ is non-empty. Now, consider the runs 
	$\xi_0 =  \pi_0$, $\xi_1 = \pi_0\pi_1$, $\ldots$, $\xi_{h} = \pi_0\pi_1\ldots \pi_h$.  Then, the increasing sequence 
	$$f_{\xi_0} \latbel f_{\xi_0} \sqcup f_{\xi_1} \latbel  f_{\xi_0}\sqcup f_{\xi_1} \sqcup f_{\xi_2} \ldots \latbel f_{\xi_0} \sqcup f_{\xi_1} \sqcup \ldots \sqcup f_{\xi_h}$$
	has at most $h$ distinct elements. Thus, there is an $i$ such that 
	$f_{\xi_0} \sqcup \ldots \sqcup f_{\xi_i} =  f_{\xi_0} \sqcup \ldots \sqcup f_{\xi_i} \sqcup f_{\xi_{i+1}}$, that is $f_{\xi_{i+1}} \latbel f_{\xi_0} \sqcup \ldots \sqcup f_{\xi_i}$. Hence, by distributivity  of composition,
	\begin{align*}
	f_{\pi} = f_{\xi_{i+1}};f_{\pi_{i+2}\ldots \pi_h\pi'}  &\latbel
	f_{\xi_0};f_{\pi_{i+2}\ldots \pi_h\pi'} \sqcup \ldots \sqcup f_{\xi_i}f_{\pi_{i+2}\ldots \pi_h\pi'} \\
	&= f_{\xi_0\pi_{i+2}\ldots \pi_h\pi'} \sqcup \ldots \sqcup f_{\xi_i\pi_{i+2}\ldots \pi_h\pi'} 
	\end{align*}
	Consider the set of runs 
	\begin{align*}
	P ~  = ~ \{\pi_0\pi_{i+2}\pi_{i+3}\ldots\pi_h\pi',~ & \pi_0\pi_1\pi_{i+2}\pi_{i+3}\ldots\pi_h\pi',~ \ldots,~ \pi_0\pi_1\ldots\pi_i\pi_{i+2}\pi_{i+3}\ldots\pi_h\pi',\\
	& ~ \ldots,~  \pi_0\pi_1\ldots\pi_i\pi_{i+2}\ldots\pi_h\pi'\}
	\end{align*}
	From the above calculation, $P$ strictly  covers $\pi$ and further
	every run in $P$ is strictly shorter than $\pi$.  By the induction hypothesis,
	each $\rho$ in this set $P$ is strictly covered by a set of runs $P_\rho$ containing
	only runs of length at most $Q.(h+1).B^K$. Thus, by Fact 1, $\bigcup_{\rho \in P} P_\rho$ is the desired strict covering set for $\pi$.

Following \cite{bozzelli2011complexity}, for any $I \subseteq \{1, \ldots, K\}$ a subset of
the counters, we define $\ol{u}^I$ to be the function which returns $\ol{u}(i)$
if $i \in I$ and $0$ otherwise. For such an $I$ and a system $G$, we define
$G^I$ to be the one obtained from $G$ where each transition is modified to
leave all counters outside $I$ untouched and operate on counters from $I$  as before.
For any run $\pi$ from $(p,\ol{u})$ to  $(q,\ol{v})$  in $G$ there is a
corresponding run $\pi^I$ from $(p,\ol{u}^I)$ to $(q,\ol{v}^I)$ constituting
a valid run in $G^I$. 

Let us fix a system $G$ and a target location $t$.  
For any configuration $s$ (i.e. $s$ is of the form $(p,\ol{u})$),
lattice function $f$, and $I \subseteq \{1, \ldots, K\}$  we define 
$\dist(I,f,s))$ as follows:
$$
\dist(I,f,s) ~=~ Min (\{0\} \cup \{ \elength(\pi) ~|~ \exists \ol{v}. s \Path{\pi} (t,\ol{v}) \mbox{ in $G^I$ and } f \latbel f_\pi \})
$$

For any $0 \leq k \leq K$, we set $g(k)$ to be 
$sup\{\dist(I,f,s))~|~ |I| = k \}$.  Thus, the function $g(k)$ provides
an upper bound on the length of runs that suffice to cover any run from
any configuration $s$ to a configuration above $(t,\ol{0})$ in any
system $G^I$ with $|I| = k$. We now derive bounds on $g(k)$.

\begin{lemma}\label{lem:compOne}
	For any $k \in \{0,1, \ldots K\}$, 
	\begin{align*}
	g(0) ~& \leq ~ Q.(h+1)  \\
	g(k) ~& \leq ~ Q.(h+1).(1 + g(k-1))^k + 1 +  g(k-1) \mbox{        if $k > 0$}
	\end{align*}
	In particular, $g(k)$ is finite for all $0 \leq k \leq K$.
\end{lemma}
\emph{Proof.}
	The proof follows an argument in the style of Rackoff (\cite{rackoff1978covering}, \cite{bozzelli2011complexity}) and proceeds by induction on $k$. 
	
	For $k = 0$ the result follows directly from Lemma \ref{lem:BBound}.
	
	We examine the inductive case next. Let $|I| = k$.  Suppose, for $s,f$, there is a 
	run $\pi$ from $s$ to $(t,\ol{v})$ in $G^I$ with $f \latbel f_\pi$. We consider
	two cases.
	
	\paragraph{Case 1:}  Suppose every configuration in the run $\pi$ is bounded
	in size by $B = g(k-1) + 1$. Then, by Lemma \ref{lem:BBound}, there is a
	set of runs $P$ in $G^I$ such that $\length(P)$ is bounded
	by $Q.(h+1).(g(k-1) + 1)^k$ which covers $\pi$, i.e., 
	$f_P \latabv f_\pi \latabv f$. This completes the proof in this case.
	
	\paragraph{Case 2:} Suppose there is a configuration in the run $\pi$ where
	at least one of the counters in $I$ exceeds $g(k-1) + 1$. Let one such counter
	be $i \in I$. Then the run $\pi$ in $G^I$ breaks up as:
	$s \Path{\pi_1} s' \Move{\delta} (q,\ol{w}) \Path{\pi_2} (t,\ol{v})$,
	where $s \Path{\pi_1} s'$ is $g(k-1) + 1$ bounded run, $\delta$ is a single
	transition and $\ol{w}(i) > g(k-1)+1$.
	
	Observe that taking $J$ to be $I \setminus \{i\}$ we also get a run 
	$\pi^J_2$ from $(q,\ol{w}^J)$ to $(t,\ol{v}^J)$. Further, by the definition
	of $G^J$ and $G^I$, $f_{\pi_2} = f_{\pi^J_2}$. By the definition of 
	$g(k-1)$, since $|J| = k-1$,  we know that there is a collection of 
	runs $P_2$ in $G^J$ from $(q,\ol{w}^J)$ that cover $\pi^J_2$ such that
	$\length(P_2) \leq g(k-1)$.
	For each run $\rho \in P_2$ let $\rho^I$ be the \emph{pseudo run} (possibly invalid, since positivity of counter $i$ may not be guaranteed) induced
	by the same sequence of transitions in $G^I$. We claim $\rho^I$ is actually
	a valid run in $G^I$. This is because, $w(i) \geq g(k-1) + 1$, the length
	of $\rho^I$ is no more than $g(k-1)$ and each transition may decrease  
	counter $i$ by at most $1$. By the definitions of $G^I$ and $G^J$, $f_{\rho^I} = f_{\rho}$. Thus, the set of runs $P'_2 = \{ \rho^I ~|~ \rho \in P_2 \}$  covers $\pi_2$ in $G^I$.
	
	Further, by Lemma \ref{lem:BBound}, we have a set of runs $P_1$ that
	strictly cover $\pi_1$ in $G^I$, and $\length(P_1) \leq Q.(h+1).(g(k-1)+1)^k$. Thus, using Fact 2, $P_1.\delta$ 
	strictly covers $\pi_1.\delta$ in $G^I$. Then using the above and Fact 3, $P_1.\delta.P'_2$ covers $\pi_1.\delta.\pi_2$ in $G^I$.
	The length of any run in this set is bounded by $Q.(h+1).(g(k-1) + 1)^k + 1 + g(k-1)$ as required.

\begin{lemma}\label{lem:recSol}
	Let $g(k)$ be as defined above.  Then 
	$$ g(k) \leq (Q.(h+1).2)^{(3k)! + 1} $$
\end{lemma}
\emph{Proof.}

	We just follow the proof in \cite{bozzelli2011complexity}. The proof proceeds by induction on $k$.
	For $k=0$, the result follows directly. In the inductive case:
	\begin{align*}
	f(k) ~& \leq ~ Q.(h+1).(1 + g(k-1))^k + 1 +  g(k-1) ~~~& \mbox{By Lemma \ref{lem:compOne}} \\
	&\leq ~ Q.(h+1).[(1 + g(k-1))^k + (1 + g(k-1))] & \\
	&\leq ~ (Q.(h+1)(1 + g(k-1)))^{k+1} & \\
	&\leq ~ (Q.(h+1)(2.g(k-1)))^{k+1} & \\
	&\leq ~ (Q.(h+1).2.g(k-1))^{k+1} & \\
	&\leq ~ (Q.(h+1).2.(Q.(h+1).2)^{(3(k-1))!+1})^{k+1} & \\
	&\leq ~ (Q.(h+1).2)^{(3(k-1))!+2})^{k+1} & \\
	&\leq ~ (Q.(h+1).2)^{(3k)!+1} & \\
	\end{align*}

Let $\lbound = (Q.(h+1).2)^{(3K)! + 1}$.
From Lemma \ref{lem:compOne} and \ref{lem:recSol} we know that for any run $\pi$ from any configuration $s = (p,\ol{u})$ to one with control state $t$ can be covered by runs of length at most $\lbound$. This allows us restrict our analysis
entirely to configurations bounded by $\lbound$.

\begin{lemma}\label{lem:confBound}
	Suppose $s = (p,\ol{u}) \Path{\pi} (t,\ol{v})$ be a run of length at most $m$.
	Let $\ol{w}$ be such that $\ol{u'}(i) = Min(\ol{u}(i),m)$.  Then, there is
	a run $(p,\ol{u'}) \Path{\pi'} (t,\ol{v'})$ where $\pi'$ and $\pi$ follow
	the same sequence of transitions, in particular, $f_\pi = f_{\pi'}$.
\end{lemma}
\emph{Proof.}
	Follows simply from the fact that the length of the run is bounded $m$ and
	each transition may reduce the value of a counter by at most $1$.

\begin{lemma}\label{lem:bound}
	Let $s$ be any configuration and $t$ a location. Then, 
	\begin{align*}
	\bigsqcup & \{ f_\pi ~|~ \exists s' \below s.~ \exists \ol{v}.~  s' \Path{\pi} (t,\ol{v})\} \\
	& ~=~ \bigsqcup \{ f_\pi ~|~ \exists \ol{v}.~ s \Path{\pi} (t,\ol{v})\}\\
	& ~=~ \bigsqcup \{f_\pi ~|~ \exists s' \below s.~ \exists \ol{v}.~ s' \mbox{ is $\lbound$-bounded},  s' \Path{\pi} (t,\ol{v}), \mbox{  with $\length(\pi) \leq \lbound$}\} 
	\end{align*}
\end{lemma}
\emph{Proof.}
	That the second and third  values are below (under $\latbel$) the first follows simply
	from the containment of the underlying sets on which the join is applied.
	Now, if $s' \below s$  and $s' \Path{\pi} (t,\ol{v})$ then clearly there is
	a run $s \Path{\pi'} (t,\ol{w})$ following the same sequence of transitions
	(so that $f_\pi = f_{\pi'}$). This ensures that the first value is below (under $\latbel$) the second.  If $s \Path{\pi} (t,\ol{v})$ then by Lemma \ref{lem:recSol} there is a covering set $P$ for $\pi$ with $\elength(P) \leq \lbound$. 
	We then apply Lemma \ref{lem:confBound} to each element of $P$ to conclude
	that the second value is below the third under $\latbel$, completing the proof the Lemma.


Consider Algorithm 1 and assume that the working set is maintained as a queue.
Then, paths are extended in increasing order of length. We think of the
algorithm as proceeding in rounds. 
Round $i$ pertains to the segment when paths of lengths $i$ 
are extended to paths of length $i+1$, added to the working list if required
and placed in the appropriate bins  i.e. $sPaths(q)$ for appropriate $q$.

Suppose $(q,\ol{u}) \Path{\pi} (target,\ol{v})$ is any run of length $i$.
Let $p$ be the path induced by the run $\pi$. Then, clearly $p$ has length $i$ and further $demand(p,\ol{0}) \leq \ol{u}$.
The proof of correctness given earlier showed  that  there are paths $C_p
\subseteq sPaths(q)$ at the end of round $i$ that cover the path $p$ (here
cover refers to paths and is used in the sense defined in the main paper). Thus,
\begin{itemize}
	\item Each path $p'$ in $C_p$ moves from control location $q$ to $target$ 
	\item For each path $p'$ in $C_p$,  $demand(p',\ol{0}) \leq demand(p,\ol{0})$.
	Thus there is a run $(p,\ol{u}) \Path{\pi_{p'}} (target,\ol{w_{p'}})$ using the
	sequence of transitions $p'$.  Let $P$ be the collection of these runs.
	\item $f_p \latbel \bigsqcup_{p' \in C_p} f_{p'}$. But $f_p = f_\pi$ and 
	for each $p' \in C_p$, $f_{p'} = f_{\pi_{p'}}$. Thus, $f_\pi ~=~ f_p ~\latbel~ 
	\bigsqcup_{p' \in C_p} f_{p'} ~=~ \bigsqcup_{\pi_{p'}\in P} f_{\pi_{p'}}$.
\end{itemize}
For a configuration $s = (q,\ol{u})$ let $\Pi^i_s$ be the set of paths 
in $sPaths(q)$ after the $i$th round  whose demand is below $\ol{u}$ under $\below$.
In other  words, $\Pi^i_s = \{x \in sPaths(q) ~|~ (q,demand(x,\ol{0})) \leq (q,\ol{u})\}$ at the end of the $i$th round. 
Then, for any run of the form $s \Path{\pi} (target,\ol{v})$ inducing a path $p$  $C_p \subseteq \Pi^i_s$ and thus have $f_\pi \latbel \bigsqcup_{p' \in \Pi^i_s} f_{p'}$.


\begin{lemma}\label{lem:correctness}
	For any configuration $s = (p,\ol{u})$ 
	$$
	\bigsqcup \{f_\pi ~|~ \exists \ol{v}.~ s \Path{\pi} (t,\ol{v})\}
	~=~ \bigsqcup \{f_p ~|~ p\in \Pi^\lbound_s\}$$
\end{lemma}
\emph{Proof.}
	This follows from Lemma \ref{lem:bound} and the fact that each run of
	length at most $\lbound$ from $s$ is subsumed by $\Pi^\lbound_s$ as shown
	above.

If $s = (q,\ol{0})$ then $\Pi^\lbound_s$ consists only of paths with demand
$\ol{0}$.  Thus,  $\jop(q,target)$ is the join of the transfer functions
defined by all the runs from $(q,\ol{0})$ to the target as required.

In addition note that the demands at the end of round $i$ of Algorithm 1  are no more than
$i$ on each counter and thus no more than $\lbound$ at the end of $\lbound$
rounds. 


Now we can complete the computation of the complexity of Algorithm 1.
We first note that instead of maintaining a path $p$ it suffices
to maintain its demand vector along with the transfer function defined
by the run.  Since all our demand vectors are $\lbound$-bounded, the
total number of demand vectors is no more than $\lbound^K$. In addition,
in any stage of the algorithm, in any bin, for any demand vector $\ol{v}$
at most $h$ different copies exist (with different associated transfer
functions).  To see this suppose $(\ol{v},f_1), \ldots (\ol{v},f_{h+1})$ 
appear in some bin and suppose this is the order in which they were added.
Then, as argued before, there is an $i$ such that $f_{i+1} \latbel f_1 \sqcup f_2 \ldots f_i$ contradicting the definition of Algorithm 1. Thus,
at any point in the algorithm, there are at most $Q$ bins, each of which
contain at most $h.\lbound^K$ demand vector - transfer function pairs.
In each round, for each possible transition $\delta$  we consider at most
$h.\lbound^K$ possible candidates for extension (all drawn from the same bin). Thus each round considers $\Delta.h.\lbound^K$ candidates.  For each such candidiate the operations required are:
\begin{enumerate}
	\item  manipulating the demand vector by combining it with $\delta$ to determine the new demand. These vectors both consist  of 
	$K$ values each of size $\lbound$ (hence can represented and manipulated using $log(\lbound)$ bits).
	\item  composing the transfer functions 
	\item  checking if the composed transfer function is subsumbed by
	the join of subset of functions selected from the same bin.
\end{enumerate}
The first part uses time proportional to $K.log(\lbound)$  . Let us suppose that composing
transfer functions take $\compComp$ time. Selecting the desired subset
in step three requires us to examine each element of the bin and compare
with its demand vector. Thus, $h.\lbound^K$ comparisons, each taking
$K.log(\lbound)$ steps is necessary. Assuming each join takes time
$\compJoin$ time, we can carry out the resulting join in time 
$h.\lbound^K.\compJoin$. Finally, we need to compare the resulting
function with the candidate taking $\compCompare$ time.
Thus, the time spent at each candidate is proportional to
$$ O(K.log(\lbound)) + \compComp + O(h.\lbound^K.K.log(\lbound) + h.\lbound^K.\compJoin + \compCompare)$$
Thus, the time spent in each round $i$ is bounded by
$$
O(\Delta.h.\lbound^K. [K.log(\lbound) + \compComp + h.\lbound^K.K.log(\lbound) + h.\lbound^K.\compJoin + \compCompare]) $$
where $\Delta$ is the total number of transitions.
Finally, there are at most $\lbound$ rounds and thus the over all complexity is bounded by
$$
\Delta.h.\lbound^{K+1}. [K.log(\lbound) + \compComp + h.\lbound^K.K.log(\lbound) + h.\lbound^K.\compJoin + \compCompare]) $$
which is 
$$ O(\Delta.h.\lbound^{K+1}[\compComp + h.\lbound^K.(K.log(\lbound)+\compJoin) + \compCompare]) $$
where $\lbound = (Q.(h+1).2)^{(3K)! + 1}$. Thus, for a fixed function lattice
where lattice operations take constant time, the complexity simplifies to
$$ O(\Delta.h^2.\lbound^{2K+1}.K.log(\lbound)) $$

\subsection{Correctness Proof for Forward DFAS}

We first present the important definitions required for the proof.

The complete lattice $D$ that we for
this purpose is defined as follows:\\[1em]
\hspace*{1cm} $ D \ \equiv \ D_{r,\forwthresh} \rightarrow \dlattice$\\

The ordering on this lattice is as follows: $(d_1 \in D) \sqsubseteq (d_2
\in D)$ iff $\forall c \in D_{r,\forwthresh}.\ d_1(c)
\sqsubseteq_\dlattice d_2(c)$.

The define the abstract trasnfer function we first define a
ternary relation $\boundedmovei$ as follows. Any triple $(p, q, s) \in \boundedmovei$
iff
$$
\begin{array}{lr}
\ (0 \leq p \leq \forwthresh) \ \wedge \\
\ ( (q \geq 0 \wedge p + q \leq \forwthresh \wedge s = p + q)\: \vee
& \hspace*{2cm} \mathit{(a)}\\ 
\ \ \: (q \geq 0 \wedge p + q > \forwthresh \wedge s = \forwthresh)\:
\vee & \hspace*{2cm} \mathit{(b)}\\
\ \ \: (q < 0 \wedge p = \forwthresh \wedge 0 \leq s \leq \forwthresh
\wedge \forwthresh - s \leq -1*q) \: \vee& \hspace*{2cm} \mathit{(c)}\\
\ \ \: (q < 0 \wedge p < \forwthresh \wedge p + q \geq 0 \wedge s = p +
q)) & \hspace*{2cm} \mathit{(d)}
\end{array}
$$

We now define a relation $\boundedmove$ on vectors. 
A triple of vectors $(c_1, c_2, c_3)$ belongs to 
relation $\boundedmove$ iff all three vectors are of the same size, and for
each index $i$, $(c_1[i], c_2[i], c_3[i]) \in \boundedmovei$.

The $D \rightarrow D$ transfer function for the VCFG edge
$q_1 \vcfgtrans{f}{w} q_2$ is given as follows:
$$ \fun(l \in D) \ \equiv \ \lambda c_2 \in D_{r,\forwthresh}. \,
\left(\bigsqcup_{c_1 \mathrm{\ such\ that\ } (c_1,w,c_2) \in \boundedmove}
f(l(c_1))\right) $$

As we intend to prove the correctness by adopting the correctness proof technique of abstract
interpretation to argue the soundness of our
approach, we fist present the ``concrete '' lattice and transfer function.  Let $D_r$ be the set of all vectors of size $r$ of
natural numbers.  Consider the ``concrete'' lattice $D_c \ \equiv\ D_r
\rightarrow \dlattice$, and the following ``concrete'' transfer function
for the VCFG edge $q_1 \vcfgtrans{f}{w} q_2$:

$$ \funconc(l \in D_c)
\ \equiv \ \lambda c_2 \in D_r. \, \left(\bigsqcup_{c_1 \in D_r
	\mathrm{\ such\ that\ } c_1 + w = c_2} f(l(c_1))\right)$$.

We now prove that the function $\fun$ is a \emph{consistent abstraction} of the function $\funconc$. For that we define the following function $\gamma$

$$\gamma(l \in D) \ \equiv \ \lambda c \in D_r. \,l(\mathit{min}(c,\vec{\forwthresh}))$$

where $\vec{\forwthresh}$ is a vector of size $r$ all of whose elements are equal to $\forwthresh$.  

We need to prove that for any
$d \in D$, $\gamma(\fun(d)) \sqsupseteq \funconc(\gamma(d))$. As discussed in the paper, this will be sufficient to prove the correctness.

Let $d$ be any bounded queue configuration in $D_{r, \forwthresh}$, and let $d_1$ be any
configuration in $\mathbb{N}^r$ such that $\mathit{min}(d_1, \forwthresh)$ is equal to $d$. Let $d_2$ be the
concrete successor of $d_1$ (if one exists) along the VCFG edge $t$. It is easy
to see from the definitions of the transfer functions that to prove consistent abstraction,  it is enough to prove that $(d, w, \textit{min}(d_2, \vec{\forwthresh}))$ is in $\boundedmove$, where $w$ is the queuing vector of t.

To prove this, let us assume for simplicity that there is a single
counter. Therefore, $d$, $d_1$, $d_2$, and $w$ all are integers.  Also, in this case $\vec{\forwthresh}$ can be written as $\forwthresh$ itself. The generalization
to multiple counters is easy, since the argument actually applies to each
counter individually.

\begin{itemize}
	\item  Case $(w \geq 0$ and $d_1 + w \leq \forwthresh)$: In this case, $d_1$ is $\leq \forwthresh$ and $d_2 = d_1 + w$. In this case, $d = d_1$, and $\mathit{min}(d_2, \forwthresh) = d_2$. Hence by definition, $(d, w, d_2) \in \boundedmove$.  Therefore, $(d, w, \mathit{min}(d_2, \forwthresh)) \in \boundedmove$
	
	\item  Case $(w \geq 0$ and  $d_1 + w > \forwthresh)$: We have $d_2 = d_1 + w$. In this case, $(d, w, \forwthresh) \in \boundedmove$. As $\mathit{min}(d_2, \forwthresh) = \forwthresh$, we have proved that $(d, w, \mathit{min}(d_2, \forwthresh)) \in \boundedmove$ .
	
	\item Case $(w < 0$ and $d_2 \geq \forwthresh)$: Clearly, we have $d_1 \geq \forwthresh$. Therefore, $d = \mathit{min}(d_1, \forwthresh) = \mathit{min}(d_2, \forwthresh) = \forwthresh$, and $(\forwthresh, w, \forwthresh) \in Pre$. Therefore, $(d, w, \mathit{min}(d_2, \forwthresh)) \in \boundedmove$.

	\item Case $(w < 0$ and $d_2 < \forwthresh$ and $ d_1 \geq \forwthresh)$: Since $d_2 = d_1 + w$, we have $\forwthresh - d_2 \leq -1*w$.  Also, $(\forwthresh, w, d_2) \in \boundedmove$, and $d = \mathit{min}(d_1, \forwthresh) = \forwthresh$ and
	$\mathit{min}(d_2, \forwthresh) = d_2$. Therefore, $(d, w, \mathit{min}(d_2, \forwthresh)) \in \boundedmove$.

	\item Otherwise $(w < 0$ and  $d_1 < \forwthresh$ and $ d_2 < \forwthresh)$: In this case  we have $(d_1, w, d_2) \in \boundedmove$, and $\mathit{min}(d_2, \forwthresh) = d_2$ and $\mathit{min}(d_1, \forwthresh) = d_1$.  Therefore, $(d, w, \mathit{min}(d_2, \forwthresh)) \in \boundedmove$.
	
	\emph{Hence Proved.}
\end{itemize}

\bibliographystyle{splncs04}
\bibliography{popl20}




\vfill

{\small\medskip\noindent{\bf Open Access} This chapter is licensed under the terms of the Creative Commons\break Attribution 4.0 International License (\url{http://creativecommons.org/licenses/by/4.0/}), which permits use, sharing, adaptation, distribution and reproduction in any medium or format, as long as you give appropriate credit to the original author(s) and the source, provide a link to the Creative Commons license and indicate if changes were made.}

{\small \spaceskip .28em plus .1em minus .1em The images or other third party material in this chapter are included in the chapter's Creative Commons license, unless indicated otherwise in a credit line to the material.~If material is not included in the chapter's Creative Commons license and your intended\break use is not permitted by statutory regulation or exceeds the permitted use, you will need to obtain permission directly from the copyright holder.}

\medskip\noindent\includegraphics{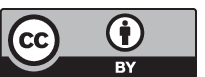}

\end{document}